\documentclass{aa} 

\usepackage{graphicx}
\usepackage{txfonts}
\usepackage[colorlinks = true,citecolor = blue, linkcolor = blue, urlcolor = blue]{hyperref}
\usepackage{tabularx,booktabs}
\usepackage{multirow}
\usepackage{subcaption}

\newcolumntype{Y}{>{\arraybackslash}X}
\newcolumntype{x}{>{\arraybackslash}p}
\newcolumntype{m}{>{\centering\arraybackslash}X}
\begin{document}

        \title{A neural network-based methodology to select young stellar object candidates from IR surveys
   \thanks{The full catalog of young stellar objects and contaminants
   is only available at the CDS via anonymous ftp to cdsarc.u-strasbg.fr 
   (130.79.128.5) or via 
   http://cdsarc.u-strasbg.fr/viz-bin/qcat?J/A+A/???/A??}}
   \titlerunning{A Neural Network-based methodology to select YSOs}


        \author{D. Cornu
            \inst{1,2}
            \and
            J. Montillaud\inst{1}
            }

   \institute{Institut UTINAM - UMR 6213 - CNRS - Univ. Bourgogne Franche Comté, OSU THETA, 41bis avenue de l'Observatoire, 25000, Besançon, France
   \and
   Sorbonne Université, Observatoire de Paris, PSL research university, CNRS, LERMA, F-75014 Paris, France}

   \date{Received 28 Mai 2020; accepted 07 December 2020}

 
  \abstract
   {Observed young stellar objects (YSOs) are used to study star formation and characterize star-forming regions. For this purpose, YSO candidate catalogs are compiled from various surveys, especially in the infrared (IR), and simple selection schemes in color-magnitude diagrams (CMDs) are often used to identify and classify YSOs.}
   {We propose a methodology for YSO classification through machine learning (ML) using Spitzer IR data. We detail our approach in order to ensure reproducibility and provide an in-depth example on how to efficiently apply ML to an astrophysical classification.}
   {We used feedforward artificial neural networks (ANN{s}) that use the four IRAC bands ($3.6,\ 4.5,\ 5.8$, and $8\ \mu m$) and the $24\ \mu m$ MIPS band from Spitzer to classify point source objects into CI and CII YSO candidates or as contaminants. We focused on nearby ($\lesssim 1$ kpc) star-forming regions including Orion and NGC 2264, and assessed the generalization capacity of our network from one region to another.}
   {We found that ANN{s} can be efficiently  applied to YSO classification with a contained number of neurons ($\sim$ 25). Knowledge gathered on one star-forming region has shown to be partly efficient for prediction in new regions. The best generalization capacity was achieved using a combination of several star-forming regions to train the network. Carefully rebalancing the training proportions was necessary to achieve good results. We observed that the predicted YSOs are mainly contaminated by under-constrained rare subclasses like Shocks and polycyclic aromatic hydrocarbons (PAHs), or by the vastly dominant other kinds of stars (mostly on the  main sequence). We achieved above 90\% and 97\% recovery rate for CI and CII YSOs, respectively, with a precision above 80\% and 90\% for our most general results. We took advantage of the  great flexibility of ANNs to define, for each object, an effective membership probability to each output class. Using a threshold in this probability was found to efficiently improve the classification results at a reasonable cost of object exclusion. With this additional selection, we reached 90\% and 97\% precision on CI and CII YSOs, respectively, for more than half of them. Our catalog of YSO candidates in Orion (365 CI, 2381 CII) and NGC 2264 (101 CI, 469 CII) predicted by our final ANN, along with the class membership probability for each object, is publicly available at the CDS.
   }
   {Compared to usual CMD selection schemes, ANNs provide a possibility to quantitatively study the properties and quality of the classification. Although some further improvement may be achieved by using more powerful ML methods, we established that the result quality depends mostly on the training set construction. Improvements in YSO identification with IR surveys using ML would require larger and more reliable training catalogs, either by taking advantage of current and future surveys from various facilities like VLA, ALMA, or Chandra, or by synthesizing such catalogs from simulations.}


   \maketitle
%

%

\section{Introduction}
\label{intro}

   Observing young stellar objects (YSOs) in stellar clusters is a common strategy to characterize star-forming regions. Their presence attests star formation activity, their spatial distribution within a molecular complex provides clues about its star formation history \citep{Gutermuth_2011}, and their surface density can be used as a measure of the local star formation rate \citep{heiderman_star_2010}. They have recently been combined with astrometric surveys like Gaia to recover the 3D structure and motion of star-forming clouds \citep{grossschedl_3d_2018}. Their identification is often summarized as a classification problem. Such a classification relies mainly on their spectral energy distribution (SED) in the infrared (IR), and { makes it possible} to distinguish evolutionary steps that range from the star-forming phase to the main sequence \citep[class 0 to III,][]{Lada87, allen_infrared_2004}. This subclassification provides additional information on the structure and evolution of star-forming regions.

   Modern observation missions produce highly challenging  datasets with an unprecedented number of objects characterized by many parameters. The most common example is Gaia DR2 \citep{brown_gaia_2018} with almost 1.7 billion  observed stars. To handle such a large amount of data, modern industrial-grade solutions were adopted, like Hadoop \citep{chansler_hadoop_2010}, that are mainly used by the  ``Big Tech'' companies (e.g., Amazon, Google, Microsoft) to manage and maintain their databases.

   Regardless of their capacity to produce the data, these large observed catalogs become almost impossible to analyze with the usual algorithm schemes because they often scale poorly with the data size and do not allow easy integration or visualization in many dimensions. In this context astronomers are increasingly getting involved in powerful and automated statistical approaches like machine learning \citep[ML; e.g.,][used to analyze the Galaxy Zoo survey]{Huertas-Company_2011}. This family of methods takes advantage of large dataset sizes to construct a generalization of the problem to solve, in any number of dimensions. They can be supervised to take advantage of a priori knowledge, or unsupervised to combine information in a new way. They need a training phase that relies on a large number of objects to learn the generalization, but they scale nicely with the number of dimensions and objects. Once trained, they are able to provide answers more quickly than most common analysis tool. They are therefore sometimes proposed as accelerators for various physical problems \citep[for example to replace a numerical solver for three-body problems in][]{breen_2020}. In addition to those properties, they are able to solve a large variety of problems like classification, regression, clustering, time series prediction, compression, image recognition. The most emblematic ML methods are certainly artificial neural networks (ANNs), which stand out because of their unique flexibility. Among their numerous advantages, they scale well with the number of dimensions, and are able to solve a large variety of problems with only slight adjustments. They are intuitive to construct and use, and even if it can take a long time to train them, once trained they are very computationally efficient when making predictions in comparison to other ML methods. ANNs have been successfully applied in astronomy for a large variety of predictions, including galaxy type classification \citep[e.g.,][]{dieleman_2015,Huertas-Company2015}, computation acceleration \citep{Grassi_2011, Mijolla_2019}, and ISM turbulent regime classification \citep{Peek_2019}. However, these powerful methods require special care as they are strongly sensitive to the construction of the training sample and to the tuning of their parameters. Estimating the quality of the results is a somewhat subtle task, and the proper tools must be used for their representation and interpretation.

   In this context {it should be possible to design a classification method for YSOs that relies} on current and future large surveys and that take advantage of ML tools. This has notably been attempted by \citet{marton_all-sky_2016, marton_2019} and \citet{miettinen_protostellar_2018}. The study by \citet{marton_all-sky_2016} uses supervised ML algorithms called support-vector machines (SVMs) applied to the mid-IR ($3 - 22\ \mu$m) all-sky data of the Wide-field Infrared Survey Explorer \citep[WISE;][]{wright_wide-field_2010}. {Overall, the SVM method offers great performance} on linearly separablde data. However, it is not able to separate more than two classes at the same time, and does not scale as well with the number of dimensions as other methods do. The full-sky approach in the Marton study produces large YSO candidate catalogs, but suffers from the uncertainty and artifacts in star-forming regions of the WISE survey \citep{lang_unwise:_2014}. Additionally, the YSO objects used for the training were identified using SIMBAD, resulting in a strong heterogeneity in the reliability of the training sample.
In their subsequent study, \citet{marton_2019} added Gaia magnitudes and parallaxes to the study. Gaia is expected to add a large statistical sample and to complete the spectral energy distribution coverage, but the necessary cross-match between Gaia and WISE excludes most of the youngest and embedded stars. The authors also compared the performance of several ML algorithms (e.g., SVMs, neural networks, random forest), and reported the random forest to be the most efficient with their training sample. This is a better solution as it overcomes the  limitations of the SVM. However, as in their previous study, the training sample compiles objects from different identification methods including SIMBAD. This adds more heterogeneity and is likely to increase the lack of reliability of the training sample, despite the use of a larger training sample.

        \citet{miettinen_protostellar_2018} adopts a different approach by compiling a large number of ML methods applied on reliably identified YSOs using ten photometric bands ranging from $3.6$ to $870\ \mu m$. For this he used the Herschel Orion Protostar Survey \citep{HOPS_2013}, resulting in just less than $300$ objects. Such a large number of input dimensions combined with a small learning sample is often highly problematic for most ML methods. Moreover, this study focuses on the subclass distinction of YSOs and does not attempt to extract them from a larger catalog that contains other types of objects. In consequence, it cannot be generalized to currently available large surveys, and relies on a prior YSO candidate selection.

        In the present study we propose a YSO identification and classification method based on ML, and capable of taking advantage of present and future large surveys. We selected ANNs for their qualities, as stated above, and because they can identify several classes at the same time. To build our training sample, we used a simplified version of the popular method by \citet{gutermuth_spitzer_2009}, a multistep classification scheme that combines data in the J, H, and K$_s$ bands from the Two Micron All Sky Survey \citep[2MASS;][]{skrutskie_two_2006} and data between 3 and 24 $\mu$m from the Spitzer space telescope \citep{werner_spitzer_2004}. By using Spitzer data we expect to cover only specific regions on the sky, but with a better sensitivity ($ \approx~1.6$ to $27\ \mu J$ for the IRAC instrument) and spatial resolution ($1.2\arcsec$) than WISE ($ \approx~80$ to $6000\ \mu J$ and $ 6.1\arcsec$ to $12\arcsec$). In this paper we show results based on three different datasets, namely, Orion \citep{megeath_spitzer_2012}, NGC 2264 \citep{rapson_spitzer_2014}, and a sample of clouds closer than 1 kpc provided by \citet{gutermuth_spitzer_2009} that excludes the first two   regions. We took advantage of ANNs to go beyond the capabilities of simple selection schemes in color-magnitude diagrams, like that by \citet{gutermuth_spitzer_2009}, by quantitatively studying the classification characteristics and quality. We adopted a bottom-up approach to slowly increase the complexity and diversity toward a more general classification.

    A secondary goal of the paper is to provide a guided example of how to efficiently apply ANNs to a classification problem, with an effort to make our results reproducible. Therefore, we present how we constructed our ANN in Sect.~\ref{sec:method}, where we also present the various difficulties that are inherent to expressing the problem in a form that is solvable by an ANN. In Sect.~\ref{sec:data_prep_chap} we detail the data preparation phase and our choice of representations for the results along with their analysis, presenting the encountered limitations. We discuss the results in detail in Sect.~\ref{sec:results_header}, using different star-forming regions, and the observed specific behavior for each of them. Our best results are publicly available at the CDS. Finally, in Sect.~\ref{sec:discussion} we discuss the caveats and potential improvements of our methodology, and propose a probabilistic characterization of our results, which is included in our public catalog.

\section{Deep learning method}
\label{sec:method}

Deep learning methods are based on ANNs, a supervised ML approach. As mentioned before, it is able to iteratively learn a statistical generalization from a previously labeled dataset. At the end of the training phase it should be able to retrieve the expected outputs from unseen data in most cases, allowing us to estimate the quality of the learned generalization. An extensive introduction can be found in \citet{Bishop:2006:PRM:1162264} or \citet{MarslandBook2} that relies on several reference papers, including \citet{rosenblatt_perceptron:_1958}, \citet{rumelhart_learning_1986}, \citet{rumelhart_parallel_1986}, and \citet{widrow_30_1990}. We summarize in this section the elements on which the present study is based.

\subsection{Deep artificial neural networks: Multilayer perceptron}
\label{deep_ann}

We adopted the widely used ANN architecture and training procedure of the multilayer perceptron (MLP). It consists of multiple layers of neurons, each of which is connected to all the neurons of the previous layers up to an input layer that contains the input features of the problem \citep{rumelhart_parallel_1986}. Each neuron performs a weighted sum of the previous layer values and computes an activation function from it, defining the value of the neuron. In the present work we use the common sigmoid activation function
\citep{rumelhart_parallel_1986}
\begin{equation}
        \centering
        g(h) = \frac{1}{1+ \exp(-\beta h)}
        \label{eq_sigm}
,\end{equation}
where $h$ is the weighted sum of the previous layer values and $\beta$ is a positive hyperparameter that defines the steepness of the curve. This $S$-shaped function, with results between $0$ and $1$, emulates the overall binary behavior of neurons but with a continuous activation.

The multilayer architecture allows the network to combine sigmoid functions in a non-linear way, each layer increasing the complexity of the achievable generalization. The combination of sigmoid functions can be used to represent any function, which means that such a network is a universal function approximator, as demonstrated by \citet{cybenko_approximation_1989}. They also showed that a single hidden layer with enough neurons is able to approximate any function as accurately as an arbitrarily deep network (universal approximation theorem). 

Training the neurons consists in finding a suitable set of weight values. This is achieved in an iterative fashion by comparing the output layer activation with the expected output regarding the current input object. This is done using an error function at the output layer that is used to correct the output layer weights. Since there is no direct comparison possible for the hidden layers, the output error must be propagated through the network using the ``backpropagation'' algorithm \citep{rumelhart_parallel_1986}, which computes an error gradient descent through the entire network. For the output we used the sum of square difference. The weight corrections for a given layer $l$ are then computed as 
        \begin{equation}
        \omega_{ij} \leftarrow \omega_{ij} - \eta \frac{\partial E}{\partial \omega_{ij}}
        \label{eq_grad_desc}
        ,\end{equation}
where $\omega_{ij}$ is the weight matrix of the present layer and $\eta$ is a learning rate that scales the updates, and where the gradient $\frac{\partial E}{\partial \omega_{ij}}$ can be expanded as
        \begin{equation}
        \frac{\partial E}{\partial \omega _{ij}} = \delta _l (j) \frac{\partial h_j}{\partial \omega_{ij}} 
        \quad \text{with} \quad
        \delta _l (j) \equiv \frac{\partial E}{\partial h_j} = \frac{\partial E}{\partial a_j}\frac{\partial a_j}{\partial h_j}.
        \label{eq_update_full_network}
        \end{equation}
In these equations the indices $i$ and $j$ run through the number of input dimensions of the current layer and its number of neurons, respectively. These equations are the same for each layer. The quantity $\delta_l$ defines a local error for each layer of neurons so that, for the hidden layer $l$, the error $E$ in eqs.~(\ref{eq_grad_desc}) and (\ref{eq_update_full_network}) is replaced by $\delta_{l+1}$. It also depends on the activation function $a=g(h)$ at each layer through the derivative $\partial a_j/\partial h_j$.
Thus, this kind of gradient can be evaluated for an arbitrary number of layers.

\subsection{Adopted ANN} 
\label{final_ann}

\begin{figure*}[t]
\centering
\includegraphics[width=0.77\hsize]{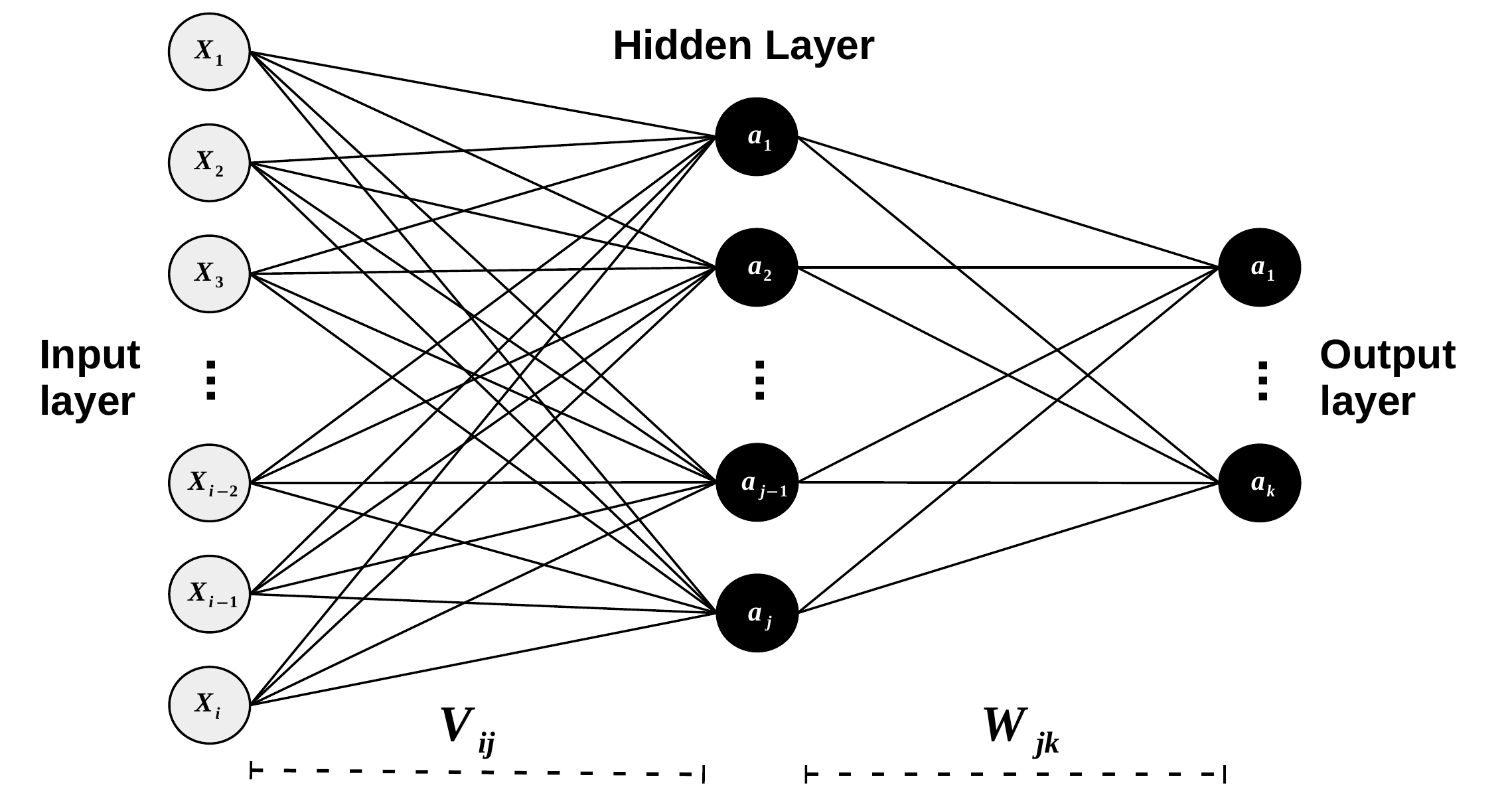}
\caption{Schematic view of a simple neural network with only one hidden layer. The light dots are input dimensions. 
The black dots are neurons with the linking weights represented as continuous lines. Learning with this network relies on eqs.~(\ref{eq_sigm}) to (\ref{eq_update_first}). $X_{[1,\dots,i]}$ are the dimensions for one input vector, $a_{[1,\dots,j]}$ are the activations of the hidden neurons, $a_{[1,\dots,k]}$ are the activations of the output neurons, while $V_{ij}$ and $W_{jk}$ represent the weight matrices between the input and hidden layers, and between the hidden and output layers, respectively.}
\label{fig_network}
\end{figure*}

In the present paper our aim is to classify young stellar objects. We detail here the choices we adopted to solve this problem for the general architecture of the network and for the activation functions.

   Because of the universal approximation theorem (Sect.~\ref{deep_ann}), we chose to use only one hidden layer. For the output layer we adopted as many output neurons as the number $o$ of classes to distinguish, so that each class is encoded by the dominant activation of one output neuron. In other words, if the activation values of the output layer are $(a_1,\dots,a_o)$ and $a_i>a_k$ for all $k\neq i$, then the class predicted by the network is the $i$-th class.

   For the hidden layer we adopted the sigmoid activation function (see previous section). For the output layer, we preferred the softmax activation function, also known as the normalized exponential activation,
        \begin{equation}
        \centering
        a_k = g(h_k) = \frac{\exp(h_k)}{\sum_{k^\prime=1}^{o}{\exp(h_{k^\prime})}}
        \label{eq_activ_softmax}
        ,\end{equation}
where $k$ is the neuron index in the output layer. Thanks to the normalization over all the output neurons, the $k$-th output neuron provides a real value between zero and one, which acts as a proxy for the membership probability of the input object in the $k$-th class. This gives  the network some attributes of a probabilistic neural network \citep[PNN;][]{specht_probabilistic_1990, stinchcombe_universal_1989}, although it often fails to represent a genuine physical probability. We discuss in Sect.~\ref{proba_discussion} how these outputs can be used to estimate the reliability of the predicted class of each object and to point out the degree of confusion between multiple classes for the algorithm.

Our final network is composed of an input layer constrained by the input dimensions $m$ of our problem; a hidden layer with a tunable number of neurons $n$; and an output layer with $o$ neurons, one for each output class, with a softmax activation. Figure~\ref{fig_network} presents a general illustration of this common architecture. The gradient descent is computed from the backpropagation equations (eqs.~(\ref{eq_grad_desc}) and (\ref{eq_update_full_network})) as follows. The local error $\delta_o(k)$ of the $k$-th output neuron, with the softmax activation, is computed using
        \begin{equation}
        \centering
        \delta_o(k) = (a_k - t_k)a_k(1-a_k).
        \label{eq_deltao}
        \end{equation}
The obtained values are used to derive the local error for neurons in the hidden layer, with the sigmoid activation
        \begin{equation}
        \centering
        \delta_h(j) =  \beta a_j(1-a_j) \sum_{k=1}^{o}{\delta_o(k)\omega_{jk}}
        \label{eq_deltah}
        ,\end{equation}
where $j$ is the index of a hidden neuron. Once the local errors are computed,
the weights of both layers are updated using
        \begin{eqnarray}
        \centering
        \omega_{jk} & \leftarrow & \omega_{jk} - \eta \delta_o(k)a_j
        \label{eq_update_second},
        \\
        v_{ij} & \leftarrow & v_{ij} - \eta \delta_h(j)x_i
        \label{eq_update_first}.
        \end{eqnarray}
Here $\omega_{jk}$ and $v_{ij}$ denote the weights between the hidden and output layers, and between the input and hidden layers, respectively; $a_j$ is the activation value of the $j$-th hidden neuron; and $x_i$ is the $i$-th input value.

Equations~(\ref{eq_deltao}) to (\ref{eq_update_first}) also show that the particular vectors $x_i=0$ or $a_i=0$ for all $i$ are pathological points; in this case the weighted sum $h$ is independent of the weights, and the weight correction is always null, regardless of the error function or its propagated value in the present layer. To circumvent this peculiarity, one approach consists in adding an extra value to the input vector, fixed to $X_{m+1}=-1$, and connected to the neuron by a free weight $\omega_{m+1}$, which behaves as any other weight. Because the input values $X_i$ are often called input nodes, this additional input dimension is generally referred to as the ``bias node.'' The additional degree of freedom provided by the bias node enables the neuron to behave normally when $X_i=0$ for $1 \le i \le m$. {Because all layers work the same way, they are all extended with an individual bias node}.

\subsection{Additional network properties and hyperparameters}
\label{optim_perf}

\subsubsection{Properties of the learning rate}

Lower values of the learning rate $\eta$ increase the stability of the learning process, at the cost of lower speed and a higher chance for the system to get stuck in a local minimum. Conversely, higher values increase the speed of the learning process and its ability to roam from one minimum to another, but values that are too large might prevent it from converging to a good but narrow minimum. In this study we adopt values in the range $\eta = 3 - 8 \times 10^{-5}$, but in our formalism the correction depends on the number of elements shown before applying the correction (see also Sect.~\ref{network_tuning}). The correction also scales with the input value $x_i$ and $a_i$ in eqs.~(\ref{eq_update_second}) and~(\ref{eq_update_first}), respectively, correcting more the weights of the inputs that are more responsible for the neuron activation.

\subsubsection{Weight initialization}

The initial state of the weights impacts the convergence speed. \citet{rumelhart_parallel_1986} proposed to set them at small random values. We used random initializations in the range $ -1/\sqrt{N} < \omega < 1/\sqrt{N}$, where $N$ is the number of nodes in the input layer. This breaks the symmetries that would occur if weights were initialized to zero, and it guarantees that the weights are large enough to learn, and small enough to avoid divergence of the weights when the error of the neuron is large. Details on the initialization and on recent and efficient alternatives  (e.g., the He-et-al or Xavier methods) are discussed by \citet{glorot_understanding_2010} and \citet{he_delving_2015}.

\subsubsection{Additional optimizations}

In order to get the best results from our network we added various optimizations. To help avoid local minima, we added a weight momentum. This is a classical speeding up method for various iterative problems \citep{polyak_methods_1964, qian_momentum_1999}. It consists in adding a fraction of the previous weight update to the next one, during the training phase. This memory of the previous steps helps keep a global direction during the training especially in the first steps. It allows a faster training even when using a lower learning rate. It also helps reduce the spread between repeated trainings. It is controlled by an hyperparameter $0 < \alpha < 1$, the usual values being between $0.6$ and $0.9$.

\subsubsection{Training strategy and performance}

An important aspect of ML is how the data are presented to the network \citep{wilson_general_2003}. Since the training set must be shown numerous times, the order of the objects and the frequency of the weight updates are two important parameters. The simplest way is to show the objects one by one and to update the weights for each object, shuffling the dataset after each epoch. Another classical way to train is the batch method, where the complete dataset is shown at once, and the weights are updated after each batch. This method is easy to implement within a matrix formalism, but since the updates are summed over the whole dataset, the dilution of rare inputs tends be a more salient issue than with other training methods. A popular alternative is the stochastic gradient descent, where input objects are randomly drawn with replacement from the learning dataset. This is a powerful way to avoid local minima and to converge more quickly with fewer objects. However, most implementations of neural networks choose the Combined method, called mini-batch, where the dataset is split into equal parts, and the weights are updated after each part. In this scheme the training set is shuffled and the batches are redefined between each epoch.

We implemented these methods so that our code can work in the following modes: basic single-thread CPU, multithread OpenMP \citep{dagum_openmp:_1998}, multithread matrix OpenBLAS \citep{xianyi_model-driven_2012}, and  GPU CUDA accelerated \citep{Nickolls:2008:SPP:1365490.1365500}. Interestingly, although in principle the batch method is the least efficient  in terms of the number of times each object is seen, in practice it led to better performance than others when run on GPU, using a GTX 780. For example, with a training set of around $3\times 10^4$ objects with ten input dimensions, 25 hidden neurons, and three outputs, the training took less than 15 minutes for more than one million epochs. The results obtained by all methods were compared at various steps of this study, but none has significantly outperformed the others. For less heavy training, all the methods are effective enough to converge in a matter of seconds. The forward on millions of objects after the training is always a matter of seconds even for large datasets.

\section{Data preparation and network settings}
\label{sec:data_prep_chap}

In this section we detail how we connected the general network presented in the previous sections with the YSO classification problem. We show how we arranged the data in a usable form for the network and describe the needed precautions for this process. We also explain how we defined the various datasets used to train our network.

\subsection{Definition of the classes and the labeled sample}
\label{data_prep}

We summarize here the construction of the training sample based on a simplified version of the method by \citet[][hereafter G09]{gutermuth_spitzer_2009}, where only Spitzer data are used. In their original method they performed the classification in several steps. In addition to Spitzer they used 2MASS data, but mainly in additional steps to refine the classification of some objects. Therefore, restricting our analysis to Spitzer data still allowed a reasonable classification, with the advantage of using a simple and homogeneous dataset. In our adapted method, we started with the four IRAC bands, at $3.6,\ 4.5,\ 5.8$, and $8\ \mu m$, applying a  pre-selection that kept only the sources with a detection in the four bands and with uncertainties $\sigma < 0.2$ mag, as in the original classification. Similarly to G09, we used the YSO classes defined by \citet{allen_infrared_2004}. Class 0 objects (C0) are starless dense cores or deeply embedded protostars, mainly visible as  blackbody spectra in the far-IR, and are quiet in the mid-IR.
Class I objects (CI) are protostars that emit as blackbodies in the mid-IR and are dominated by the emission of an infalling envelope that induces a strong excess in the far-IR. Class II objects (CII) are pre-main sequence stars with a thick disk that flattens the emission in the far-IR. Class III objects (CIII) are pre-main sequence stars with or without a faint disk that are devoid of far-IR emission.

\begin{figure*}[t]
        \centering
        \begin{subfigure}[t]{0.33\textwidth}
        \includegraphics[width=\textwidth]{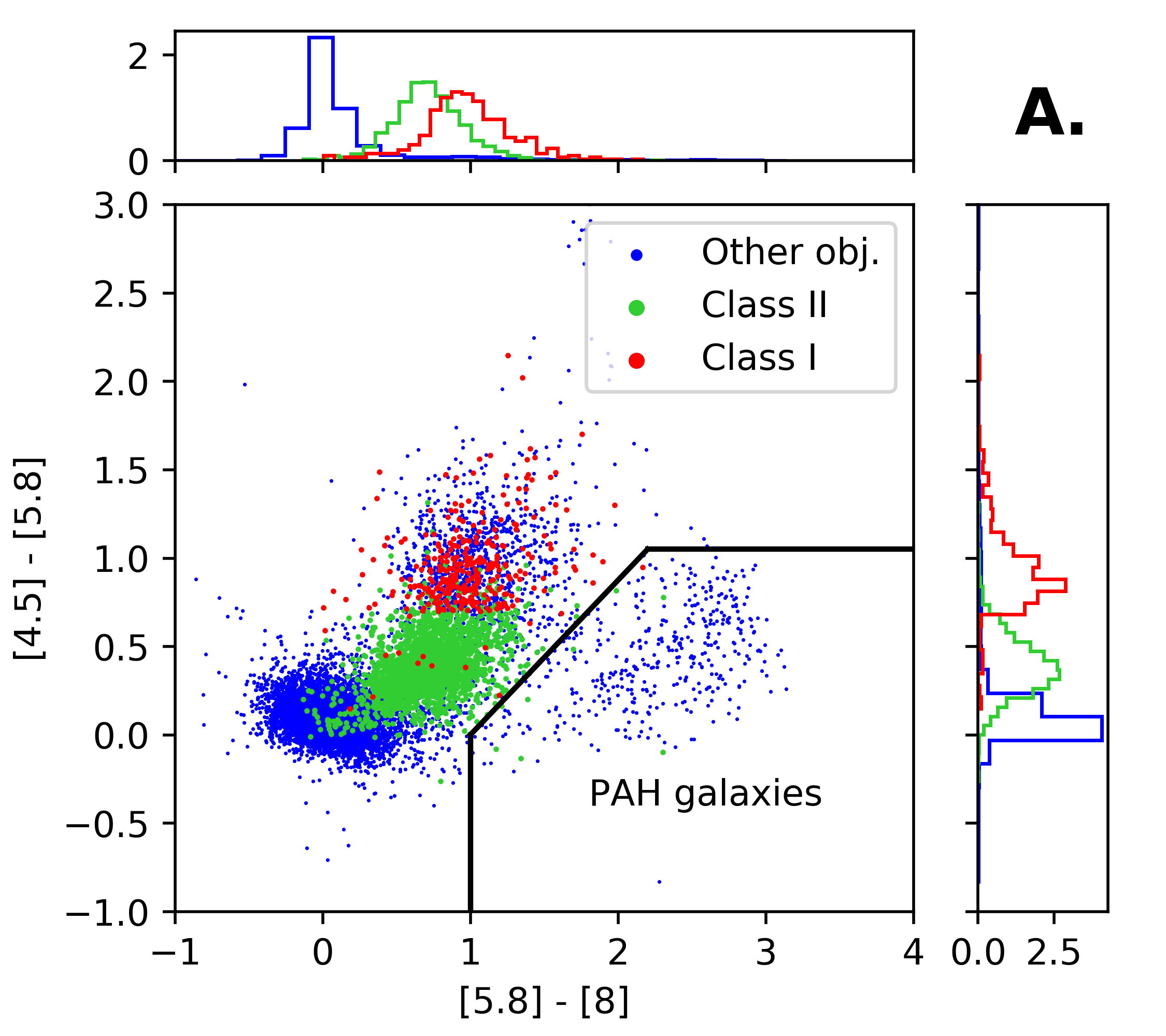}
        \end{subfigure}
        \begin{subfigure}[t]{0.33\textwidth}
        \includegraphics[width=\textwidth]{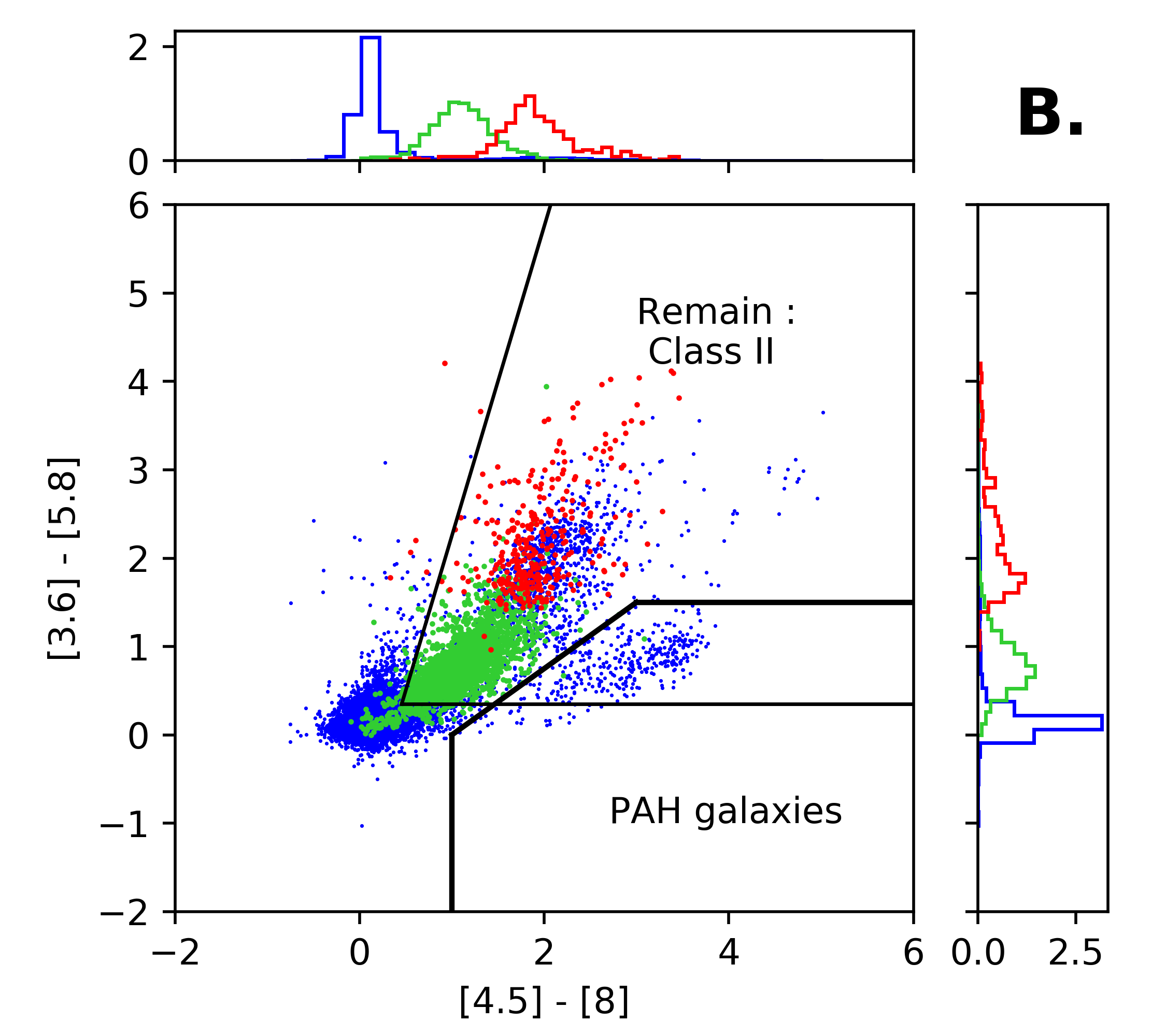}
        \end{subfigure}
        \begin{subfigure}[t]{0.33\textwidth}
        \includegraphics[width=\textwidth]{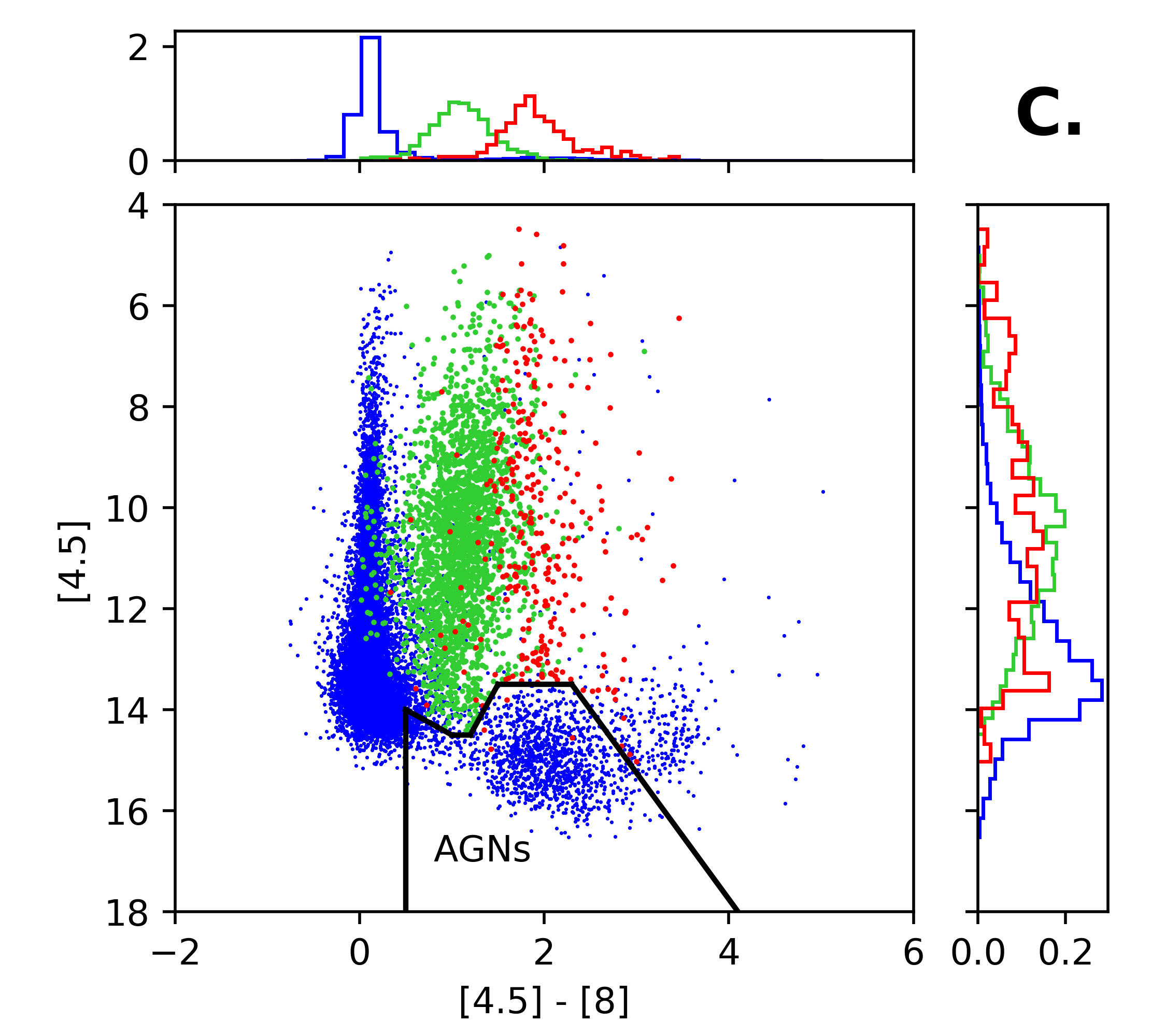}
        \end{subfigure}\\
        \begin{subfigure}[t]{0.33\textwidth}
        \includegraphics[width=\textwidth]{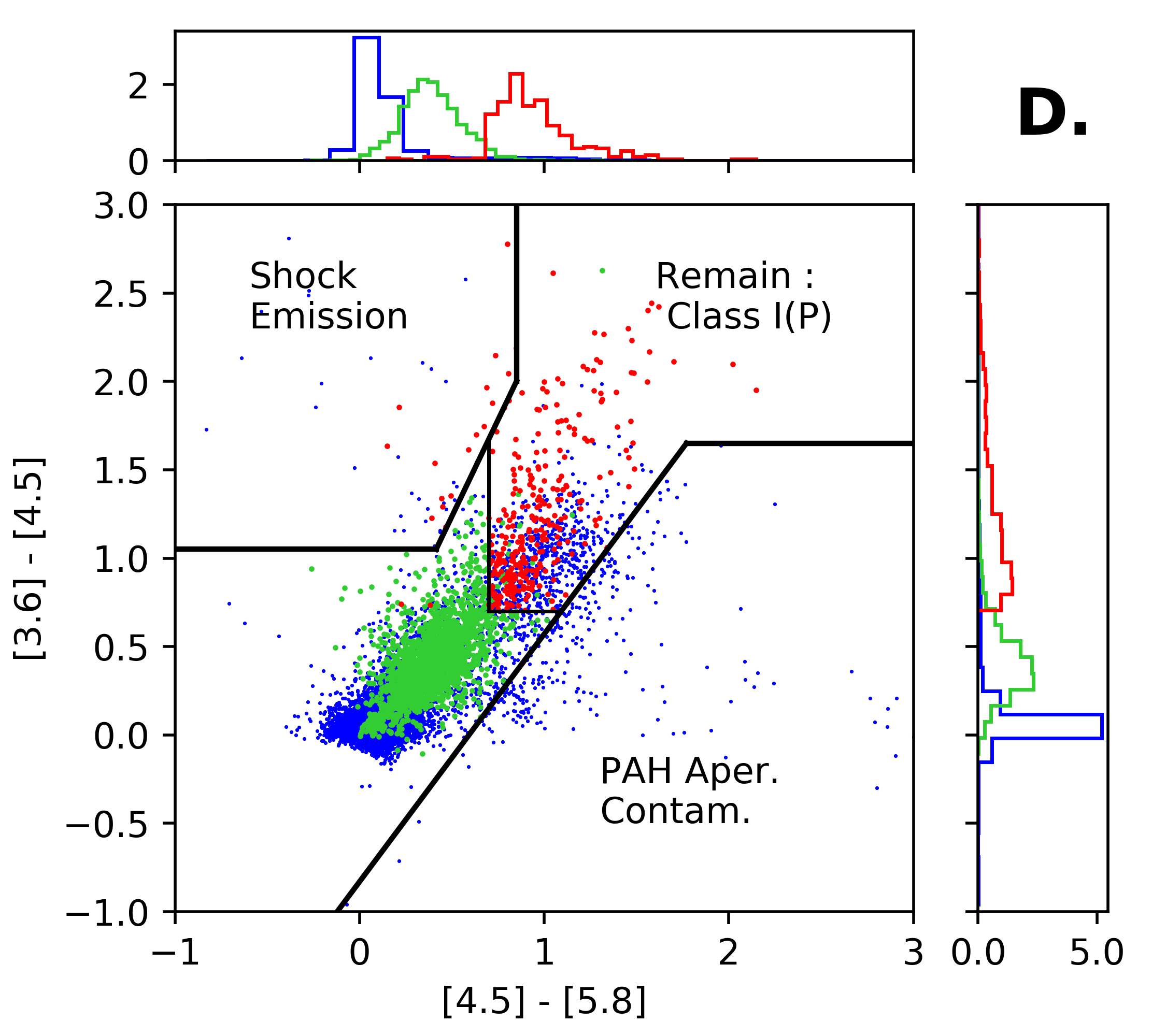}
        \end{subfigure}
        \begin{subfigure}[t]{0.33\textwidth}
        \includegraphics[width=\textwidth]{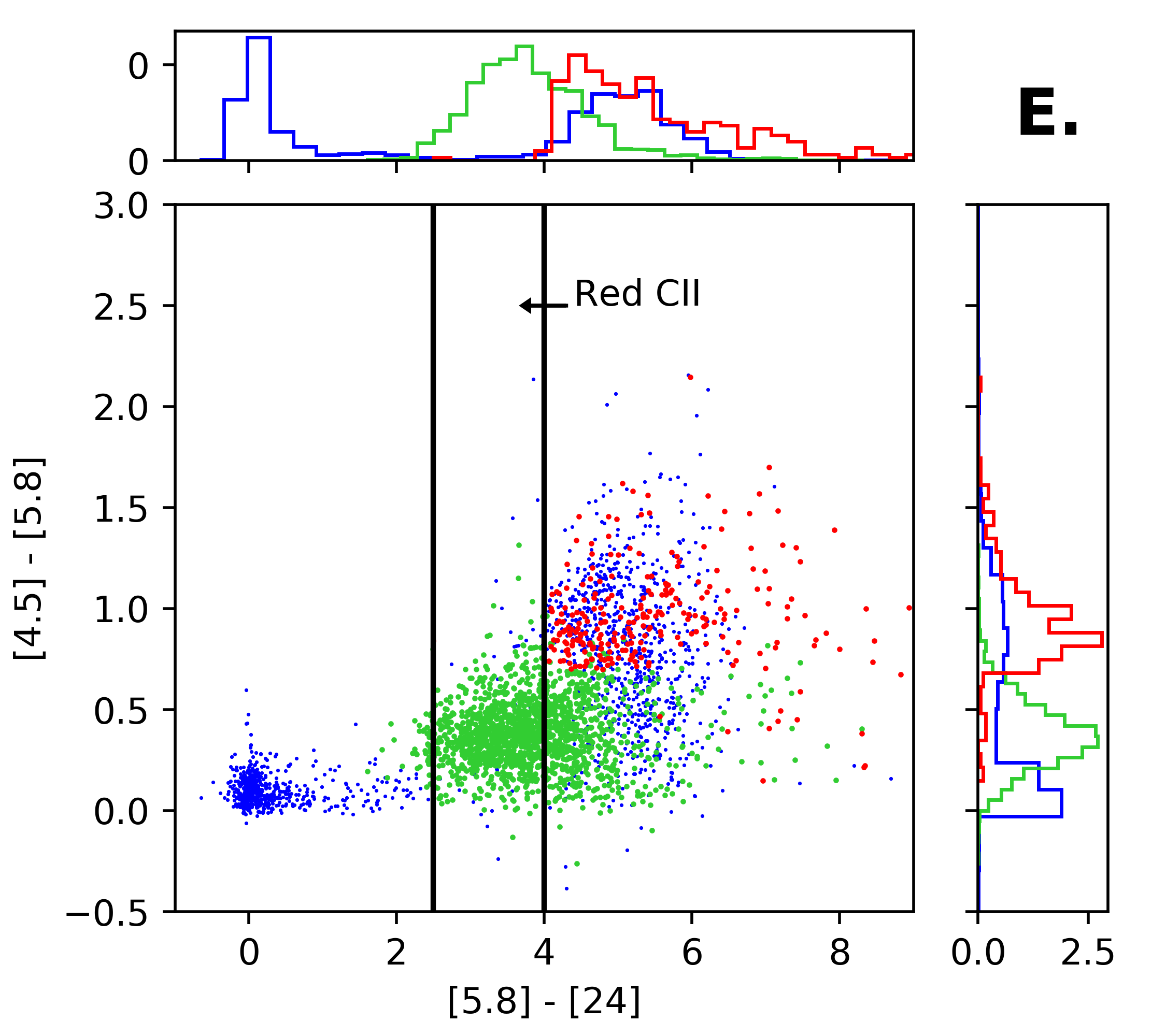}
        \end{subfigure}
        \caption{Selection of color-color and color-magnitude diagrams from our simplified multistep G09 classification. The data used in this figure correspond to the Orion labeled dataset in Table~\ref{tab_selection}. The contaminants, CII YSOs, and CI YSOs are shown in blue, green, and red, respectively. They are plotted in that order and partly screen one another, as revealed by the histograms in the side frames. The area of each histogram is normalized to one. In frame A, some PAH galaxies are excluded. In frame B, leftover PAH galaxies are excluded based on other criteria. It also shows the criteria of class II extraction that is in a later step. In frame C, AGNs are excluded. In frame D, Shocks and PAH aperture contaminants are excluded. It also shows the last criteria of class I extraction. In frame D, one of the criteria from the Multiband Imaging Photometer (MIPS) $24\ \mu m$ band is shown, which identifies reddened Class~II in the previously classified Class~I.}
        \label{fig_gut_method} 
\end{figure*}

Using solely IRAC data prevented us from identifying class 0 objects since they do not emit in the IRAC wavelength range. Similarly, because of Spitzer uncertainties, the class III objects are too similar to main sequence stars to be distinguished. For these reasons we limited our objectives to the identification of CI and CII YSOs. We then proceeded to ``phase~1'' from G09 (their Appendix~A) to successively extract different contaminants using specified straight cuts into color-color and color-magnitude diagrams (CMDs) along with their respective uncertainties. This step enabled us to exclude star-forming galaxies, active galactic nuclei (AGNs), shock emission, and resolved polycyclic aromatic hydrocarbon (PAH) emission. It ends by extracting the class I YSO candidates from the leftovers, and then again extracting the remaining class II YSO candidates from more evolved stars. The cuts used on these steps are shown in Fig.~\ref{fig_gut_method}, with the final CI and CII YSO candidates from the Orion region (Sect.~\ref{data_setup}).

For the sake of simplicity, we adopted only three categories: CI YSOs, CII YSOs, and contaminants, which we also refer to as ``Others'' in our tables, which forced the network to focus on the separation of the contaminant class from the YSOs, rather than between different contamination classes. Therefore, we defined the output layer of our network with three neurons using a softmax activation function, {corresponding to} one neuron per class that returned a membership probability (see Sect.~\ref{final_ann}).

Because we chose not to use 2MASS, we skipped ``phase~2'' of the G09 classification scheme. However, G09 also proposed a ``phase 3'', which  uses the MIPS $24\,\mu$m band and which might be useful for our classification. In this last phase some objects that were misclassified in the previous two phases are rebranded. Although this can raise difficulties, as discussed in Sect.~\ref{sec:mips24}, we use it in our analysis because it relies only on Spitzer data.
Since MIPS $24\,\mu$m data are only used to refine the classification, we did not exclude objects without detection in this band. We only used it in phase 3 when it had an uncertainty $\sigma_{24} < 0.2$ mag. This additional phase ensured that the features identified in the SED with the four IRAC bands are consistent with longer wavelength data. It allowed us to investigate the following: (i) to test the presence of a transition disk emission in objects classified as field stars, rebranding them as class II; (ii) to test the presence of a strong excess in this longer wavelength that is characteristic of deeply embedded class I objects, potentially misclassified as AGNs or Shocks; and  (iii) to refine the distinction between class I and II by testing whether the SED still rises at wavelengths longer than $8\,\mu$m for class I, otherwise rebranding them as reddened class II. These refinements explain the presence of objects beyond the boundaries in almost all frames in Fig.~\ref{fig_gut_method}. For example in frame B, some class II objects, shown in green, are located behind the boundary at the bottom left part in a region dominated by more evolved field stars. In this figure all the steps of this refinement are not shown, only the criteria on reddened class II identification is illustrated in frame E. Our adapted classification scheme was therefore composed of five bands (four IRAC, one MIPS), complemented by their five respective uncertainties, which gives a total of ten different input dimensions (or features).

In summary, our labeled dataset was structured as a list of (input, target) pairs, one per point source, where the input was a vector with ten values ($[3.6]$, $[\sigma_{3.6}]$, $[4.5]$, $[\sigma_{4.5}]$, $[5.8]$, $[\sigma_{5.8}]$, $[8]$, $[\sigma_8]$, $[24]$, $[\sigma_{24}]$), and the target was a vector of three values ($P(\text{CI})$, $P(\text{CII})$, $P(\text{Contaminant})$). Here $P()$ denotes the membership probability normalized over the three neurons of the output layer.

\subsection{Labeled datasets in Orion, NGC 2264, 1\,kpc and combinations}
\label{data_setup}

As introduced in Sect.~\ref{intro} we chose to use well-known and well-constrained star-forming regions, where YSO classification was already performed using Spitzer data. The main idea was to test the learning process on individual regions, and then compare it with various combinations of these regions. It is expected that, due to the increased diversity in the training set, the combination of regions should improve the generalization by the trained network and make it usable on other observed regions with good confidence. We selected regions analyzed in three studies, all using the original G09 method. However, some differences remain between the parameters adopted by the authors (e.g., the uncertainty cuts). Using our simplified G09 method, as presented in Sect. \ref{data_prep}, allowed us not only to base our study solely on Spitzer data, but also to built a homogeneous dataset with the exact same criteria for all regions.

The first region we used was the Orion molecular cloud with the dataset from \citet{megeath_spitzer_2012}. This work contains all the elements we needed with the four IRAC bands and the MIPS $24\ \mu$m band, and relies on the G09 method. The authors provide the full point source catalog they used to perform their YSO candidate extraction. This is one of the most important elements in our study since the network needs to see both the YSOs and the other types of objects to be able to learn the differences between them.

For the second dataset we used the catalog by \citet{rapson_spitzer_2014}, who analyzed Spitzer observations of NGC 2264 in the Mon OB1 region using the same classification scheme. In contrast to the Orion dataset they do not provide the full point source catalog, but a pre-processed object list compiled after performing band selection and magnitude uncertainty cuts. It should not affect the selection {because} we used the exact same uncertainty cuts as they did.

We then defined a dataset that is the combination of the previous two catalogs, which we call the ``Combined'' dataset. We used it to test the impact of combining different star-forming regions in the training process because distance, environment, and star formation history can impact the statistical distributions of YSOs in CMDs. We pushed this idea further by defining an additional catalog, the ``1\,kpc'' catalog, directly from \citet{gutermuth_spitzer_2009}. It contains a census of the brightest star-forming regions closer than 1\,kpc, excluding both Orion and NGC 2264. However, this catalog only contains the extracted YSO candidates and not the original point source catalog with the corresponding contaminants. This is an important drawback since it cannot be used to add diversity information in this category; however, it can be used to increase the number of class I and II and increase their specific diversity. We refer to the dataset that combines the three previous datasets Orion, NGC 2264 and 1\,kpc, as the ``Full 1\,kpc'' dataset.

This first classification provided various labeled datasets that were used for the learning process. The detailed distribution of the resulting classes for all our datasets are presented in Table~\ref{tab_selection}.

\begin{table*}[t]
        \centering
        \vspace{0.2cm}
        \caption{Results of our simplified G09 method for our various datasets.}
        \label{tab_selection}
\vspace{-0.2cm}
        \begin{tabularx}{1.0\hsize}{ l *{2}{x{0.06\hsize}} @{\hskip 0.065\hsize} *{5}{Y} @{\hskip 0.065\hsize} *{3}{x{0.075\hsize}}}
        \toprule
        \toprule
        \vspace{-0.3cm}\\
        \multirow{2}{*}{Dataset} & \multicolumn{2}{c}{\hspace{-0.6cm}Pre-selection} & \multicolumn{5}{c}{\hspace{-0.6cm}Detailed contaminants} & \multicolumn{3}{c}{\hspace{-0.6cm}Labeled classes}\\
        \cmidrule(l){2-11}
        & Total & Selected & Gal. & AGNs & Shocks & PAHs & Stars & CI YSOs & CII YSOs & Others\\
        \vspace{-0.2cm}\\
        \midrule
        \vspace{-0.2cm}\\
        Orion & 298405 & 19114 & 407 & 1141 & 28 & 87 & 14903 & 324 & 2224 & 16566\\
        \vspace{-0.2cm}\\
        NGC 2264 & 10454 & 7789 & 114 & 250 & 6 & 1 & 6893 & 90 & 435 & 7264\\
        \vspace{-0.2cm}\\
        Combined & 308859 & 26903 & 521 & 1391 & 34 & 88 & 21796 & 414 & 2659 & 23830\\
        \vspace{-0.2cm}\\
        1\,kpc* & 2548 & 2171 & 1 & 57 & 0 & 1 & 3 & 370 & 1735 & 67\\
        \vspace{-0.2cm}\\
        Full 1\,kpc & 311407 & 29074 & 522 & 1448 & 34 & 89 & 21799 & 784 & 4396 & 23897\\
        \vspace{-0.2cm}\\
        \bottomrule
        \bottomrule
        \end{tabularx}
        \tablefoot{
        The third group of columns gives the labels used in the learning phase. The last column is the sum of the columns in the ``Detailed contaminants'' group.
        *The 1\,kpc sample contains only pre-identified YSO candidates. We still classified some of them as contaminants because of the simplifications in our method.}
\end{table*}

We highlight the discrepancies between our results and those provided in the respective publications. In the case of Orion from \citet{megeath_spitzer_2012}, we merged their various
subclasses and found 488 class I and 2991 class II; no details were provided for the distribution of contaminant classes. This is consistent with our simplified G09 method, considering that the absence of the 2MASS phase prevented us from recovering objects that lack detection in some IRAC bands, and that the authors also applied additional customized YSO identification steps. For the NGC 2264 region, \citet{rapson_spitzer_2014} report 308 sources that present an IR excess, merging class 0/I, II, and transition disks. However, they used more conservative criteria than in the G09 method to further limit the contamination, which partly explains why our sample of YSOs is larger in this region. The authors do not provide all the intermediate numbers, but they mention that they excluded 5952 contaminant stars from the Mon OB1 region, a number roughly consistent with our own estimate (6893). Finally, the 1\,kpc dataset only contains class I and II objects, which means that every object that we classified as a contaminant is a direct discrepancy between the two classifications. This is again due to the absence of some refinement steps in our simplified G09 method. \citet{gutermuth_spitzer_2009} report 472 class I and 2076 class II extracted, which is also consistent with our results taking into account the absence of the 2MASS phase.

From these results, the strong imbalance between the three labeled classes is striking. This is an important characteristic of this problem because it makes it fall in the category of ``imbalanced learning.'' This is a situation known to be difficult \citep{he_learning_2009}. It requires special attention in the interpretation of the results (see Sect. \ref{sec:results_header}) and in the preparation of the training and test datasets.

\subsection{Building the training{, test, and forward} sets}
\label{train_process}

The learning process requires  a training dataset to update the weights (Sect. \ref{sec:method}), and also a test set and a validation set, which contain sources that were not shown to the network during the learning process. The test set is used after the training phase to assess the quality of the generalization. The validation set is used regularly during the training phase to compute an error that enables one to monitor the evolution of the training process. Generally, the error on the validation set decreases slowly during training, and starts increasing when the network begins to over-fit the training sample. This is a criterion to stop the learning process. Most objects are usually included in the training set because having a strong set of  statistics is particularly critical for the training phase;  fewer objects are kept for the testing and validation steps. 

The class proportions in these datasets can be kept, as in the labeled dataset, or can be rebalanced to have an even number of objects per class. However, our sample suffers from two limitations: its small size and its strong imbalance. To optimize the quality of our results we needed to carefully define our training and test datasets. Since one of the classes that we are interested in, namely CI YSOs, is represented by a relatively small number of objects, the efficiency of the training strongly depends on how many of them we kept in the training sample. Therefore, we adopted a widespread strategy where the same dataset is used for both validation and test steps \citep[e.g.,][]{lecun-98,Bishop:2006:PRM:1162264}. It is efficient to track over-fitting, but it increases the risk of stopping the training in a state that is abnormally favorable for the test set. As discussed in Sect.~\ref{sec:results_header}, discrepancies between results on the training and test datasets can be used to diagnose remaining over-training. Even with this strategy, the labeled dataset has only a few objects to be shared between the training and the test set.

In addition to the previous point, to evaluate the  quality of the result it was necessary for the test set to be representative of the true problem. As before, this was difficult mainly because our case study is strongly imbalanced. Therefore, we needed to keep observational proportions for the test set. We defined a fraction $\theta$ of objects from the labeled dataset that was taken to form the test dataset. This selection was made independently for each of the seven subclasses provided by the modified G09 classification. It ensured that the proportions were respected even for highly diluted classes of objects (e.g., for Shocks). The effect of taking such proportions is discussed in Sect.~\ref{sec:results_header} for various cases.

In contrast, the training set does not need to have observational proportions. It needs to have a greater number of objects from the classes that have a greater intrinsic diversity and occupy a larger volume in the input parameter (or feature) space. It is also necessary to have greater accuracy for the most abundant classes since even a small statistical error induces a large contamination of the diluted classes. As the CI YSOs are the most diluted class of interest, we used them to scale the number of objects from each class as follows. We shared all the CI objects between the training and the test samples as fixed by the fraction $\theta$ (i.e.,  $N^{\rm train}_{\rm CI} = (1-\theta) \times N^{\rm tot}_{\rm CI}$ and $N^{\rm test}_{\rm CI} = \theta \times N^{\rm tot}_{\rm CI}$, respectively, where $N^{\rm tot}_{\rm CI}$ is the total number of CI objects). Then, we defined a new hyperparameter, the factor $\gamma_i$, as the ratio of the number of selected objects from a given subclass $N^{\rm train}_i$ to the number $N^{\rm train}_{\rm CI}$ of CI YSOs in the same dataset:
\begin{equation}
\gamma_i = \frac{N^{\rm train}_i}{N^{\rm train}_{\rm CI}}.
\end{equation}

If $N^{\rm train}_i$ is computed directly from this formula, it may exceed $(1-\theta)\times N^{\rm tot}_i$ in some cases, a situation incompatible with keeping $N^{\rm test}_{\rm CI} = \theta \times N^{\rm tot}_{\rm CI}$ in the test set. Thus, we limited the values of $N^{\rm train}_i$ as follows:
\begin{equation}
N^{\rm train}_i = \min(\gamma_i\times (1-\theta) \times N_{\rm CI}^{\rm tot}, (1-\theta) \times N_i^{\rm tot}).
\label{eq:Ntrainmin}
\end{equation}
{ The values of the $\theta$ and $\gamma_i$ factors were determined manually by trying to optimize the results on each training set. Appendix~\ref{app:tuning} illustrates this optimization in the case of $\gamma_{\rm Stars}$.}
We note that for the most populated classes, this approach implies that only part of the sample was used to build the training and test sets. As discussed below, this was a motivation to repeat the training with various random selections of objects, and thus assess the impact of this random selection on the results.

\begin{table*}[t]
        \centering
        \vspace{0.2cm}
        \caption{Composition of the training and test datasets for each labeled dataset.}
        \label{sat_factors}
        \begin{tabularx}{1.0\hsize}{l l@{\hskip 0.05\hsize} x{0.085\hsize} x{0.105\hsize} *{5}{Y}  @{\hskip 0.07\hsize} c}
        \toprule
        \toprule
        \vspace{-0.3cm}\\
        & & CI & CII & Gal. & AGNs & Shocks & PAHs & Stars & Total\\
        \vspace{-0.3cm}\\
        \toprule
        \vspace{-0.3cm}\\
        \multicolumn{10}{c}{ Orion - $\theta = 0.3$}\\
        \cmidrule(lr){1-10}
        Test: & & 97 & 667 & 122 & 342 & 8 & 26 & 4470 & 5732\\
        \cmidrule(lr){2-10}
        \multirow{2}{*}{Train:} & $\gamma_i$ & 1.0 & 3.35 & 0.6 & 1.3 & 0.1 & 0.3 & 4.0 & \\
        & $N_i$ & 226 & 757 & 135 & 293 & 19 & 60 & 904 & 2394\\
        \vspace{0.0cm}\\
        \multicolumn{10}{c}{ NGC 2264 - $\theta = 0.3$}\\
        \cmidrule(lr){1-10}
        Test: & & 27 & 130 & 34 & 75 & 1 & 0 & 2067 & 2334\\
        \cmidrule(lr){2-10}
        \multirow{2}{*}{Train:} & $\gamma_i$ & 1.0 & 2.5 & 0.3 & 0.6 & 0.1 & 0.3 & 3.5 & \\
        & $N_i$ & 62 & 155 & 18 & 37 & 4 & 0 & 217 & 493 \\
        \vspace{0.0cm}\\
        \multicolumn{10}{c}{ Combined - $\theta = 0.2$}\\
        \cmidrule(lr){1-10}
        Test:& & 82 & 531 & 104 & 278 & 6 & 17 & 4359 & 5377\\
        \cmidrule(lr){2-10}
        \multirow{2}{*}{Train:} & $\gamma_i$ & 1.0 & 3.45 & 0.7 & 1.6 & 0.1 & 0.3 & 3.8 &\\
        & $N_i$ & 331 & 1141 & 231 & 529 & 27 & 70 & 1257 & 3586\\
        \vspace{0.0cm}\\
        \multicolumn{10}{c}{ Full 1\,kpc - $\theta = 0.2$}\\
        \cmidrule(lr){1-10}
        Test**:& & 82 & 531 & 104 & 278 & 6 & 17 & 4359 & 5377\\
        \cmidrule(lr){2-10}
        \multirow{2}{*}{Train:} & $\gamma_i$ & 1.0/1.0* & 3.3/3.0* & 1.0 & 1.4 & 0.1 & 0.3 & 8.0 &\\
        & $N_i$ & 331/331* & 1092/993* & 331 & 463 & 27 & 70 & 2648 & 6286\\
        \vspace{-0.2cm}\\
        \bottomrule
        \bottomrule
        \end{tabularx}
        \tablefoot{*The first and second values of $\gamma_i$ are for YSOs from the Combined and 1\,kpc datasets, respectively. \newline **The 1\,kpc dataset does not add contaminants, therefore the Full 1\,kpc test set is the same as the Combined test dataset to keep realistic observational proportions.}
\end{table*}

The adopted values of $\theta$ and $\gamma_i$, and the corresponding number of objects in the training sample are given in Table~\ref{sat_factors} for each dataset. It shows that with larger labeled datasets we can use smaller values of $\theta$ because it corresponds to a large enough number of objects in the associated test set. { For the training set of NGC 2264,} the number of objects is significantly smaller than in the other datasets, which impacted the results for the associated training. The fine tuning of the $\gamma_i$ values is discussed for each region in Sect.~\ref{sec:results_header} and a deeper explanation of their impact on the results is given at the end of Sect.~\ref{repres_qualit}. 

The results presented in Sects.~\ref{sec:results_header} and \ref{sec:discussion} were obtained using two types of forward datasets. Our main approach was to use the test dataset to perform the forward. In the context of this study, where a label is available for every object and where we are limited by the size of the samples, this allows us to maximize the number of objects in the training set.  It does not cause any legitimacy issues because, as explained above, the test set is built with the purpose of being independent of the training dataset and is in observational proportions. To complete our analysis, we also show results obtained by forwarding the complete labeled dataset, to address the effect of the small size of our samples, and to search for hints of over-fitting. We detail this strategy further in Sect.~\ref{orion_results}.

Finally, to ensure that our results are statistically robust, each training was repeated several times with different random selections of the testing and training objects based on the $\theta$ and $\gamma_i$ factors. This allowed us to estimate the variability of our results, as discussed in Sect.~\ref{sec:results_header}. We checked the variability after each change in any of the hyperparameters. In the case of subclasses with many objects, some objects were not included   in the training or in the test set. This ensured that the random selection could pick up various combinations of them at each training. In contrast, in the case of the rare subclasses, since they are entirely included in either the training or the test set, it is more difficult to ensure a large diversity in their selection to test the stability against selection. For each result presented in Sect.~\ref{sec:results_header} we took care to also dissociate this effect from the one induced by the random initialization of the network weights by doing several trainings with the same data selection, which is  an indication of the intrinsic stability of the network for a specific set of hyperparameters.

\subsection{Tuning the network hyperparameters}
\label{network_tuning}

We adjusted most of the network hyperparameters manually to find appropriate values for our problem, in a similar way as illustrated in Appendix~\ref{app:tuning}. To ease the research of optimal values, we started with values from general recipes. 

To start, we defined the number of neurons in the hidden layer. The number of neurons can be roughly estimated with the idea that each neuron corresponds at least to a continuous linear separation in the input feature space (Sect.~\ref{deep_ann}). Based on Fig.~\ref{fig_gut_method} at least $n = 10$ neurons should be necessary since this figure does not represent all the possible combinations of inputs. We then progressively raised the number of neurons and tested whether the overall quality of the classification was improving (Sect.~\ref{repres_qualit}). In most cases it improved continuously and then fluctuated around a maximum value. The corresponding number of neurons and the maximum value can vary with the other network hyperparameters. The chosen number of neurons is then the result of a joint optimization of the different parameters. We observed that, depending on the other parameters, the average network reached its maximum value for $n \geq 15$ hidden neurons when trained on Orion. However, the network showed better stability with a slightly larger value. We adopted $n = 20$ hidden neurons for almost all the datasets, and increased it to $n = 30$ for the largest dataset, because it slightly improved the results in this case (Table~\ref{tab_hyperparam}). Increasing this number too much   could lead to less stability and increases the computation time. 

The optimum number of neurons and the maximum quality of the classification also depends on the number of objects in the training dataset. A widely used empirical rule prescribes that the number of objects for each class in the training sample must be an order of magnitude larger than the number of weights in such a network. This means that the minimum size of the training dataset increases with the complexity of the problem using ML algorithms. In our case, including the bias nodes, we would need $(m+1)\times n + (n+1)\times o \times 10$ objects in our training set, with the same notations as in Sect.~\ref{final_ann}. This gives us a minimum of $2830$ objects in the whole training set using our network structure with $n = 20$, assuming a balanced distribution among the output classes. As shown in Table~\ref{sat_factors}, some of our training samples are too small for the class I YSOs and critically small for various subclasses of contaminants. Still, each class does not get the same number of neurons from the network. Some classes have a less complex distribution in the parameter space and can be represented by a small number of weights, therefore with fewer training examples. The extra representative strength can then be used to better represent more complex classes that may be more abundant. Thus, it is a matter of balance between having a sufficient number of neurons to properly describe our problem and the maximum amount of available data. This is a strong limitation on the quality of the results.

\begin{table}
        \small
        \centering
        \caption{Non-structural network hyperparameter values used in training for each dataset. }
        \vspace{-0.1cm}
        \begin{tabularx}{\hsize}{l @{\hskip 0.1\hsize} *{4}{Y}}
        \toprule
         & Orion & NGC 2264 & Combined & Full 1\,kpc \\
        \vspace{-0.3cm}\\
        Size & 2394 & 493 & 3586 & 7476 \\
        \midrule
        $\eta$ & $3 \times 10^{-5}$ & $2 \times 10^{-5}$ & $4 \times 10^{-5}$ &  $8 \times 10^{-5}$ \\
        $\alpha$ & 0.7 & 0.6 & 0.6 & 0.8 \\
        $n$ & 20 & 20 & 20 & 30 \\ 
        $n_e$ & 5000 & 5000 & 5000 & 3000 \\
        \bottomrule
        \end{tabularx}
        \vspace{-0.1cm}
        \label{tab_hyperparam}
        \tablefoot{The size of the corresponding training set is put for comparison. $\eta$ is the learning rate, $\alpha$ the momentum, $n$ the number of neurons in the hidden layer, and $n_e$ the number of epochs between two control steps.}
\end{table}

Our datasets were individually normalized in an interval of $-1$ to $1$, with a mean close to zero for each input feature. This was done for each dimension individually by subtracting the mean value and then dividing by the new absolute maximum value. This evened out the numerical values of the input dimensions, which avoided, at the beginning of the training,   inappropriately giving a stronger impact to the dimensions with larger numerical values. Therefore, we set the steepness $\beta$ of the sigmoid activation of the hidden neurons to $\beta=1$, which worked well with the adopted normalization.

As described in Sect.~\ref{optim_perf}, we only used the batch CUDA method to train our network throughout  this study. We adopted learning rates in the range $\eta = 3 - 8 \times 10^{-5}$ and a momentum ranging from $\alpha = 0.6$ to $ \alpha = 0.8$ depending on the dataset, as shown in Table~\ref{tab_hyperparam}. { We obtained these values by optimizing the results as illustrated in Appendix~\ref{app:tuning}.} We note that during training we summed the weight update contributions from each object in the training set (as in, e.g., \citealt{rumelhart_parallel_1986}); in order  to keep the update values of similar order, the learning rate {should be decreased according} to the size of the training set (eqs.~\ref{eq_update_second},~\ref{eq_update_first}). A variation of this approach could have been employed, where the contributions from each object in the training set would be averaged (as in, e.g., \citealt{glorot_understanding_2010}), sparing the user the necessity of adapting the learning rate to the sample size. {This would not change the results since the two strategies are strictly identical with the appropriate choice of learning rate value}. Interestingly, we observed  for this specific study that the learning rate could instead be slowly increased when the training dataset was larger. This indicates that, in small training sets, the learning process is dominated by the lack of constraints, causing a less stable value of convergence. This translates into a convergence region in the weight space that contains numerous narrow minima due to the larger description granularity of the objects in a smaller dataset. The network can only properly resolve it with a slower learning rate and will be less capable of generalization. This is an expected issue because we intentionally included small datasets in the analysis to assess the limits of the method with few objects.

Finally, one less important hyperparameter is the number of epochs between two monitoring steps, which was set from $n_e = 3000$ to $ n_e = 5000$. It defines at which frequency the network state is saved and checked, leaving the opportunity to decrease the learning rate $\eta$ if necessary.

\subsection{Representation and quality estimators}
\label{repres_qualit}

In this last part we define the concepts necessary to present our results statistically, and to characterize their quality. For this we use the ``confusion'' matrix. It is defined as a two-dimensional table with rows corresponding to the original class distribution (targets), and columns corresponding to the classes given to the same objects by our network classification (output). As an example, Table~\ref{tab:OO} shows the $3 \times 3$ confusion matrix for the Orion test set using observational proportions. { This representation directly provides a visual indication of the quality of the network classification. It allows us to define quality estimators for each class.} The ``recall'' represents the proportion of objects from a given class that were correctly classified. The ``precision'' is a purity indicator of an output class. It represents the fraction of correctly classified objects in a class as predicted by the network. And finally, the ``accuracy'' is a global quantity that gives the proportion of objects that are correctly classified with respect to the total number of objects. In our confusion matrices we show it at the intersection of the recall column and the precision row. Limiting the result analysis to this latter quantity may be misleading because it would hide class-specific qualities and would be strongly impacted by the imbalance between the output classes. The matrix format is particularly well-suited to reveal the weaknesses of a classification. It could, for example, reveal that the vast majority of a subclass goes mistakenly into a specific other subclass, which is informative about any degeneracy between the two classes.

\begin{table*}[!t]
        \centering
        \caption{List of case studies regarding the dataset used to train the network and the dataset to which it was applied to provide predictions.}
        \vspace{-0.1cm}
        \def\arraystretch{1.3}
        \begin{tabularx}{0.7\hsize}{r l |*{4}{m}}
        \multicolumn{2}{c}{} & \multicolumn{4}{c}{{\large Forward dataset}}\\
        \cmidrule[\heavyrulewidth](lr){2-6}
        \parbox[l]{0.2cm}{\multirow{6}{*}{\rotatebox[origin=c]{90}{{\large Training dataset}}}} & & Orion & NGC 2264 & Combined & Full 1\,kpc \\
        \cmidrule(lr){2-6}
         & Orion & O-O & \hspace{0.15cm}O-N* & & \\
         & NGC 2264 & \hspace{0.15cm} N-O* & N-N & & \\ 
         & Combined & & & C-C & \\
         & Full 1\,kpc & & & F-C & \\
        \cmidrule[\heavyrulewidth](lr){2-6}
        \end{tabularx}
        \vspace{-0.05cm}
        \label{results_cases}
        \tablefoot{*These cases were only forwarded on the full corresponding dataset with no need for a test set. There was no forward on the Full 1\,kpc dataset since, as a combination of a complete catalog and a YSO-only catalog, it is not in observational proportions.}
\end{table*}

One difficulty highlighted by the use of a confusion matrix is the absence of a global quality estimator since it depends on the end objective. As for any classification problem, one must choose the appropriate balance between reliability and completeness.
As our aim we chose  maximizing the precision for CI, while keeping a large enough value in recall (ideally $>90\%$ for both), and a good precision for CII as well. This choice strongly impacts the tuning of the $\gamma_i$ values, since they directly represent the emphasis given to a class against the others during the training phase, hence biasing the network toward the class that needs the most representative strength. This will lower the quality of the most dominant objects. A typical example of emphasis on CI YSOs is presented in Appendix~\ref{app:tuning} for the tuning of $\gamma_{\rm Stars}$.

\subsection{Convergence criteria}
\label{conv_crit}

Since training the network is an iterative process, a convergence criteria must be adopted.
In principle, this criteria should enable one to identify an iteration where the training has sufficiently converged for the network to capture the information in the training sample, but is not yet subject to over-training. It is customary to monitor the global error on both the training and test set during the process. While the error of the training set will slowly converge to a minimum value, the error on the test set will follow the same curve only for part of the training and then rise when over-fitting begins. However, this global error is affected by the proportions in the training sample and does not necessarily reflect the underlying convergence of each subclass. Our approach to this issue was to let the network learn for an obviously unnecessary amount of steps and regularly save the state of the network. This allowed us to better monitor the long-term behavior of the error, and to compare the confusion matrix at regular steps. In most cases the error of the training and test sets both converged to a stable value and stayed there for many steps before the second one started to rise. During this apparently stable moment, the prediction quality of the classes oscillated, switching the classes that get the most representativity strength from the network. Because we want to put the emphasis on CI YSOs, we then manually selected a step that was near the maximum value for CI YSOs precision, with  special attention to avoid the ones that would be too unfavorable to CII YSOs.

We observed that the convergence step changed significantly with the network weight random initialization, even with the exact same dataset and network, ranging from 100 to more that 1000, where each step corresponds to several thousands epochs (Table~\ref{tab_hyperparam}). Most of the time, the error plateau lasted around 100 steps. We note that the number of steps needed to converge has no consequences on the quality of the results; it only reflects the length of the particular trajectory followed by the network during the training phase.

\section{Results}
\label{sec:results_header}
This section presents the YSO classification obtained for the various labeled datasets described in Sect.~\ref{sec:data_prep_chap}. To ease the reading of this section, we summarize all the cases in Table~\ref{results_cases}.

\subsection{Results for the Orion molecular cloud}
\label{orion_results}

In this section we consider the case where both the training and forward datasets were built from the Orion labeled dataset, hereafter denoted the O-O case. The network hyperparameters used for Orion are described in Sect.~\ref{sec:data_prep_chap} and Table~\ref{tab_hyperparam}. The resulting confusion matrix is shown in Table~\ref{tab:OO}, {where the forward set is} the test set from Table~\ref{sat_factors} {, which is in observational proportions}.
The optimal $\gamma_i$ factors found for Orion show a stronger importance of the CII YSOs ($\gamma_{\text{CII}} = 3.35$) and of the Stars subclass ($\gamma_{\text{Stars}} = 4.0$) than for any other subclass ($\gamma_i \lesssim 1$). In contrast, the optimal values for Shocks and PAHs are saturated in the sense that in eq.~(\ref{eq:Ntrainmin}), $N_i^{\rm train} = (1-\theta) \times N_i^{\rm tot}$, but they appear to have a negligible impact on the classification quality in this case. Galaxies and PAHs appear to be easily classified with a rather small number of them in the training sample. This is convenient since adding too many objects of any class hampers the capacity of the network to represent CI objects {(i.e., the most diluted class of interest)} in the network, degrading the reliability of their identification. Therefore, Stars and CII objects could be well represented with a large fraction of them in the training sample, still limiting their number to avoid an excessive dilution of CI YSOs. We note that we have explored different values for the $\theta$ parameter. It revealed that the network predictions improve continuously when increasing the number of objects in the training sample. However, to keep enough objects in the test dataset, we had to limit $\theta$ to 0.3 (Table~\ref{sat_factors}). The only classes for which the number of objects in the training sample is limited by the $\theta$ value rather than their respective $\gamma_i$ values are CI YSO, Shocks, and PAHs. Since Shocks and PAHs are rare in the observational proportions, they are unlikely to have a significant impact on the prediction quality. This leads to the outcome that the following results on Orion are currently limited by the number of CI YSOs in the dataset.

The global accuracy of this case is 98.4\%, but the confusion matrix (Table~\ref{tab:OO}) shows that this apparently good accuracy is unevenly distributed among the three classes. The best represented class is the contaminant class, with an excellent precision of 99.7\% and a very good recall of 98.6\%. The results are slightly less satisfying for the two classes of interest,  with recalls of 90.7\% and 97.6\%, and precisions of 83.0\% and 91.3\% for the CI and CII YSOs, respectively. In spite of their very good recall, due to their widely dominant number, objects from the Others class are the major contaminants of both CI and CII YSOs, with 11 out of 18, and 58 out of 62 contaminants, respectively. Therefore, improving the relatively low precision of CI and CII objects mainly requires us to better classify the objects labeled Others. In addition, less abundant classes are more vulnerable to contamination. This is clearly illustrated by the fact that the seven CII YSOs misclassified as CI YSOs account for a loss of 7\% in precision for CI objects, while the nine CII YSOs misclassified as Others account for a loss of only 0.2\% in the Others precision. Those properties are typical of classification problems with a diluted class of interest, where it is essential to compute the confusion matrix using observational proportions. Computing it from a balanced forward sample would have led to apparently excellent results, which would greatly overestimate the quality genuinely obtained in a real use case. { It} illustrates the necessity of a high $\gamma_i$ value for dominant classes regardless of their interest (e.g., Stars) as we need to maximize the recall of these classes to enhance the precision of the diluted classes.

To illustrate the interest of selecting our training and test composition with the $\theta$ and $\gamma_i$ factors, we made a test with a balanced training set where all three classes were represented by an equal number of objects. The best we could achieve this way was not more than $\sim 55\%$ precision on CI YSOs and $\sim 87\%$ on CII YSO. This was mostly due to the small size of the training sample, which was constrained by the less abundant class, and to the poor sampling of the Others class compared to its great diversity. In contrast, when using our more complex sample definition, despite the reduced proportion of YSOs in the training sample, the precision and recall quantities for both CI and CII remained above $80\%$ and  $90\%$, respectively. This means that we found an appropriate balance between the representativity   of each class and their dilution in the training sample.

As discussed in Sect.~\ref{train_process}, we tested the stability of those results regarding (i) the initial weight values using the exact same training dataset, and (ii) the random selection of objects in the training and test set. For point (i) we found that in Orion the weight initialization has a weak impact with approximately $\pm 0.5\%$ dispersion in almost all the quality estimators. For point (ii) we found the dispersion to average around $\pm 1\%$ for the recall of YSO classes. Contaminants were found to be more stable with a recall dispersion under $\pm 0.5\%$. Regarding the precision value, there is more instability for the CI YSOs because they are weakly represented in the test set and one misclassified object changes the precision value by typically $1\%$. Overall, we observed values ranging from $77\%$ to $83\%$ for the CI YSOs precision. For the better represented classes we obtained much more stable values with dispersions of $\pm 0.5$ to $\pm 1\%$ on class II, and less than $\pm 0.5\%$ on the Others objects. This relative stability is strongly related to the proper balance between classes, controlled by the $\gamma_i$ parameters, since strong variations between runs imply that selection effects are important, and that there are not enough objects to represent the input parameter space properly.

We also looked at the detailed distribution of classified objects regarding their subclasses from the labeled dataset. These results are shown for Orion in Table~\ref{tab:OO-sub}. It is particularly useful to detail the distribution of contaminants across the three network output classes. For CI YSOs, the contamination appears to originate evenly from various subclasses, while for CII there is a strong contamination from non-YSO stars, though this represents only a small fraction ($\sim 1\%$) of the Stars population. The distribution of Others objects among the subclasses is very similar to the original (Table~\ref{tab_selection}). Interestingly, the Shocks subclass is evenly scattered across the three output classes, which we interpret as a failure by the network to find a proper generalization for these objects. More generally, Table~\ref{tab:OO-sub} shows that the classes that are sufficiently represented in the training set like AGNs or Stars are well classified, while the Galaxies, Shocks, and PAHs are less well predicted. This is {mostly because} the training dataset does not fully cover the respective volume in the input parameter space or {because} they are too diluted in the dataset. Additionally, Stars and Galaxies mainly contaminate the CII class. This is a direct consequence of the proximity of these classes in the input parameter space, as can be seen in Fig.~\ref{fig_gut_method}.

\begin{table}[!t]
        \small
        \centering
        \caption{Confusion matrix for the O-O case for a typical run.}
        \vspace{-0.1cm}
        \begin{tabularx}{\hsize}{r l |*{3}{m}| r }
        \multicolumn{2}{c}{}& \multicolumn{3}{c}{{Predicted}}&\\
        \cmidrule[\heavyrulewidth](lr){2-6}
        \parbox[l]{0.2cm}{\multirow{6}{*}{\rotatebox[origin=c]{90}{{Actual}}}} & Class & CI YSOs & CII YSOs & Others & Recall \\
        \cmidrule(lr){2-6}
         &  CI YSOs    & 88     & 4       & 5       & 90.7\% \\
         &  CII YSOs   & 7      & 651     & 9       & 97.6\% \\
         &  Others     & 11     & 58      & 4899    & 98.6\% \\
        \cmidrule(lr){2-6}
         &  Precision & 83.0\% & 91.3\% & 99.7\% & 98.4\% \\
        \cmidrule[\heavyrulewidth](lr){2-6}
        \end{tabularx}
        \vspace{-0.05cm}
        \label{tab:OO}
\end{table}

\begin{table}[!t]
        \small
        \centering
        \caption{Subclass distribution for the O-O case.}
        \vspace{-0.1cm}
        \begin{tabularx}{\hsize}{r l | *{6}{Y} l}
        \multicolumn{2}{c}{}& \multicolumn{7}{c}{{Actual}}\\      
        \cmidrule[\heavyrulewidth](lr){2-9}
        \parbox[l]{0.0cm}{\multirow{5}{*}{\rotatebox[origin=c]{90}{{Predicted}}}} & & CI & CII & Gal & AGNs & Shocks & PAHs & Stars\\
        \cmidrule(lr){2-9}
          & CI & 88 & 7 & 1 & 2 & 3 & 3 & 2 \\
          & CII & 4 & 651 & 5 & 0 & 2 & 4 & 47 \\
          & Others & 5 & 9 & 116 & 340 & 3 & 19 & 4421 \\
        \cmidrule[\heavyrulewidth](lr){2-9}
        \end{tabularx}
        \vspace{-0.1cm}
        \label{tab:OO-sub}
\end{table}

\begin{table}[!t]
        \small
        \centering
        \caption{Confusion matrix for the O-O case forwarded on the full dataset.}
        \begin{tabularx}{\hsize}{r l |*{3}{m}| r }
        \multicolumn{2}{c}{}& \multicolumn{3}{c}{{Predicted}}&\\
        \cmidrule[\heavyrulewidth](lr){2-6}
        \parbox[l]{0.2cm}{\multirow{6}{*}{\rotatebox[origin=c]{90}{{Actual}}}} & Class & CI YSOs & CII YSOs & Others & Recall \\
        \cmidrule(lr){2-6}
         &  CI YSOs    & 305     & 11      & 8        & 94.1\% \\
         &  CII YSOs   & 34      & 2157    & 33       & 97.0\% \\
         &  Others     & 34      & 201     & 16331    & 98.6\% \\
        \cmidrule(lr){2-6}
         &  Precision & 81.8\% & 91.1\% & 99.7\% & 98.3\% \\
        \cmidrule[\heavyrulewidth](lr){2-6}
        \end{tabularx}
        \vspace{-0.1cm}
        \label{tab:OO_all}
\end{table}

\begin{table}[!t]
        \small
        \centering
        \caption{Subclass distribution for the O-O case forwarded on the full dataset.}
        \vspace{-0.1cm}
        \begin{tabularx}{\hsize}{r l | *{6}{Y} l}   
        \multicolumn{2}{c}{}& \multicolumn{7}{c}{{Actual}}\\      
        \cmidrule[\heavyrulewidth](lr){2-9}
        \parbox[l]{0.0cm}{\multirow{5}{*}{\rotatebox[origin=c]{90}{{Predicted}}}} & & CI & CII & Gal & AGNs & Shocks & PAHs & Stars\\
        \cmidrule(lr){2-9}
         &  CI & 305 & 34 & 2 & 11 & 11 & 7 & 3 \\
         &  CII & 11 & 2157 & 10 & 9 & 9 & 18 & 155 \\
         &  Others & 8 & 33 & 395 & 1121 & 8 & 62 & 14745 \\
        \cmidrule[\heavyrulewidth](lr){2-9}
        \end{tabularx}
        \vspace{-0.1cm}
        \label{tab:OO_all-sub}
\end{table}

To circumvent the limitations due to the small size of our test set, we also applied our network to the complete Orion dataset. The corresponding confusion matrix is in Table~\ref{tab:OO_all}, and the associated subclass distribution is in Table~\ref{tab:OO_all-sub}. It may be considered  a risky practice because it includes objects from the training set that could be over-fitted, so it should not be used alone to analyze the results. Here, we use it jointly with the results on the test set as an additional over-fitting test. If the classes are well constrained, then the confusion matrix should be stable when switching from the test to the complete dataset. For Orion we see a strong consistency between Tables~\ref{tab:OO} and \ref{tab:OO_all} for the Others and CII classes, both in terms of recall and precision. For CI YSOs, the recall  increases by 3.4\%, and the precision  decreases by 1.2\%. These variations are of the same order as the variability observed when changing the training set random selection, indicating that over-fitting is unlikely here. If there is  over-fitting it should be weak and restricted to CI YSOs. Therefore, the results obtained from the complete Orion dataset appear to be reliable enough to take advantage of their greater statistics. Table~\ref{tab:OO_all} gives slightly more information than Table~\ref{tab:OO}, and mostly confirms the previous conclusions on the contamination between classes. Table~\ref{tab:OO_all-sub} provides further insight. AGNs, which seemed to be almost perfectly classified, are revealed to be misclassified as YSOs in 1.8\% of cases. It also shows that the missed AGNs are equally distributed across the CI and CII YSO classes. Shocks are still evenly spread across the three output classes. Regarding PAHs, Table~\ref{tab:OO_all-sub} reveals that there is more confusion with the CII YSOs than with the CI YSOs.

\subsection{NGC 2264 open cluster}
\label{NGC2264_results}

\begin{table}
        \small
        \centering
        \caption{Confusion matrix for the N-N case for a typical run.}
        \vspace{-0.1cm}
        \begin{tabularx}{\hsize}{r l |*{3}{m}| r }
        \multicolumn{2}{c}{}& \multicolumn{3}{c}{{Predicted}}&\\
        \cmidrule[\heavyrulewidth](lr){2-6}
        \parbox[l]{0.2cm}{\multirow{6}{*}{\rotatebox[origin=c]{90}{{Actual}}}} & Class & CI YSOs & CII YSOs & Others & Recall \\
        \cmidrule(lr){2-6}
         &  CI YSOs    & 26      & 1       & 0      & 96.3\% \\
         &  CII YSOs   & 1       & 121     & 8      & 93.1\% \\
         &  Others     & 2       & 31      & 2144   & 98.5\% \\
        \cmidrule(lr){2-6}
         &  Precision & 89.7\% & 79.1\% & 99.6\% & 98.2\% \\
        \cmidrule[\heavyrulewidth](lr){2-6}
        \end{tabularx}
        \vspace{-0.1cm}
        \label{tab:NN}
\end{table}

\begin{table}
        \small
        \centering
        \caption{Subclass distribution for the N-N case.}
        \vspace{-0.1cm}
        \begin{tabularx}{\hsize}{r l | *{6}{Y} l}       
        \multicolumn{2}{c}{}& \multicolumn{7}{c}{{Actual}}\\      
        \cmidrule[\heavyrulewidth](lr){2-9}
        \parbox[l]{0.0cm}{\multirow{5}{*}{\rotatebox[origin=c]{90}{{Predicted}}}} & & CI & CII & Gal & AGNs & Shocks & PAHs & Stars\\
        \cmidrule(lr){2-9}
         &  CI  & 26 & 1 & 0 & 2 & 0 & 0 & 0 \\
         &  CII  & 1 & 121 & 4 & 5 & 1 & 0 & 21 \\
         &  Others & 0 & 8 & 30 & 68 & 0 & 0 & 2046 \\
        \cmidrule[\heavyrulewidth](lr){2-9}
        \end{tabularx}
        \vspace{-0.1cm}
        \label{tab:NN-sub}
\end{table}

\begin{table}
        \small
        \centering
        \caption{Confusion matrix for the N-N case forwarded on the full dataset.}
        \vspace{-0.1cm}
        \begin{tabularx}{\hsize}{r l |*{3}{m}| r }
        \multicolumn{2}{c}{}& \multicolumn{3}{c}{{Predicted}}&\\
        \cmidrule[\heavyrulewidth](lr){2-6}
        \parbox[l]{0.0cm}{\multirow{6}{*}{\rotatebox[origin=c]{90}{{Actual}}}} & Class & CI YSOs & CII YSOs & Others & Recall \\
        \cmidrule(lr){2-6}
         &  CI YSOs    & 88      & 2       & 0        & 97.8\% \\
         &  CII YSOs   & 7       & 406     & 22       & 93.3\% \\
         &  Others     & 12      & 77      & 7175     & 98.8\% \\
        \cmidrule(lr){2-6}
         &  Precision & 82.2\% & 83.7\% & 99.7\% & 98.4\%\\
        \cmidrule[\heavyrulewidth](lr){2-6}
        \end{tabularx}
        \vspace{-0.1cm}
        \label{tab:NN_all}
\end{table}

\begin{table}
        \small
        \centering
        \caption{Subclass distribution for the N-N case forwarded on the full dataset.}
        \label{tab:NN-sub_all}
        \vspace{0.1cm}
        \begin{tabularx}{\hsize}{r l | *{6}{Y} l}     
        \multicolumn{2}{c}{}& \multicolumn{7}{c}{{Actual}}\\      
        \cmidrule[\heavyrulewidth](lr){2-9}
        \parbox[l]{0.0cm}{\multirow{5}{*}{\rotatebox[origin=c]{90}{{Predicted}}}} & & CI & CII & Gal & AGNs & Shocks & PAHs & Stars\\
        \cmidrule(lr){2-9}
         &  CI & 88 & 7 & 0 & 8 & 3 & 0 & 0 \\
         &  CII & 2 & 406 & 8 & 10 & 1 & 0 & 58 \\
         &  Others & 0 & 22 & 106 & 232 & 2 & 0 & 6835 \\
        \cmidrule[\heavyrulewidth](lr){2-9}
        \end{tabularx}
        \vspace{-0.1cm}
        \label{tab:NN_all-sub}
\end{table}

For this section we used the training and forward datasets for NGC 2264 described in Table~\ref{sat_factors} with the corresponding hyperparameters (Table~\ref{tab_hyperparam}). The results for this region alone, obtained by a forward on the test set, are shown in Table~\ref{tab:NN}, with the subclass distribution in Table~\ref{tab:NN-sub}. We refer to this case as the N-N case. The major differences with Orion are expected to come from the differences in input parameter space coverage and from the different proportions of each subclass. This N-N case is also useful to see how difficult it is to train our network with a small dataset. { The} recall and precision of CI YSOs are greater ($96.3\%$ and $89.7\%$, respectively) than in Orion, but the corresponding number of objects is too small to draw firm conclusions. For CII YSOs, the recall and precision are less than in Orion by approximately $4\%$ and $10\%$, respectively. The Others class shows similar values to those in Orion.

We highlight here how having a small learning sample is problematic for this classification. First of all, the training set contains only 62 CI YSOs, which is far from enough in regard of the size of the network (Sect. \ref{network_tuning}). This difficulty is far worse than for Orion because, to avoid dilution, we had to limit the number of objects in the two other classes, leading to the small size of the training sample (493 objects), and consequently to worse results for all classes. To mitigate these difficulties and because the dilution effect occurs quickly, we adopted lower $\gamma_i$ values for CII YSOs and Stars, thus reducing their relative strength. It results in too small training set sizes for all the subclasses compared to the number of weights in the network. However, we observed that a decrease in the number of neurons still reduced the quality of the results. Although a lower number of hidden neurons tended to increase stability, we chose to keep them at $n = 20$ to get the best results and to reduce the learning rate to achieve better stability. We note that, due to the use of batch training, the smaller size of the dataset than for the O-O case is equivalent to an additional lowering of the learning rate because the typical magnitude of the weight updates scales with the batch size (Sect.~\ref{optim_perf}). For this dataset slight changes in the $\gamma_i$ values happened to lead to great differences in terms of results and stability, which is a hint that the classification lacks constraints.

Even for a given good $\gamma_i$ set, there is a large scatter in the results when changing the training and test set random selection. It leads to a dispersion of about $\pm 4\%$ in both recall and precision for the CI YSOs. This can be due to a lack of representativity of this class in our sample, but it can also come from small-number effects in the test set that are stronger than in Orion. These two points show that the quality estimators for CI YSOs are not trustworthy with such a small sample size. The results shown in Tables~\ref{tab:NN} and \ref{tab:NN-sub} correspond to one of the best trainings on NGC 2264, that achieves nearly the best values for CI quality estimators. The CII precision dispersion is about $\pm 2\%$, and its average value is around $80\%$, which is higher than in the specific result given in Table~\ref{tab:NN}, but still significantly lower than for Orion. In contrast, the CII recall is fairly stable with less than $\pm 1\%$ dispersion. Contaminants seem as stable as for Orion using these specific $\gamma_i$ values. However, it could come from the artificial simplification of the problem due to the quasi-absence of some subclasses (Shocks and PAHs; see Table~\ref{sat_factors}) in the test set. We note that the network would not be able to classify objects from these classes if this training were applied to any other region that contained such objects.

As in the previous section, we studied the effect of the random initialization of the weights. We found that both precision and recall of YSO classes are less stable than for the O-O case with a dispersion of $\pm 1.5\%$ to $\pm 2.5\%$. The Others class shows a similar stability to that for Orion, with up to $\pm 0.5\%$ dispersion on precision and recall, which could again be biased by the fact that the absence of some subclasses simplifies the classification. These results indicate as before that our network is not sufficiently constrained using this dataset alone with respect to the architecture complexity that is needed for YSO classification.

The forward on the complete NGC 2264 dataset is crucial in this case since it may overcome small-number effects for many subclasses. The corresponding results are shown in Tables~\ref{tab:NN_all} and \ref{tab:NN-sub_all}. It is more difficult in this case than for O-O  to be sure that there is no over-training, even with a careful monitoring of the error convergence on the test set during the training, because the small-number effects are important. As a precaution, in all the results for the N-N case, we chose to stop the training slightly earlier in the convergence phase in comparison to Orion, for which we found over-training to be negligible or absent (Sect.~\ref{orion_results}). We expect this strategy to reduce over-training, at the cost of a higher noise.
With this assumption, the results show more similarities to the Orion case than those obtained with the test set only (comparing Tables~\ref{tab:OO_all} and \ref{tab:NN_all}). Because NGC 2264 contains fewer CI and CII YSOs than Orion, their boundaries with the contaminants in the parameter space are less constrained. This results in a lower precision for YSO classes, which is mainly visible for the CII YSOs with a drop in precision  to $83.7\%$. For NGC 2264, we have smaller optimal $\gamma_i$ values for the contaminants (especially the Stars) than in Orion. Since it implicitly forces the network to put the emphasis on CI and CII, it should result in better, or at least equivalent, values for recall on these classes than on Orion. This appears to be the case for CI ($\approx 98\%$). It is less clear for CII ($93.3\%$), possibly because of their lower $\gamma_i$ value than for the Orion case. For the sub-contaminant  distributions the statistics is more robust than in Table~\ref{tab:NN-sub}, and the Galaxies and AGNs are properly represented. Even so, it appears that the AGN classification quality is not sufficient and has a stronger impact on the CI precision than in the case of Orion. The other behaviors are similar to those identified in Orion.

\subsection{Cross forwards}
\label{cross_forward}

\begin{figure*}[t]
        \centering
        \begin{subfigure}[t]{0.43\textwidth}
        \caption*{{Orion}}
        \includegraphics[width=\textwidth]{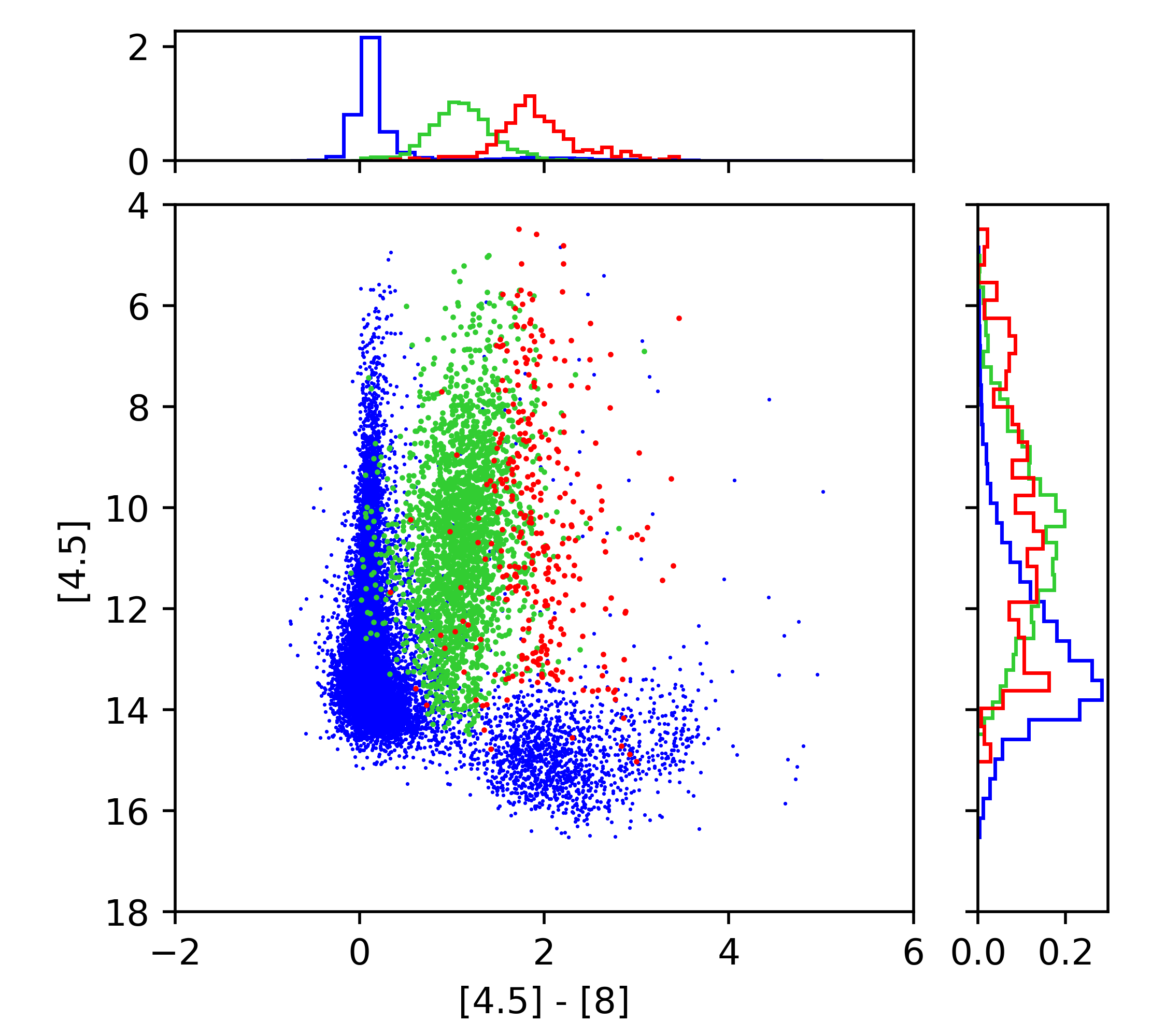}
        \end{subfigure}
        \hspace{0.2cm}
        \begin{subfigure}[t]{0.43\textwidth}
        \caption*{{NGC 2264}}
        \includegraphics[width=\textwidth]{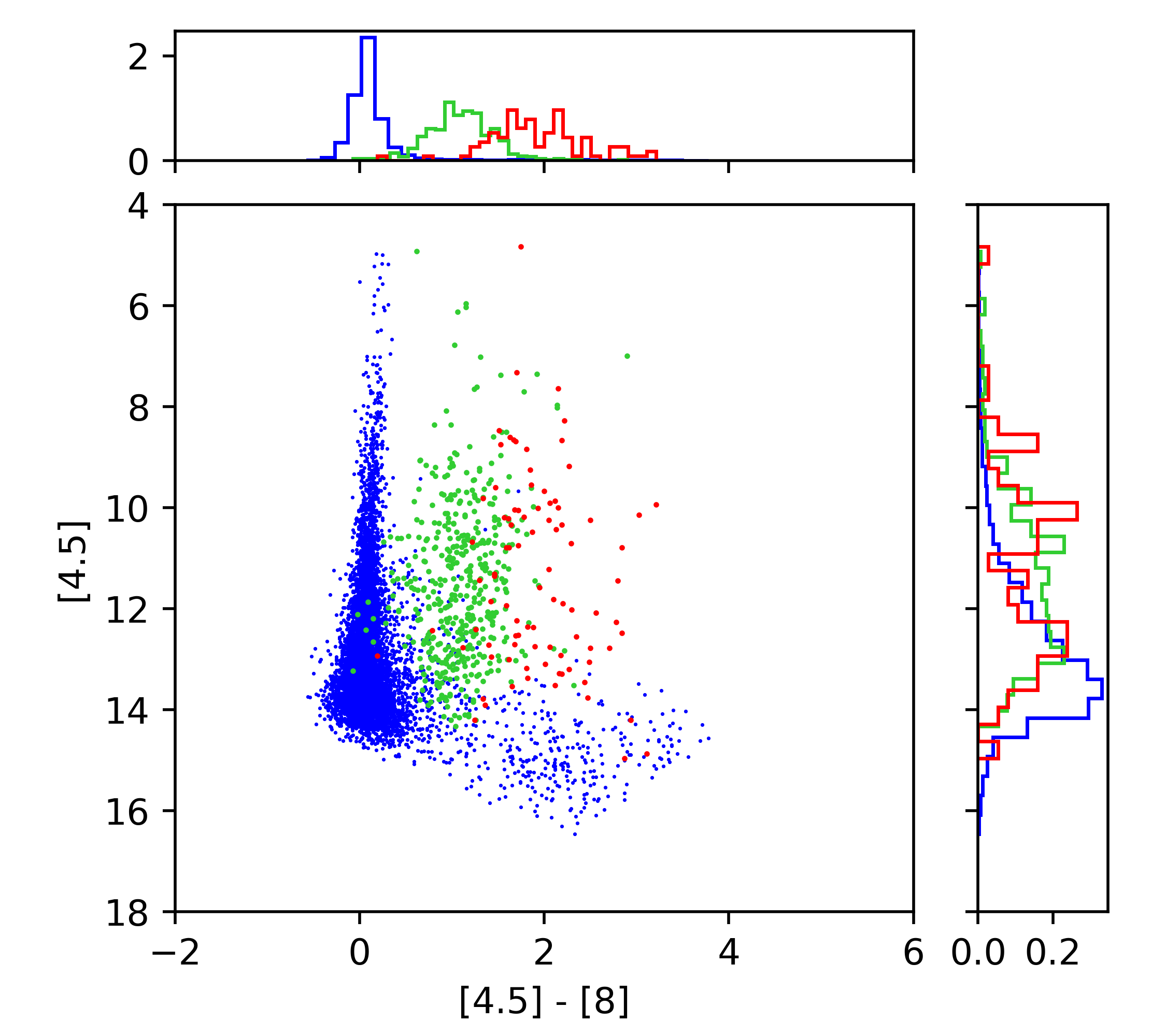}
        \end{subfigure}
        \vspace{0.1cm}\\
        \begin{subfigure}[t]{0.43\textwidth}
        \caption*{{Combined}}
        \includegraphics[width=\textwidth]{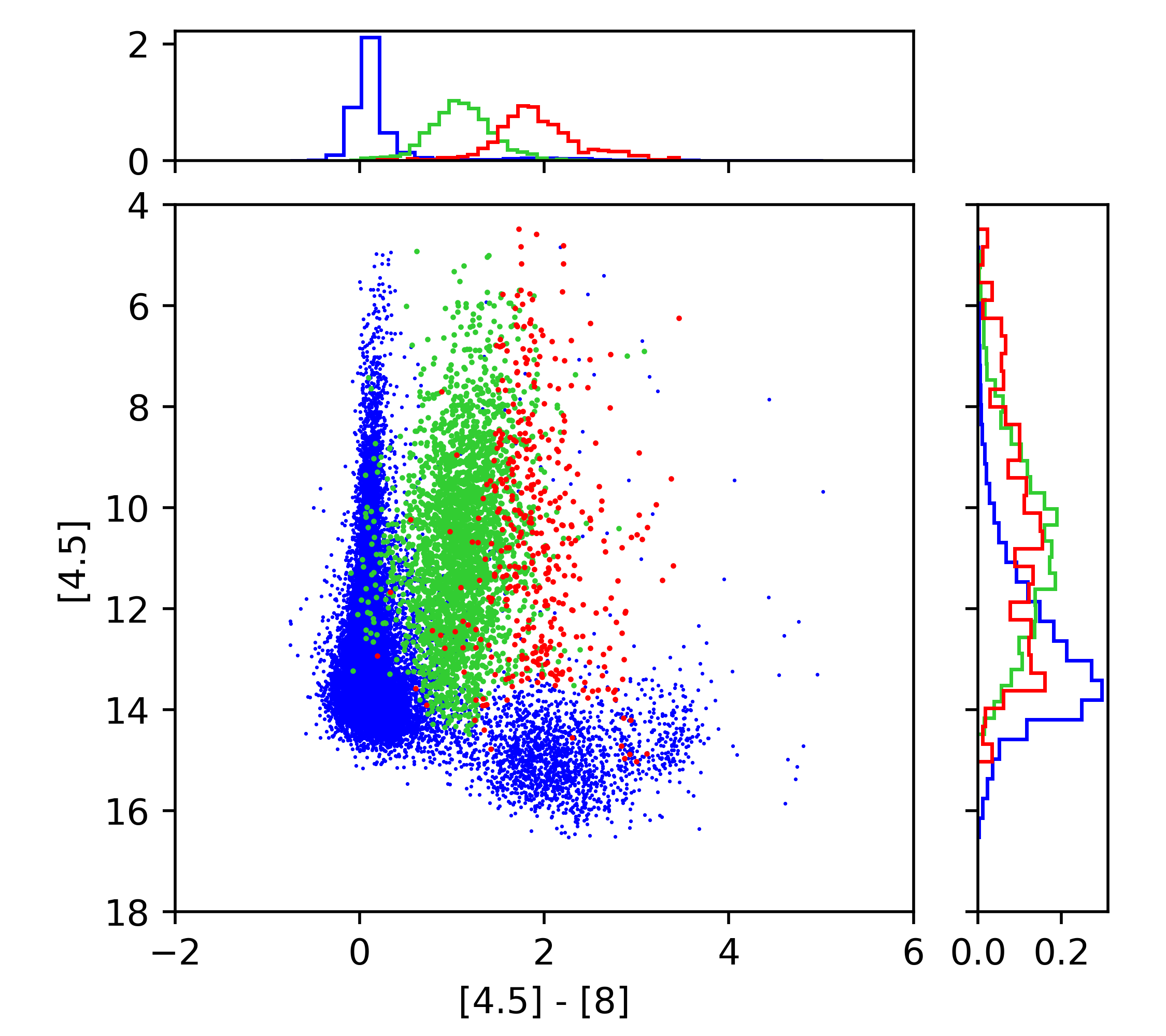}
        \end{subfigure}
        \hspace{0.2cm}
        \begin{subfigure}[t]{0.43\textwidth}
        \caption*{{Combined + 1\,kpc}}
        \includegraphics[width=\textwidth]{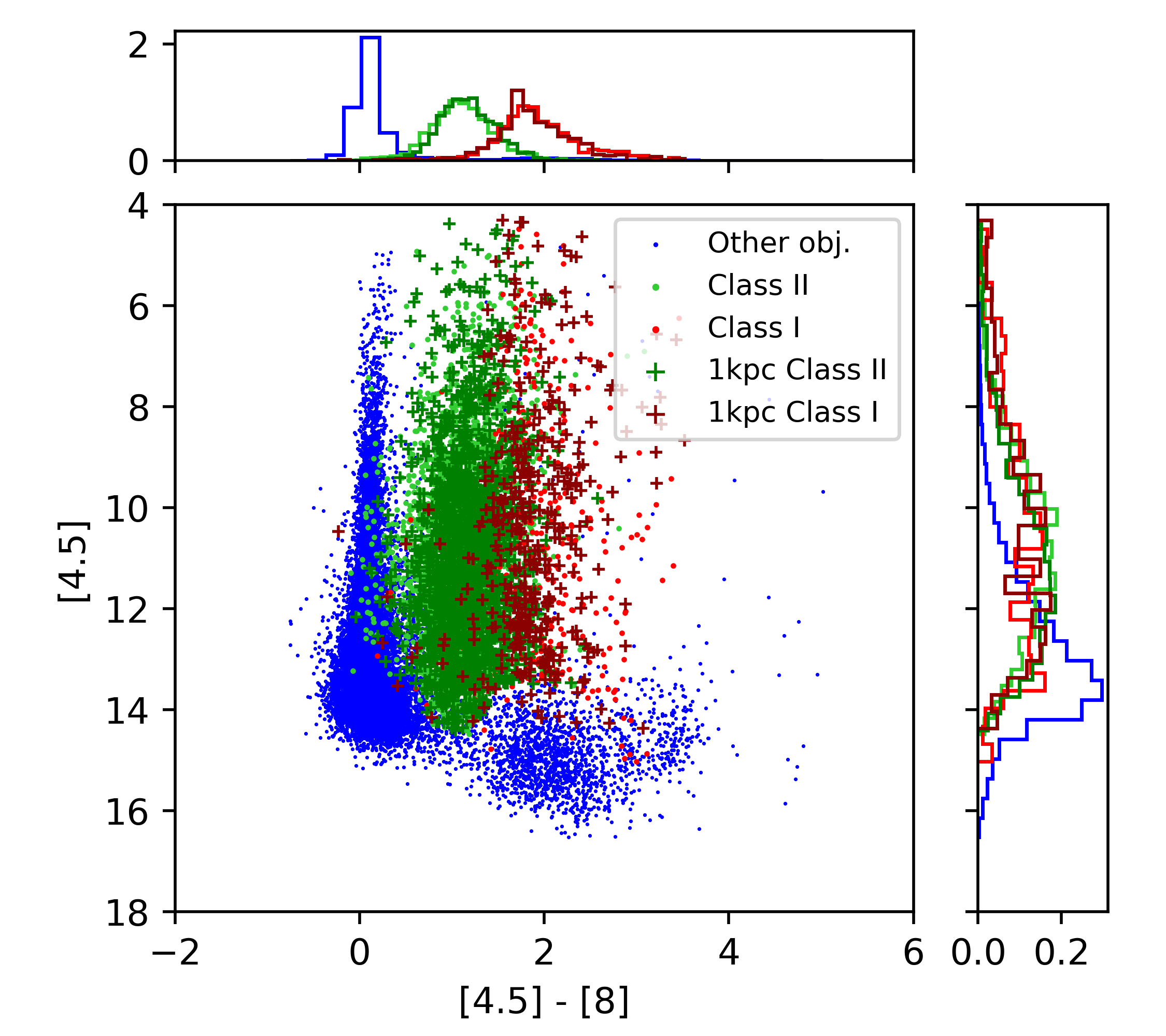}
        \end{subfigure}
        \caption{Differences in feature space coverage for our datasets. The CI YSOs, CII YSOs, and contaminants are shown in red, green, and blue, respectively, according to the simplified G09 classification scheme. The crosses in the last frame show the YSOs from the 1\,kpc sample. In the side frames, the area of each histogram is normalized to one.
}
        \label{datasets_space_coverage}
\end{figure*}

\begin{table}[!t]
        \small
        \centering
        \caption{Confusion matrix for the O-N case forwarded on the full NGC 2264 dataset.}
        \vspace{-0.1cm}
        \begin{tabularx}{\hsize}{r l |*{3}{m}| r }
        \multicolumn{2}{c}{}& \multicolumn{3}{c}{{Predicted}}&\\
        \cmidrule[\heavyrulewidth](lr){2-6}
        \parbox[l]{0.2cm}{\multirow{6}{*}{\rotatebox[origin=c]{90}{{Actual}}}} & Class & CI YSOs & CII YSOs & Others & Recall \\
        \cmidrule(lr){2-6}
         &  CI YSOs    & 74      & 2       & 14      & 82.2\% \\
         &  CII YSOs   & 6       & 402     & 27      & 92.4\% \\
         &  Others     & 9       & 52      & 7203    & 99.2\% \\
        \cmidrule(lr){2-6} 
         &  Precision & 83.1\% & 88.2\% & 99.4\% & 98.6\% \\
        \cmidrule[\heavyrulewidth](lr){2-6}
        \end{tabularx}
        \vspace{-0.1cm}
        \label{tab:ON_all} 
\end{table}

\begin{table}[!t]
        \small
        \centering
        \caption{Subclass distribution for the O-N case forwarded on the NGC 2264 dataset.}
        \vspace{-0.1cm}
        \begin{tabularx}{\hsize}{r l | *{6}{Y} l}  
        \multicolumn{2}{c}{}& \multicolumn{7}{c}{{Actual}}\\      
        \cmidrule[\heavyrulewidth](lr){2-9}
        \parbox[l]{0.0cm}{\multirow{5}{*}{\rotatebox[origin=c]{90}{{Predicted}}}} & & CI & CII & Gal & AGNs & Shocks & PAHs & Stars\\
        \cmidrule(lr){2-9}
          & CI & 74 & 6 & 0 & 3 & 5 & 0 & 1 \\
          & CII & 2 & 402 & 6 & 2 & 0 & 0 & 44 \\
          & Others & 14 & 27 & 108 & 245 & 0 & 1 & 6848 \\
        \cmidrule[\heavyrulewidth](lr){2-9}
        \end{tabularx}
        \vspace{-0.1cm}
        \label{tab:ON_all-sub}
\end{table}

In this section we present our test of the generalization capacity of the trained networks; we used the network trained on one region to classify the sources of the other one. This test is important because this is a typical use case: training the network on well-known regions, and using it on a new region. This is also a way to highlight more discrepancies between the datasets. For this we used the obtained trained networks from the O-O and N-N cases described in Sects.~\ref{orion_results} and \ref{NGC2264_results}. Since they are both built from the same original classification scheme (Sect.~\ref{data_prep}), we applied  one training directly to the other labeled dataset, which resulted in the two new cases O-N and N-O (see Table~\ref{results_cases}). However, the forwarded dataset must be normalized in the same way as the training set (Sect.~\ref{network_tuning}). Omitting this step would lead to deviations and distortions of our network class boundaries in the input parameter space, with a strong impact on the network prediction. One difficulty is that some objects end up with parameters outside the $[-1;1]$ range, corresponding to areas of the feature space where the network is not constrained. This effect could be partly hidden by excluding those out-of-boundary objects. However, they give an additional information about what kind of objects are missing in the respective training datasets and about the corresponding input feature space areas. Therefore, we preferred to keep them in the forward samples. It is legitimate here to use the full dataset directly to test the networks because none of its objects were used during the corresponding training. It also means that we forwarded datasets with different proportions than the ones they were trained with, but this is the expected end use of such networks. Moreover, both datasets are the results of observations, which means that our tests measured the effective performance of the trained network on a genuine observational use case with the corresponding proportions of classes.

It should be noted that in order to properly compare the results we needed to keep the exact same networks that produced the results in Tables~\ref{tab:OO}, \ref{tab:OO_all}, \ref{tab:NN}, and \ref{tab:NN_all}. Therefore, we did not estimate the dispersion of the prediction regarding the weight initialization, and the training set random selection on the O-N and N-O cases.

\begin{table}[!t]
        \small
        \centering
        \caption{Confusion matrix for N-O case forwarded on the full Orion dataset.}
        \vspace{-0.1cm}
        \begin{tabularx}{\hsize}{r l |*{3}{m}| r }
        \multicolumn{2}{c}{}& \multicolumn{3}{c}{{Predicted}}&\\
        \cmidrule[\heavyrulewidth](lr){2-6}
        \parbox[l]{0.2cm}{\multirow{6}{*}{\rotatebox[origin=c]{90}{{Actual}}}} & Class & CI YSOs & CII YSOs & Others & Recall \\
        \cmidrule(lr){2-6}
         &  CI YSOs    & 285     & 33      & 6       & 88.0\% \\
         &  CII YSOs   & 54      & 1967    & 203     & 88.4\% \\
         &  Others     & 98      & 293     & 16175   & 97.6\% \\
        \cmidrule(lr){2-6} 
         &  Precision & 65.2\% & 85.8\% & 98.7\% & 96.4\%\\
        \cmidrule[\heavyrulewidth](lr){2-6}
        \end{tabularx}
        \vspace{-0.1cm}
        \label{tab:NO_all} 
\end{table}
        
\begin{table}[!t]
        \small
        \centering
        \vspace{-0.1cm}
        \caption{Subclass distribution for the N-O case forwarded on the full Orion dataset.}
        \begin{tabularx}{\hsize}{r l | *{6}{Y} l}       
        \multicolumn{2}{c}{}& \multicolumn{7}{c}{{Actual}}\\      
        \cmidrule[\heavyrulewidth](lr){2-9}
        \parbox[l]{0.0cm}{\multirow{5}{*}{\rotatebox[origin=c]{90}{{Predicted}}}} & & CI & CII & Gal & AGNs & Shocks & PAHs & Stars\\
        \cmidrule(lr){2-9}
         &  CI & 285 & 54 & 8 & 37 & 12 & 39 & 2 \\
         &  CII & 33 & 1967 & 18 & 34 & 15 & 27 & 199 \\
         &  Others & 6 & 203 & 381 & 1070 & 1 & 21 & 14702 \\
        \cmidrule[\heavyrulewidth](lr){2-9}
        \end{tabularx}
        \vspace{-0.1cm}
        \label{tab:NO_all-sub}
\end{table}

Regarding the results from O-N in Tables~\ref{tab:ON_all} and \ref{tab:ON_all-sub}, we see that the recall for CI YSOs is lower by approximately $8\%$ than that for the O-O case (Table~\ref{tab:OO}) and lower by approximately $12\%$ when compared to the Orion full dataset results (Table~\ref{tab:OO_all}). Similarly, CII YSOs have a recall lower by approximately $5\%$. This difference is much greater than the dispersion of our results on the O-O case, which indicates that the Orion data lack some specific information that is contained in NGC 2264 for these classes. This should correspond to differences in feature space coverage, but these differences might be subtle in the limited set of CMDs considered in the G09 method, whereas the network works directly in the ten-dimensional space composed of the five bands and five errors. For example, as shown in Fig.~\ref{datasets_space_coverage}, it is striking that both YSO classes cover less the upper part of the diagram ([4.5] < 9) in the NGC 2264 case than for Orion. The slopes of the normalized histograms in this figure also illustrate that the density distributions are different between Orion and NGC 2264, especially for CI YSOs. For this population Orion presents a virtually symmetrical peaked distribution of [4.5]-[8] centered near [4.5]-[8] = 1.9 mag, while NGC 2264 shows a flatter and more skewed distribution. Although subtle, this specificity of the parameter space coverage is in line with the drop in the CI YSO recall in the O-N case, since in Orion the area located at [4.5]-[8] $>2$ is less constrained than for [4.5]-[8] $\approx 1.9$, while in NGC 2264 the area at [4.5]-[8] $>2$ contains a larger fraction of CI YSOs. This interpretation is also consistent with the fact that in the O-N case, CI are mostly confused with objects from the Others class, in contrast with the O-O and N-N cases, suggesting a lack of constraint for the boundary between the CI and Others classes in the lower right area of the CI distribution in Fig.~\ref{datasets_space_coverage}, although the differences in class proportions may also contribute (see next paragraph). From the perspective of the network, it is likely that the weight values were more influenced by the more abundant weight updates from objects near the CI peak at [4.5]-[8] = 1.9 mag.
Physically, the observed differences in this CMD are likely to come from the different star formation histories and from the difference in distance between the two regions,  between $\sim$ 420 pc for Orion \citep{megeath_spitzer_2012} and $\sim$ 760 pc for NGC 2264 \citep{rapson_spitzer_2014}. In contrast, the Others class appears to be well represented, suggesting that the Orion training set contains enough objects to represent properly the inherent distribution of this class also in NGC 2264.

The changes in precision are less significant than those in recall, due to the differences in class proportions between the two datasets. For example, there is a  $1.58$ factor in the CI-to-Others ratio between Orion and NGC 2264. The number of Others misclassified  as CIs is then expected to rise, with a consequent impact on CI precision. However, for this case the improved Others recall between the O-O and O-N case of 0.6\% seems to overcome this effect partly. In contrast, the CII YSOs, for which the proportions are lowered by a $2.24$ factor, indeed suffer a $\sim 8\%$ drop in precision. This strong interplay between proportions and changes in recall for each class makes the differences in precision less prone to analysis.

Concerning the results from N-O in Tables~\ref{tab:NO_all} and \ref{tab:NO_all-sub}, the precision of CI YSOs dropped to $65.2\%$, in spite of the number of objects, sufficient not to be affected by small-number effects. This is the worst quality estimator value we observed in the whole study. The precision drop in CII YSOs is less important, and only $2\%$ lower than the NGC 2264 full dataset results. The impact of the differences in feature space coverage is even stronger than for the O-N case since there are almost no YSOs brighter than [4.5]=9 mag in NGC 2264; therefore, a large part of the feature space where many Orion objects lie is left unconstrained. The NGC 2264 dataset { also} lacks shocks and PAHs that are present in non-negligible proportions in the Orion dataset. Therefore, the NGC 2264 trained network did not constrain them, as confirmed in Table~\ref{tab:NO_all-sub}, where PAHs are evenly scattered in all output classes and where shocks are completely misclassified as YSOs. In addition to these flaws, the number of objects in the training set is too small to properly constrain the overall network architecture that suits this problem (Sect.~\ref{network_tuning}).

\subsection{Combined training}
\label{cross_train}

The two major limitations identified in the cases of Orion and NGC 2264 are (i) the lack of CI YSOs in the training datasets to be properly constrained by the network, with the associated reduction of other types of objects to avoid dilution, and (ii) the differences in feature space coverage for the two different regions, which induces a lack of generalization capacity toward new star-forming regions. A simple solution to overcome those limitations is to perform a combined training with the two clouds (Fig.~\ref{datasets_space_coverage}). We refer to this case as the C-C case, where we merged the labeled samples from Orion and NGC 2264, and used it  to train the network and to perform the forward step. Since the two labeled datasets were obtained with our modified G09 classification, they formed a homogeneous dataset and it was straightforward to combine them. We normalized this new Combined dataset as explained in Sect.~\ref{network_tuning}. The detailed subclass distribution of the target sample for this dataset is presented in Table~\ref{tab_selection}. Thanks to the larger number of CI YSOs in the labeled dataset, we were able to adopt a lower value of $\theta$ ($\theta = 0.2$) to build the test set, which proved to be large enough to mitigate the small-number effects for our output classes. It conserved most data in the training set, where they were needed to improve the classification quality. We note that merging the datasets led to slightly different observational proportions.

Table~\ref{sat_factors} shows the optimal $\gamma_i$ values obtained with the Combined dataset. The $\gamma_i$ values are very similar to those of Orion, as a result of Orion providing two to five times more objects than NGC 2264 to the Combined dataset. The dataset is globally larger, so that the optimal number of neurons could have been raised to represent the expected more complex boundaries in the parameter space. However, increasing the number of hidden neurons did not show any improvement of the end results. Thus, we kept 20 hidden neurons for this C-C case. Nevertheless, the larger size of the training set tended to stabilize the convergence of the network during the training, which allowed us to increase the learning rate to $\eta = 4\times 10^{-5}$. As shown in Sect.~\ref{network_tuning}, this is counterintuitive. { Since} the weight updates are computed as a sum over the objects in the training sample, they should be greater here than in previous cases, increasing the probability that the network misses local minima. On the other hand, the larger statistics improves the weight space resolution, mitigating those local minima that originate in the limited number of objects. It appears that the latter effect was dominant in this test, which allowed us to increase the learning rate even more. We kept the momentum value at $\alpha = 0.6$ (Sect.~\ref{optim_perf}) because a greater value  made the network diverge in the first steps of training when the weight corrections were too large.

        \begin{table}
        \small
        \centering
        \caption{Confusion matrix for the C-C case for a typical run.}
        \vspace{-0.1cm}
        \begin{tabularx}{\hsize}{r l |*{3}{m}| r }
        \multicolumn{2}{c}{}& \multicolumn{3}{c}{{Predicted}}&\\
        \cmidrule[\heavyrulewidth](lr){2-6}
        \parbox[l]{0.2cm}{\multirow{6}{*}{\rotatebox[origin=c]{90}{{Actual}}}} & Class & CI YSOs & CII YSOs & Others & Recall \\
        \cmidrule(lr){2-6}
         &  CI YSOs    & 77     & 2       & 3       & 93.9\% \\
         &  CII YSOs   & 9      & 514     & 8       & 96.8\% \\
         &  Others     & 9      & 49      & 4706    & 98.8\% \\
        \cmidrule(lr){2-6} 
         &  Precision & 81.1\% & 91.0\% & 99.8\% & 98.5\%\\
        \cmidrule[\heavyrulewidth](lr){2-6}
        \end{tabularx}
        \vspace{-0.1cm}
        \label{tab:CC}
        \end{table}
        
        \begin{table}
        \small
        \centering
        \caption{Subclass distribution of the C-C case.}
        \vspace{-0.1cm}
        \begin{tabularx}{\hsize}{r l | *{6}{Y} l}     
        \multicolumn{2}{c}{}& \multicolumn{7}{c}{{Actual}}\\      
        \cmidrule[\heavyrulewidth](lr){2-9}
        \parbox[l]{0.0cm}{\multirow{5}{*}{\rotatebox[origin=c]{90}{{Predicted}}}} & & CI & CII & Gal & AGNs & Shocks & PAHs & Stars\\
        \cmidrule(lr){2-9}
         &  CI & 77 & 9 & 1 & 3 & 3 & 2 & 0 \\
         &  CII & 2 & 514 & 0 & 3 & 3 & 4 & 39 \\
         &  Others & 3 & 8 & 103 & 272 & 0 & 11 & 4320 \\
        \cmidrule[\heavyrulewidth](lr){2-9}
        \end{tabularx}
        \vspace{-0.1cm}
        \label{tab:CC-sub}
\end{table}

        \begin{table}
        \small
        \centering
        \caption{Confusion matrix for the C-C case forwarded on the full dataset.}
        \vspace{-0.1cm}
        \begin{tabularx}{\hsize}{r l |*{3}{m}| r }
        \multicolumn{2}{c}{}& \multicolumn{3}{c}{{Predicted}}&\\
        \cmidrule[\heavyrulewidth](lr){2-6}
        \parbox[l]{0.2cm}{\multirow{6}{*}{\rotatebox[origin=c]{90}{{Actual}}}} & Class & CI YSOs & CII YSOs & Others & Recall \\
        \cmidrule(lr){2-6}
         &  CI YSOs    & 389    & 14      & 11      & 94.0\% \\
         &  CII YSOs   & 53     & 2570    & 36      & 96.7\% \\
         &  Others     & 50     & 254     & 23526   & 98.7\% \\
        \cmidrule(lr){2-6}  
         &  Precision & 79.1\% & 90.6\% & 99.8\% & 98.4\% \\
        \cmidrule[\heavyrulewidth](lr){2-6}
        \end{tabularx}
        \label{tab:CC_all} 
        \vspace{-0.1cm}
        \end{table}

\begin{table}
        \small
        \centering
        \caption{Subclass distribution for the C-C case forwarded on the full dataset.}
        \vspace{-0.1cm}
        \begin{tabularx}{\hsize}{r l *{7}{Y} }
        \multicolumn{2}{c}{}& \multicolumn{7}{c}{{Actual}}\\      
        \cmidrule[\heavyrulewidth](lr){2-9}
        \parbox[l]{0.0cm}{\multirow{5}{*}{\rotatebox[origin=c]{90}{{Predicted}}}} & & CI & CII & Gal & AGNs & Shocks & PAHs & Stars\\
        \cmidrule(lr){2-9}
         &  CI & 389 & 53 & 2 & 10 & 22 & 11 & 5 \\
         &  CII & 14 & 2570 & 4 & 16 & 11 & 15 & 208 \\
         &  Others & 11 & 36 & 515 & 1365 & 1 & 62 & 21583 \\
        \cmidrule[\heavyrulewidth](lr){2-9}
        \end{tabularx}
        \vspace{-0.1cm}
        \label{tab:CC_all-sub}
\end{table}

The results of this C-C case, presented in Table~\ref{tab:CC} and \ref{tab:CC-sub}, { were obtained by a forward on the test set of the Combined sample. They} are very close to those on the O-O case with the full Orion dataset. The largest difference is  $0.7\%$ for the precision of CI YSOs. The other differences are $\leq 0.2\%$. The stability of the results regarding both the weight initialization and the random selection of the test and training sets is also very similar to that of the O-O case (Sect.~\ref{orion_results}), with recall and precision values scattered by typically $\pm 0.5\%$, except for CI precision which is scattered by about $\pm 1\%$. These fluctuations exceed the differences between the O-O and C-C case, as observed from their confusion matrices, when considering the full dataset forward. This stability was not guaranteed since, on the one hand, the Combined training set is more general than previous training sets and, on the other hand, the Combined training is a more complex problem than a single-cloud training, due to the expected more complex distribution of objects in the input parameter space, especially for YSOs.
If the latter effect dominates the results could be expected to be less good than both the O-O and N-N results individually, or than any linear combination of them. We illustrate this idea with the following conservative reasoning. If, when using the Combined training dataset, the network had only learned from Orion objects, as might be argued due to their dominance in the Combined sample, then the state of the network should be very similar to that obtained in the O-O case. The C-C confusion matrix should then be a linear combination of those of the O-O (Table~\ref{tab:OO_all}) and O-N cases (Table~\ref{tab:ON_all}), weighted by the respective abundances of Orion and NGC 2264 in the forward sampling. The recall of CI YSOs in the O-O and O-N was 94.1\% and 82.2\%, respectively. Since, in the Combined dataset, 78.3\% of CI YSOs come from Orion, the expected recall from an Orion dominated network would be 91.5\%. Considering the obtained value of 93.9\% in the C-C case (Table~\ref{tab:CC}), with a $\pm 1\%$ dispersion, the network has indisputably learned information from the NGC 2264 objects, and the increased complexity of the problem was more than balanced by the increased generality of the sample. In other words, the fact that the results of the C-C test are as good as those of the O-O test in spite of the increased complexity implies that the network managed to take advantage of the greater generality of the Combined sample to find a better generalization.
The analysis of the other two classes does not contradict these conclusions, although the improvement for CII objects is only marginal since the same reasoning applied to CII YSOs leads to a recall of 96.2\%, to be compared to the C-C value of 96.8\%, with $\pm 0.5\%$ dispersion. This is in line with the fact that the CII YSO coverage in Orion was already close to that of NGC 2264, as highlighted by the less than $1\%$ difference between the CII YSO recall in O-N and N-N. Finally, contaminants are dominated by subclasses that were already nicely constrained in the O-O and N-N cases.

The fact that the network results for the C-C case are as good as or better than for the Orion case despite the added complexity confirms that the number of objects was a strong limitation in the O-O and N-N cases. It also confirms that the O-O training might have provided better results with more observed objects in the same region, which was already established from the improvement of results with lower $\theta$ values in Sect.~\ref{orion_results}. Moreover, the absence of positive effect when raising the number of neurons demonstrates that the network efficiently combined their respective input parameter space coverage and that $n = 20$ does not limit these results. The change in observational proportions that occurred by merging the two datasets seems to have a negligible impact as they are still close to the Orion values, but adding more regions with fewer YSOs is expected to decrease the precision values for YSOs by increasing their dilution by the Others class.

The results for the complete Combined dataset are presented in Tables~\ref{tab:CC_all} and \ref{tab:CC_all-sub}. As before, the results appear to be free from over-training since there is no noticeable increase in recall for any of our classes. These results are very similar to the previous ones, with differences in quality estimators of the same order as the dispersion observed with random weight initialization. The slight decrease in precision of CI YSOs is also of the same order as the dispersion obtained from the random selection of our training and test samples. The contaminants that are not sufficiently constrained, like Shocks, could also be affected by selection effects between the two sets, which could lead to such a dispersion in precision for CI YSOs. This seems to be confirmed by the fact that two-thirds of the shocks were misclassified as CI YSOs. Interestingly, this suggests a change in the network behavior compared to the O-O case, where shocks were almost evenly distributed among the three output classes. We interpret the difference in shock distributions as a consequence of the difference in the relative abundance of this subclass compared to the rest of the training set, and to its strong dependency to the MIPS rebranding step. The special location of Shocks in the feature space, close to CII YSOs and mixed with the MIPS-identified CI YSOs (Fig.~\ref{fig_gut_method}\,D), makes this subclass identification sensitive to its small relative abundance during the learning process. Thus, in the O-O case, the number of shocks in the sample enabled the network to place the boundary in the vicinity of the Shocks region, but in an inaccurate way, hence the even distribution. Conversely, the lower fraction of shocks in the C-C sample probably made the network find an optimum where most of its representative strength was used for other parts of the feature space. In this situation the majority of shocks are likely to be included in one specific output class, which can vary according to the random training set selection, but is more likely to be a YSO class, and even more likely to be CI due to the MIPS rebranding step.

To summarize the results of this Combined training, we showed that combining two star-forming clouds  improved the underlying diversity of our prediction, and therefore the generalization capability of our network over possible new regions. The added complexity was largely overcome by the increased statistics on our classes of interest, CI and CII YSOs, which allowed us to {maintain} very good accuracy and precision for them. However, some rare contaminant subclasses suffered from their increased dilution.

\subsection{Multiple-cloud training}
\label{1kpc_train}

In this section we present the advantages of the 1\,kpc dataset to further improve the network generalization capacity by increasing the underlying diversity of the object sample. As discussed in Sect.~\ref{data_setup}, this dataset only contains YSOs. This is not a major issue because most of our contaminant subclasses are already well constrained, while we have shown that it is not the case for YSOs since adding more of them led to a better generalization. As the dataset contains several regions, it should ensure an even better diversity and input parameter space coverage for YSOs than the previous C-C case, but it might also increase again the underlying distribution complexity (Fig.~\ref{datasets_space_coverage}). In this section, we study the F-C case, that is a training on the Full 1\,kpc dataset (Combined + 1\,kpc YSOs) and a forward on the test set of the full Combined dataset, to keep a realistic test dataset with almost observational proportions. As before, the Full 1\,kpc dataset is normalized as described in Sect.~\ref{network_tuning}.

The detailed $\gamma_i$ selection for this more complicated dataset is presented in Table~\ref{sat_factors}. Because we added YSOs, we had to increase the number of contaminants to preserve their dominant representation in the training sample. However, some subclasses of contaminants were already too few in the C-C case and already included in the training set as much as possible. Therefore, we did not add all the CI YSOs at our disposal to avoid a too strong dilution of these subclasses of contaminants. For objects from the Combined dataset, we kept $\theta = 0.2$. In the same manner as for the other datasets, we tried various numbers of neurons in the hidden layer, and {for the first time the optimum value is higher with around $n=30$}. This means that we might have sufficiently raised the number of objects to break previously existing limitations regarding the size of the network. We also took advantage of the larger dataset and adopted greater values for $\eta = 8\times 10^{-5}$ and $\alpha = 0.8$.

The results for this F-C case are presented in Tables~\ref{tab:FC} and \ref{tab:FC-sub}. The precision of CI YSOs has dropped by $2.5\%$, but all the other precisions have slightly improved. Compared to the C-C case (Table~\ref{tab:CC}), the precision of CI YSOs raised by $2.8\%$, but the recall is significantly lowered with a drop of nearly $5\%$. In contrast, the precision of CII YSOs dropped by $1.2\%$, and the recall improved by $0.8\%$. Overall, these results are similar to the previous C-C case, despite the increase in complexity coming from the addition of YSOs from new star-forming regions. Similarly to the combination of Orion and NGC 2264, we could have observed a stronger drop in quality estimators because the problem becomes more general and therefore more difficult to constrain. It is worth noting that the stability of the network somewhat decreased in comparison to the O-O and C-C cases. We observed a dispersion of recall regarding the weight random initialization of about $\pm 1\%$ for CI and $\pm 0.7\%$ for CII YSOs. This dispersion affects  the Others class less, with a value of approximately $\pm 0.15\%$. The precision is less reliable with a dispersion of nearly $\pm 1.5\%$ for CI YSOs. The precision dispersion for CII YSOs is around $\pm 0.5\%$ and is less than $\pm 0.1\%$ for the Others class.

More generally, the sources of contamination of the YSO classes have not changed; however,  their overall effect is just greater. The fact that increasing the number of neurons from $20$ to $30$ in the network leads to better results is certainly an indication of the increased complexity of this problem. This means that the network uses more refined splittings in the input parameter space. However, there might not be enough objects in our dataset to perfectly constrain this larger network, despite the added YSOs. This naturally leads to a greater sensitivity to the weight initialization. In contrast, the dispersion over the training set random selection is similar to that observed in the C-C case and is of the same order as the weight initialization dispersion. As in the previous cases, the results show that the main source of contamination of CI YSOs are the CII YSOs, while the latter are mostly contaminated by the Others class. This is, as {in} the previous cases, an indication of the respective proximity of the three classes in the input parameter space.

The results of a forward of the complete Combined dataset using this network are shown in Table~\ref{tab:FC_all}, with the subclass distributions in Table~\ref{tab:FC_all-sub}. These results show a $2.3\%$ increase in the CI YSO recall compared to Table~\ref{tab:FC}, and a $2.8\%$ drop in precision for the same class. { Similarly to} all the previous cases, the Others class remained almost identical. For CII YSOs and Others the variations in precision and recall are within the weight initialization dispersion. The case of CI YSOs is less clear as their recall increase is greater than their dispersion, which could mean that there is a slight over-training. However, when searching for the optimum set of $\gamma_i$ values, we observed that the sets leading to  less over-training of CI YSOs also degraded the overall quality of the results. Even so, it suggests that the genuine CI YSO recall is between the values listed in  Table~\ref{tab:FC} and Table~\ref{tab:FC_all}.

The increased number of objects provided us with more details on the subclass distribution across the output classes. Similarly to the C-C case, the Shocks behave as completely unconstrained since they end up in mostly one class, which changes randomly when the training is repeated. Compared to the C-C case this effect is stronger, most likely because we did not add any Shocks in the training sample, therefore increasing their dilution. For almost any of the other subclasses, the variations are  within the dispersion;  overall there is  a slight trend for contaminant subclasses (galaxies, AGNs, shocks, PAH) to be less well classified, and CII YSOs and Stars to be better classified. These results  are expected  because we increased the number of YSOs and Stars in the training sample. On the other hand, we also  increased the YSO distribution complexity, which could lead to worse overall results. This is possibly what induced the slight drop in CI YSO recall observed from C-C to F-C, whereas CII YSOs and Others kept their quality indicators stable, either due to the increased statistics or because their input parameter space was already properly constrained by the Combined dataset (C-C case).

\begin{table}[t]
        \small
        \centering
        \caption{Confusion matrix for the F-C case for a typical run.}
 \vspace{-0.1cm}
        \begin{tabularx}{\hsize}{r l |*{3}{m}| r }
        \multicolumn{2}{c}{}& \multicolumn{3}{c}{{Predicted}}&\\
        \cmidrule[\heavyrulewidth](lr){2-6}
        \parbox[l]{0.2cm}{\multirow{6}{*}{\rotatebox[origin=c]{90}{{Actual}}}} & Class & CI YSOs & CII YSOs & Others & Recall \\
        \cmidrule(lr){2-6}
         &  CI YSOs    & 73      & 4       & 5       & 89.0\% \\
         &  CII YSOs   & 9       & 518     & 4       & 97.6\% \\
         &  Others     & 5       & 55      & 4704    & 98.7\% \\
        \cmidrule(lr){2-6}
         &  Precision & 83.9\% & 89.8\% & 99.8\% & 98.5\% \\
        \cmidrule[\heavyrulewidth](lr){2-6}
        \end{tabularx}
        \vspace{-0.1cm}
        \label{tab:FC} 
\end{table}

\begin{table}[t]
        \small
        \centering
        \caption{Subclass distribution for the F-C case.}
        \vspace{-0.1cm}
        \begin{tabularx}{\hsize}{r l | *{6}{Y} l}       
        \multicolumn{2}{c}{}& \multicolumn{7}{c}{{Actual}}\\      
        \cmidrule[\heavyrulewidth](lr){2-9}
        \parbox[l]{0.0cm}{\multirow{5}{*}{\rotatebox[origin=c]{90}{{Predicted}}}} & & CI & CII & Gal & AGNs & Shocks & PAHs & Stars\\
        \cmidrule(lr){2-9}
         &  CI & 73 & 9 & 0 & 0 & 1 & 2 & 2 \\
         &  CII & 4 & 518 & 1 & 6 & 5 & 6 & 37 \\
         &  Others & 5 & 4 & 102 & 272 & 0 & 9 & 4321 \\
        \cmidrule[\heavyrulewidth](lr){2-9}
        \end{tabularx}
        \vspace{-0.1cm}
        \label{tab:FC-sub}
\end{table}

\begin{table}[t]
        \small
        \centering
        \caption{Confusion matrix for the F-C case forwarded on the full Combined dataset.}
        \vspace{-0.1cm}
        \begin{tabularx}{\hsize}{r l |*{3}{m}| r }
        \multicolumn{2}{c}{}& \multicolumn{3}{c}{{Predicted}}&\\
        \cmidrule[\heavyrulewidth](lr){2-6}
        \parbox[l]{0.2cm}{\multirow{6}{*}{\rotatebox[origin=c]{90}{{Actual}}}} & Class & CI YSOs & CII YSOs & Others & Recall \\
        \cmidrule(lr){2-6}
         &  CI YSOs    & 378     & 22      & 14      & 91.3\% \\
         &  CII YSOs   & 45      & 2584    & 30      & 97.2\% \\
         &  Others     & 43      & 244     & 23543   & 98.8\% \\
        \cmidrule(lr){2-6}
         &  Precision & 81.1\% & 90.7\% & 99.8\% & 98.5\% \\
        \cmidrule[\heavyrulewidth](lr){2-6}
        \end{tabularx}
        \vspace{-0.1cm}
        \label{tab:FC_all}
\end{table}

\begin{table}[t]
        \small
        \centering
        \caption{Subclass distribution for the F-C case forwarded on the full Combined dataset.}
        \vspace{-0.1cm}
        \begin{tabularx}{\hsize}{r l | *{6}{Y} l}       
        \multicolumn{2}{c}{}& \multicolumn{7}{c}{{Actual}}\\      
        \cmidrule[\heavyrulewidth](lr){2-9}
        \parbox[l]{0.0cm}{\multirow{5}{*}{\rotatebox[origin=c]{90}{{Predicted}}}} & & CI & CII & Gal & AGNs & Shocks & PAHs & Stars\\
        \cmidrule(lr){2-9}
         &  CI & 378 & 45 & 0 & 15 & 8 & 15 & 5 \\
         &  CII & 22 & 2584 & 6 & 22 & 25 & 14 & 177 \\
         &  Others & 14 & 30 & 515 & 1354 & 1 & 59 & 21614 \\
        \cmidrule[\heavyrulewidth](lr){2-9}
        \end{tabularx}
        \vspace{-0.1cm}
        \label{tab:FC_all-sub}
\end{table}

\subsection{Final classification and public catalog}
\label{sec:publiccat}
{ 
Among all our attempts, the multiple cloud training (F-C, Sect.~\ref{1kpc_train}) is clearly the most reliable one for  general use.
We are confident that our Full 1\,kpc trained network contains a sufficient diversity of subclasses to be efficiently applied to most nearby ($\lesssim 1$ kpc) star-forming regions. Our results show that one can expect nearly $90\%$ of the CI YSOs to be properly recovered with a precision above $80\%$, while nearly $97\%$ of the CII YSOs are expected to be recovered with a $90\%$ precision.

The table containing the YSO candidates from the Orion and NGC 2264 regions is made public and is available at the CDS. It includes all objects from the catalogs by \citet{megeath_spitzer_2012} and \citet{rapson_spitzer_2014}, as described in Sect.~\ref{data_setup}, and Table~\ref{yso_catalog} shows an excerpt of our catalog. As a preview to Sect.~\ref{proba_discussion}, we note here that a membership probability is also provided for each source.
}

\section{Discussion}
\label{sec:discussion}

\subsection{Current limitations to the classification scheme}
\label{sec:limits}

\begin{figure*}[!ht]
        \centering
        \begin{subfigure}[t]{0.4\textwidth}
        \caption*{{Missed}}
        \includegraphics[width=\textwidth]{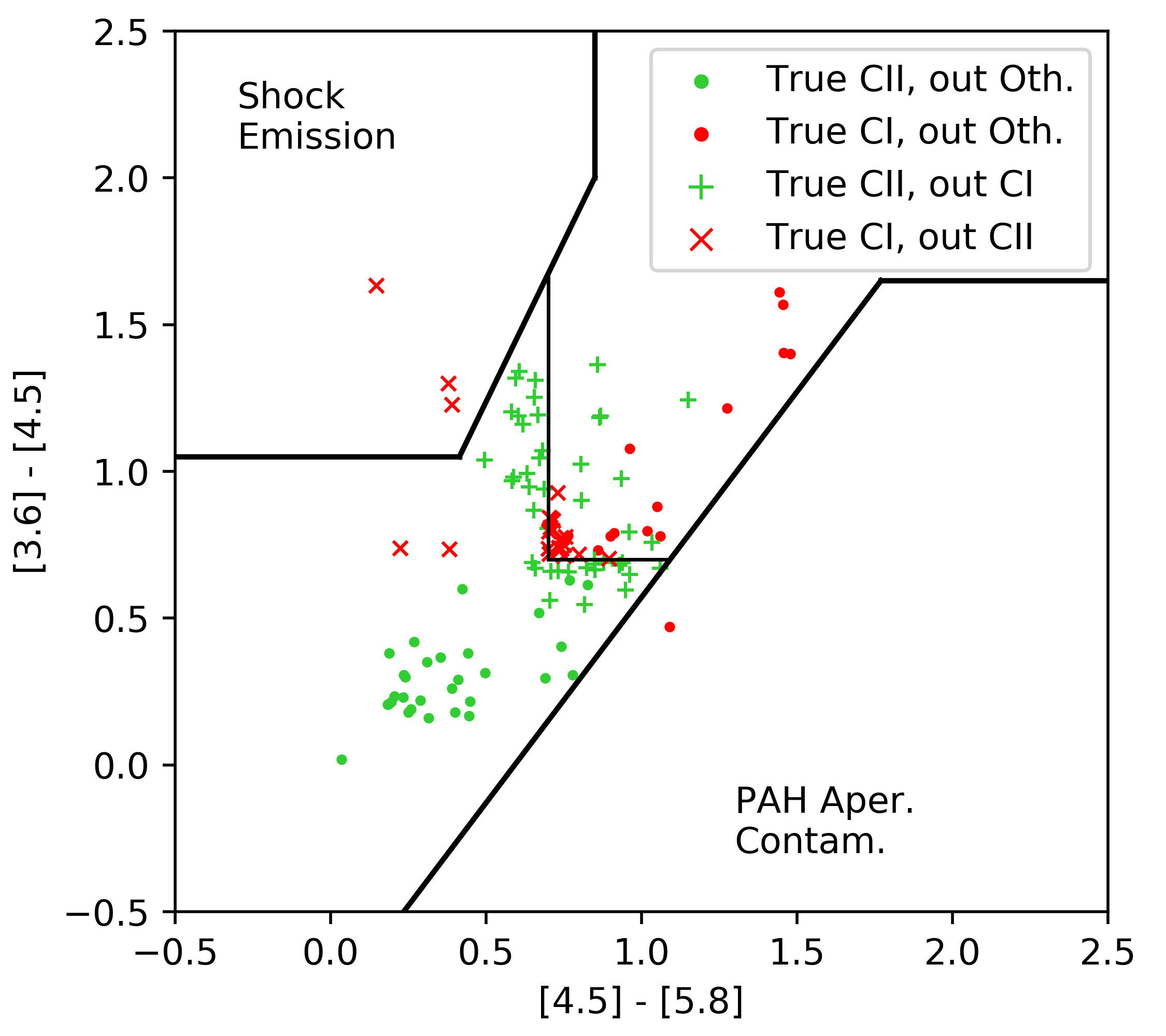}
        \end{subfigure}
        \begin{subfigure}[t]{0.4\textwidth}
        \caption*{{Wrong}}
        \includegraphics[width=\textwidth]{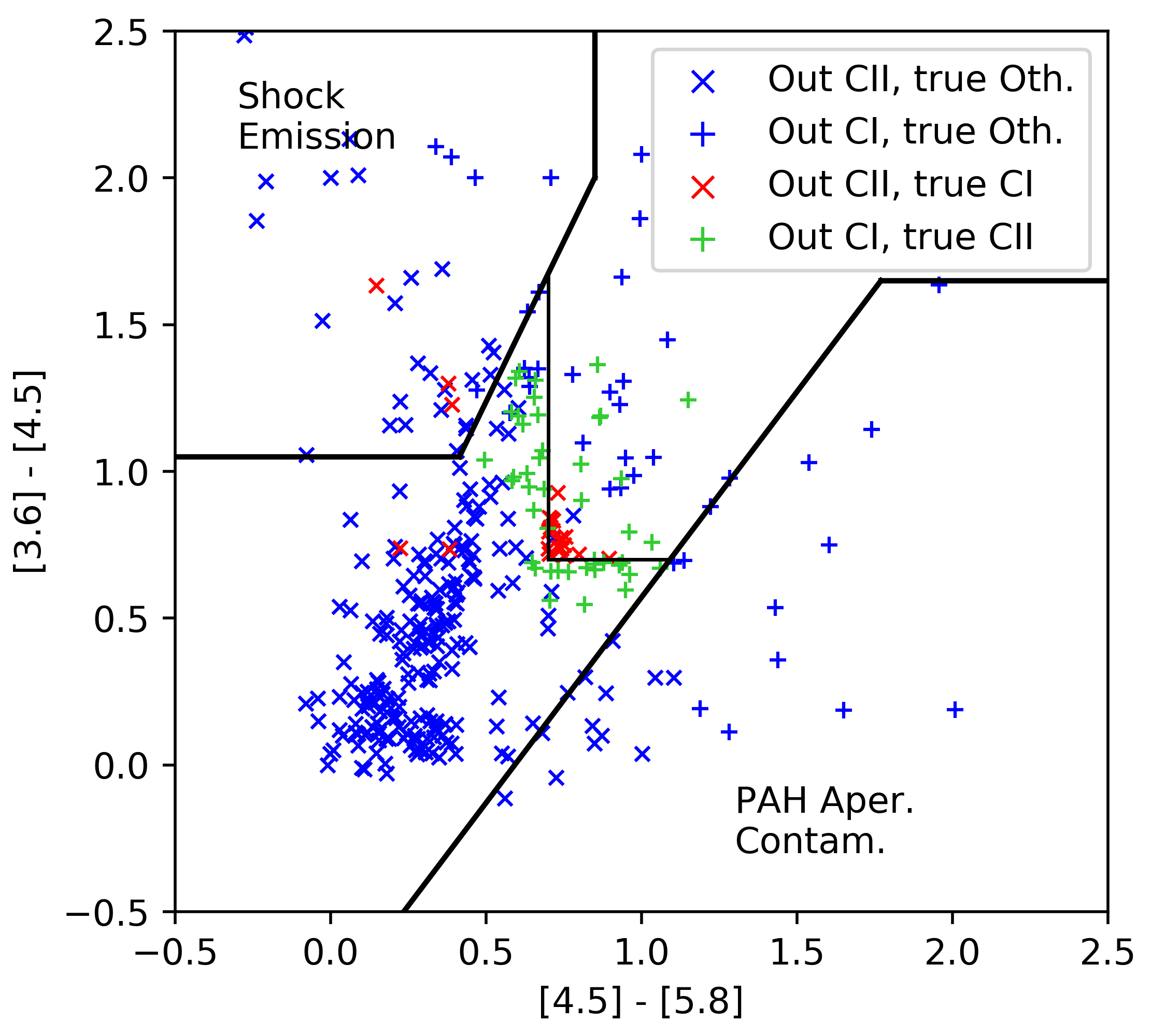}
        \end{subfigure}
        \caption{Zoom-in on the $[4.5] - [5.8]\, \text{vs.}\, [3.6] - [4.5]$ graph, for misclassified objects in the F-C case. {\textit{Missed:} Genuine CI and CII YSOs according to the labeled dataset that were misclassified by the network. \textit{Wrong:} Predictions of the network that are known to be incorrect based on the labeled dataset. In both frames, green is for genuine CII YSOs, red for genuine CI YSOs, and blue is for genuine contaminants. The symbol shapes indicate the predicted output class as specified in the legends.}}
        \label{missed_wrong_zoom}
\end{figure*}

\begin{figure*}[!ht]
        \centering
        \begin{subfigure}[t]{0.24\textwidth}
        \caption*{{Output}}
        \includegraphics[width=\textwidth]{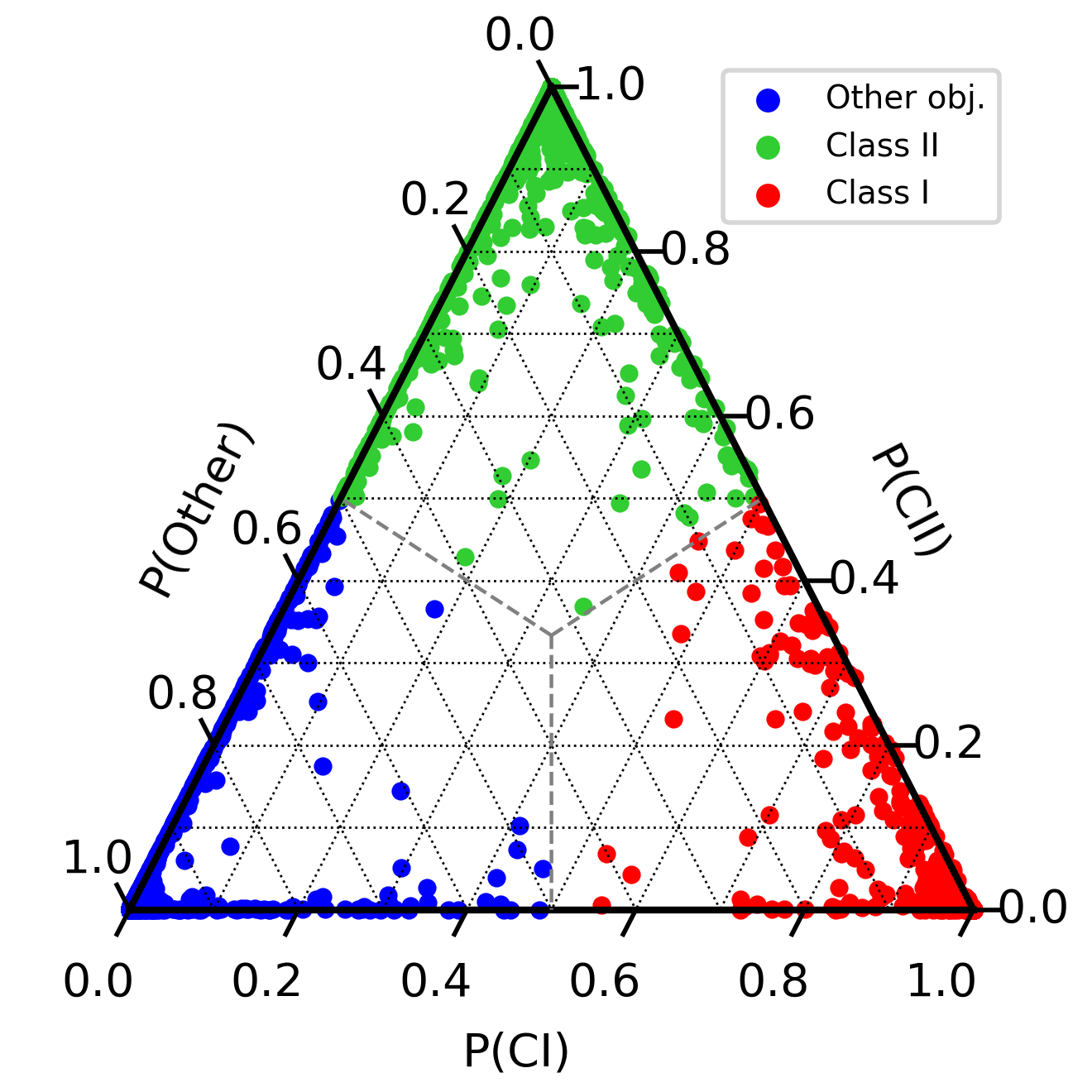}
        \end{subfigure}
        \begin{subfigure}[t]{0.24\textwidth}
        \caption*{{Correct}}
        \includegraphics[width=\textwidth]{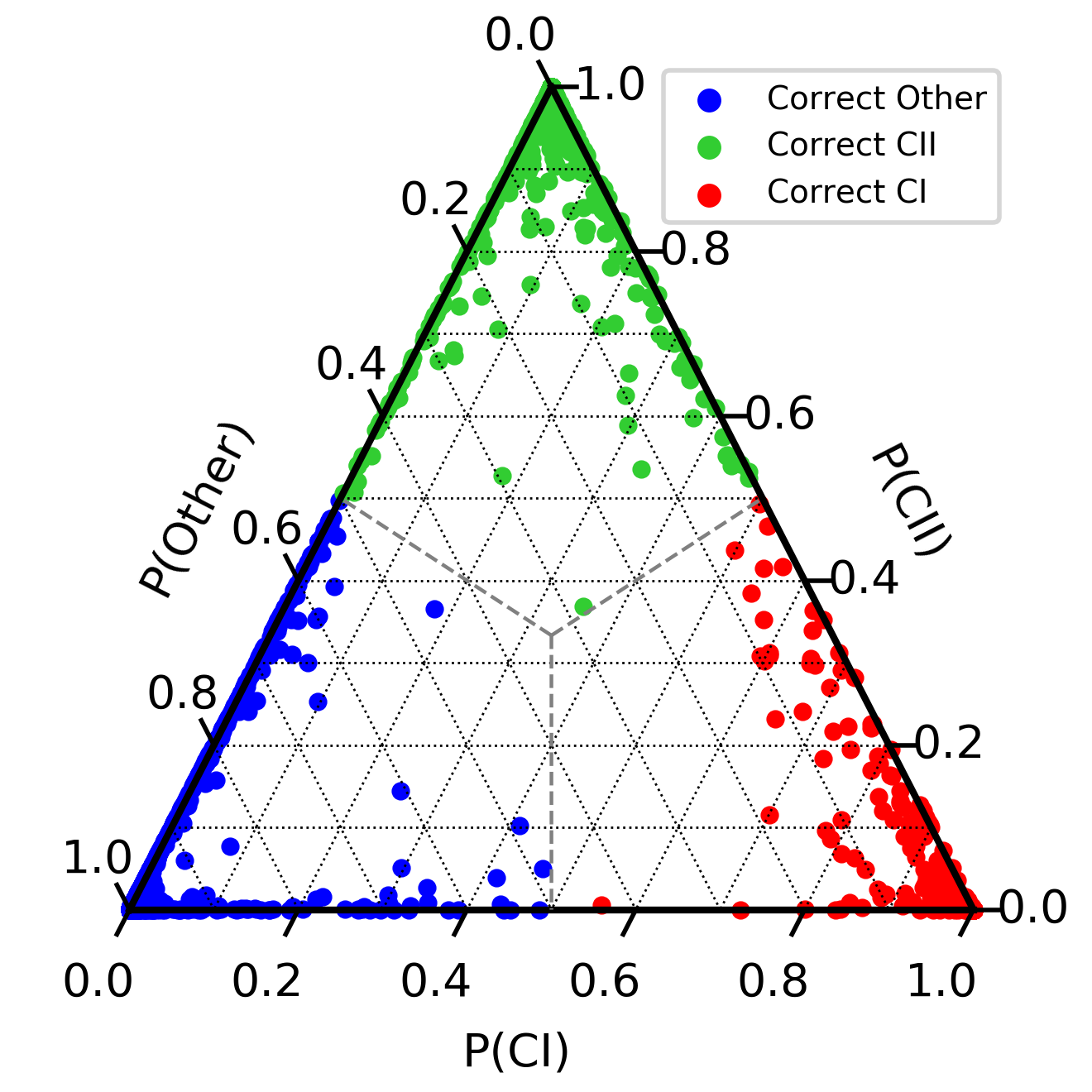}
        \end{subfigure}
        \begin{subfigure}[t]{0.24\textwidth}
        \caption*{{Missed}}
        \includegraphics[width=\textwidth]{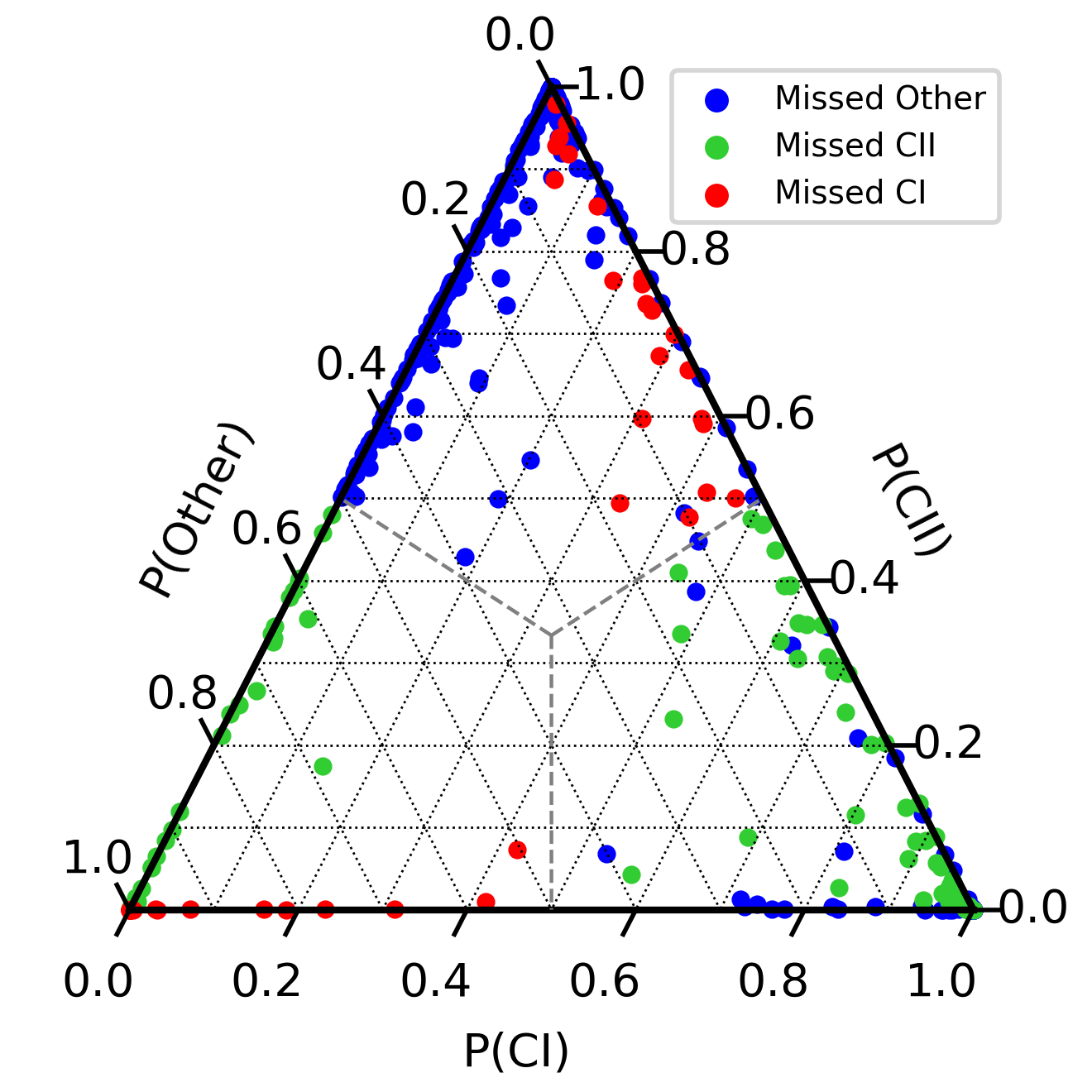}
        \end{subfigure}
        \begin{subfigure}[t]{0.24\textwidth}
        \caption*{{Wrong}}
        \includegraphics[width=\textwidth]{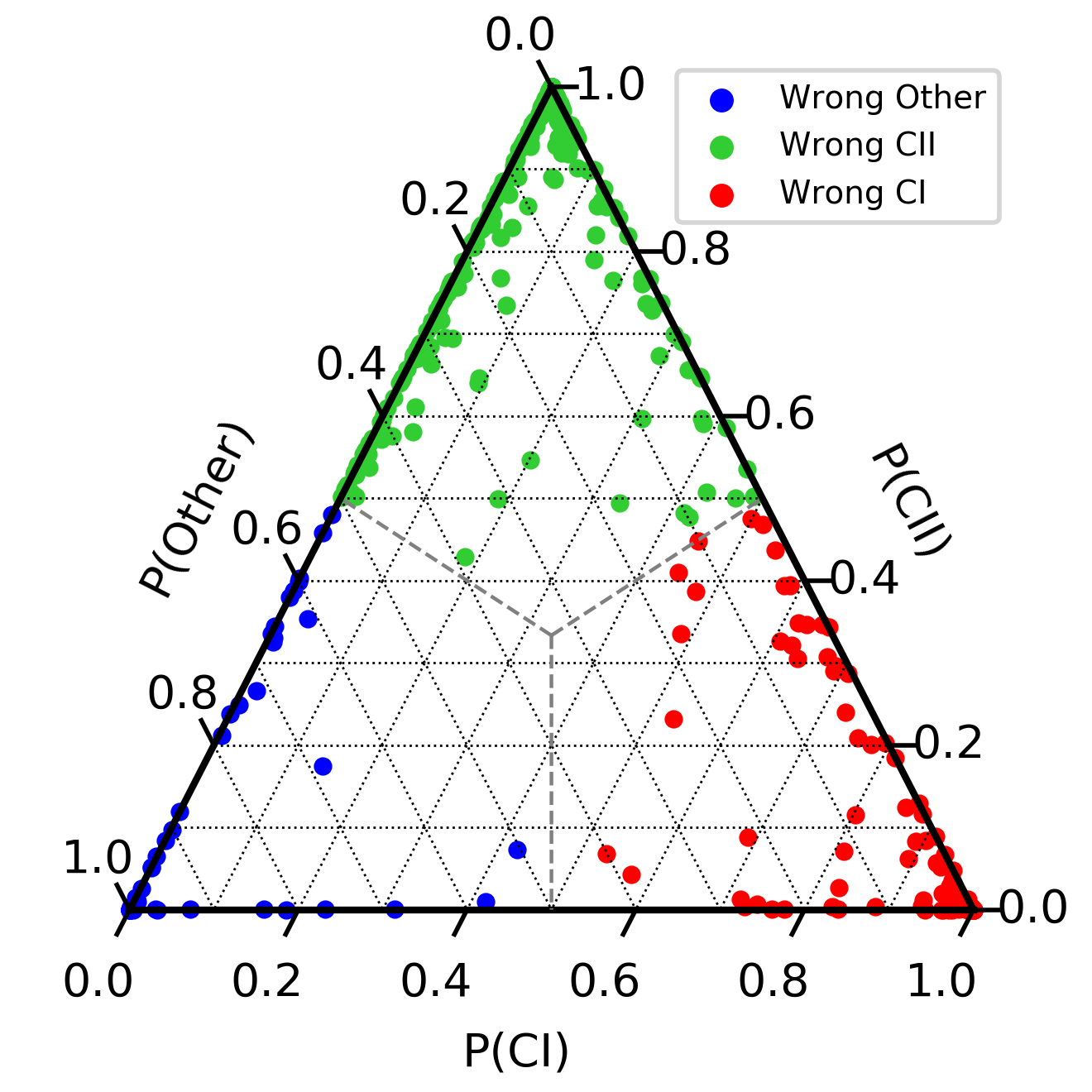}
        \end{subfigure}
        \caption[]%
        {Ternary plots of output membership probability for each class in the F-C case forwarded on the full dataset. \textit{Output:} All objects. \textit{Correct:} Genuine and predicted classes are identical. \textit{Missed:} Misclassified objects colored according to their genuine class. \textit{Wrong:} Misclassified objects colored according to their predicted class.}
        \label{ternary_plots}
\end{figure*}

\begin{figure*}[!ht]
        \centering
        \begin{subfigure}[t]{0.24\textwidth}
        \caption*{{Output}}
        \includegraphics[width=\textwidth]{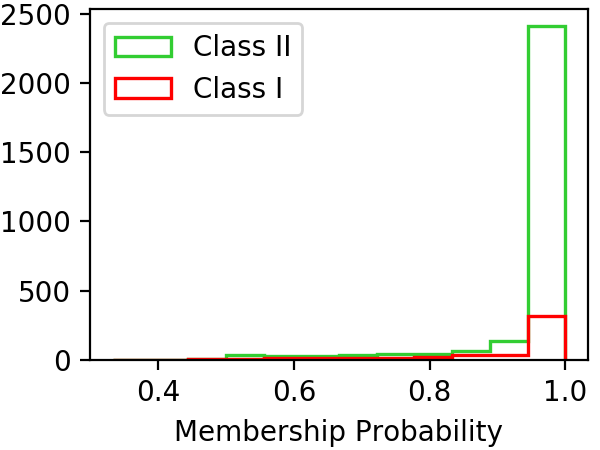}
        \end{subfigure}
        \begin{subfigure}[t]{0.24\textwidth}
        \caption*{{Correct}}
        \includegraphics[width=\textwidth]{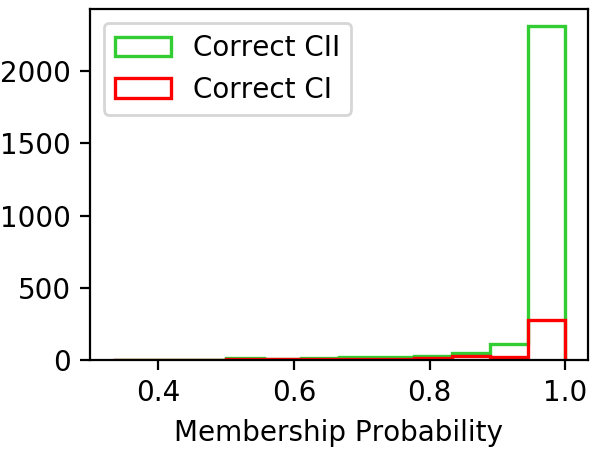}
        \end{subfigure}
        \begin{subfigure}[t]{0.24\textwidth}
        \caption*{{Missed}}
        \includegraphics[width=\textwidth]{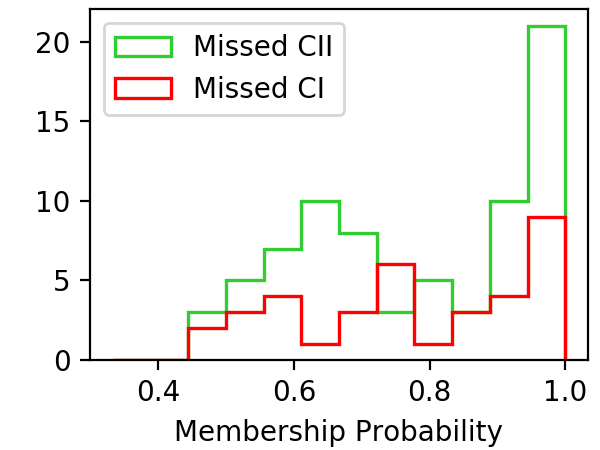}
        \end{subfigure}
        \begin{subfigure}[t]{0.24\textwidth}
        \caption*{{Wrong}}
        \includegraphics[width=\textwidth]{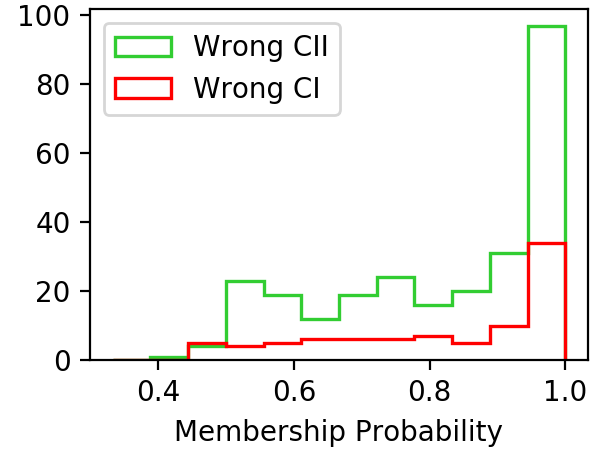}
        \end{subfigure}
        \caption{Histograms of membership probability for YSO classes regarding different populations in the F-C case forwarded on the full dataset. \textit{Output:} All objects. \textit{Correct:} Genuine and predicted classes are identical. \textit{Missed:} Misclassified objects colored according to their genuine class. \textit{Wrong:} Misclassified object colored according to their predicted class.}
        \label{hist_proba}
\end{figure*}

With the dataset selected for this paper the quality of our results is mostly dependent on the proper choice of the $\gamma_i$ factors, that is to say that the main limitation comes from the construction of our labeled dataset. It is indeed expected to be the most critical part of any ML application because the network only provides results that are  as good as the input data. One of our major issues is that some subclasses of rare contaminants remain poorly constrained, like Shocks or PAHs, which leads to an important contamination of the YSO classes. The non-homogeneity between the 1\,kpc small cloud dataset and the other datasets worsen this effect by increasing the dilution of these rare subclasses (Sect.~\ref{cross_train}). They are almost evenly distributed across output classes in the O-O case, revealing that the network was not able to identify enough constraints on those objects. In contrast, for the C-C and F-C cases, they are randomly assigned to an output class. This means that they are completely unconstrained by the network, which failed to disentangle them from the noise of another class. This effect appeared in those specific cases due to the increased dilution of those subclasses of contaminants.

The main source of contamination for CII YSOs is the Star subclass. Adding more of them has proven to improve their classification quality (Sects.~\ref{orion_results} and \ref{cross_train}), but at the cost of even more dilution of all the other subclasses, which has a stronger negative impact on the global result. Similarly, YSO classes themselves should be more present to further improve their recall, but again at the cost of an increased dilution of the contaminant subclasses. The confusion between CI and CII YSOs is illustrated by Fig.~\ref{missed_wrong_zoom}, where the misclassified YSOs of both CI and CII accumulate at the boundary between them in the input parameter space. This figure also illustrates the CII contamination from Others with the same kind of stacking, where the two classes are close to each other. A similar representation for all the CMDs is provided in Fig.~\ref{missed_wrong_space}.

Overall, we lack the data to get better results. Large Spitzer point source catalogs are available, but the original classification from \citet{gutermuth_spitzer_2009} was tailored for relatively nearby star-forming regions where YSOs are expected to be observed. Therefore, using a non-specific large Spitzer catalog would mostly add non-star-forming regions, which would create a significant number of false positive YSOs. In practice, these false positive YSOs would overwhelmingly contaminate the results, and the network performance would drop to the point where more than $50\%$ of CI YSOs are false positive. However, since one of our main limitations is the number of contaminants, a large Spitzer catalog could be used to increase the number of rare contaminants in the training sample by selecting areas that are known to be clear of YSOs. Unfortunately, this approach would mostly provide us with more Stars, Galaxies, and AGNs, which are already well constrained, while the two most critical contaminant subclasses, Shocks and PAHs, originate mostly from star-forming regions.

\subsection{Effect of the $24\ \mu m$ MIPS band}
\label{sec:mips24}
We investigate here the impact of the MIPS $24\ \mu m$ band on the original classification, and therefore on the results of the network. As stated in Sect.~\ref{data_prep}, this band is used as a refinement step of the G09 method. Considering the classification performed using the four IRAC bands, it ensures that it is consistent with the $24\ \mu m$ emission where available, for example by testing whether the SED still rises at long wavelength to better distinguish between different YSO classes. However, it adds heterogeneity in the classification scheme since objects that do not present a MIPS emission cannot be refined. It makes the results harder to interpret and gives more work to the network as it has to learn an equivalent of this additional step. Moreover, the effect of this band on the end classification strongly affects some subclasses that are very rare in the dataset. For example, almost half of the objects initially classified as Shocks are reclassified as CI YSOs after this refinement step. Therefore, as it corresponds to a significant increase in complexity on very few objects, it is difficult to get the network to constrain them, considering the other limitations. It results in a strong contamination of the CI YSOs, as highlighted multiple times in our results.

On the other hand, most of the Spitzer large surveys miss a $24\ \mu m$ MIPS band measurement, preventing us from generalizing our network to those datasets. Nevertheless, we chose to keep this band in this paper to have the most complete view of its effect on our network. To quantify this effect, we  trained networks that did not  include either the MIPS refinement step or the $24\ \mu m$ in input. These networks   show  a small increase in performance, especially for CI YSOs with $2\%$ to $3\%$ improvement in recall and precision in the F-C case. This can mainly be explained by the simplification of the problem, but also by the greater number of objects in rare subclasses like Shocks. { Such} results could be generalized over larger datasets. In this case a MIPS refinement step could still be performed {a posteriori} on the network output for objects where this band is available. Interestingly, although the absence of the MIPS refinement step could be expected to degrade the absolute reliability of the classification, the potentially large increase in the number of rare subclasses may improve the overall network performance sufficiently for the net effect on the absolute accuracy of the classification to be positive.

\subsection{Probabilistic prediction}
\label{proba_discussion}

\begin{figure}[h]
        \centering
        \begin{subfigure}[t]{0.24\textwidth}
        \caption*{{Above 0.9}}
        \includegraphics[height=0.64\paperheight]{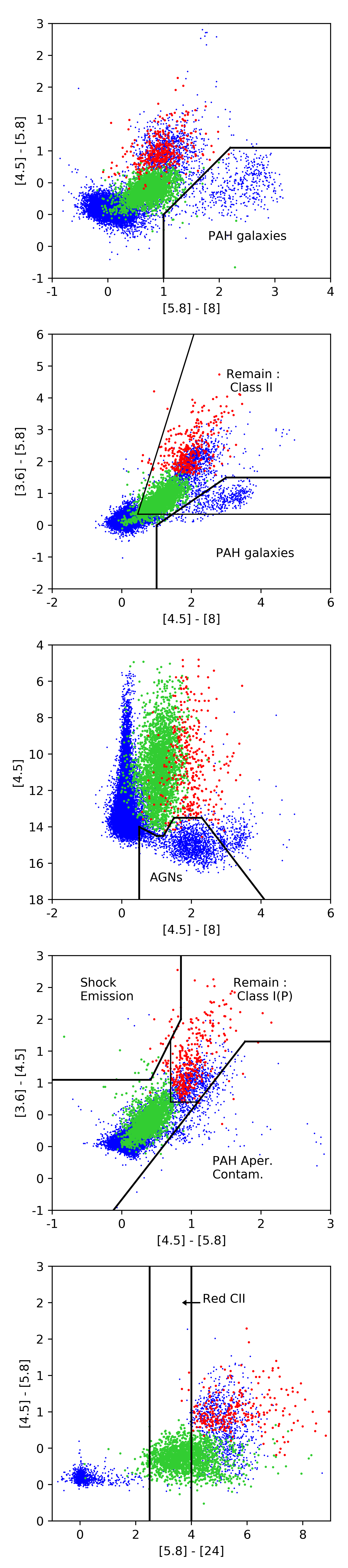}
        \end{subfigure}
        \begin{subfigure}[t]{0.24\textwidth}
        \caption*{{Below 0.9}}
        \includegraphics[height=0.64\paperheight]{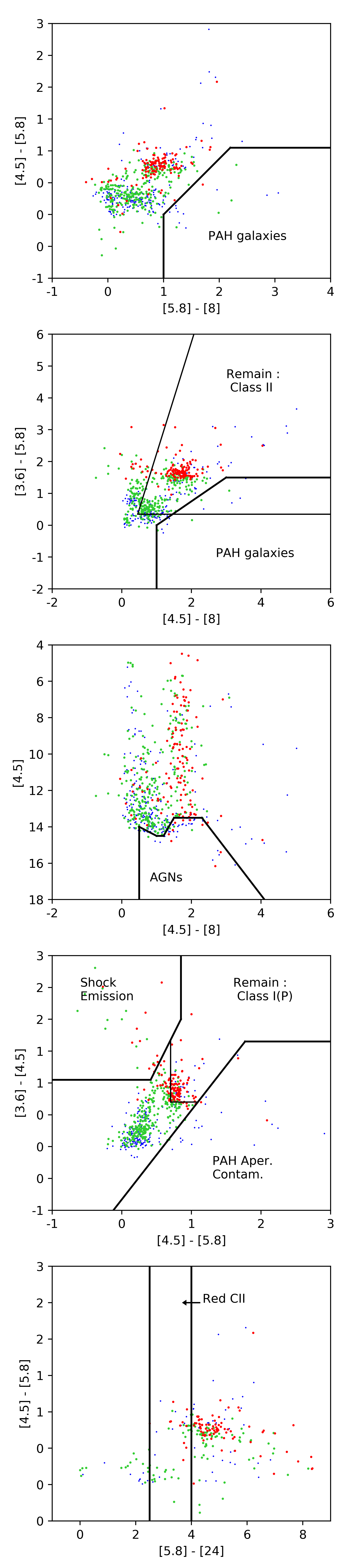}
        \end{subfigure}
        \caption{Input parameter space coverage using the usual G09 diagrams in the F-C case on the full dataset regarding their predicted membership probability. CI YSOs are in red, CII YSOs are in green, and Others are in blue. \textit{Left:} Objects with membership probability greater than $0.9$. \textit{Right:} Objects with membership probability less than $0.9$.}
        \label{membership_threshold_comparison}
\end{figure}

In this section we discuss the inclusion of a membership probability prediction in our network. If we assumed that the original classification were absolutely correct, the discrepancies would only correspond to errors. However, as illustrated by the effect of the MIPS band, the original classification has its own limitations. Therefore, the objects misclassified by our network might highlight that they were already less reliable in the original classification, or may even have been misclassified. { With a membership probability it is possible} to refine this idea by quantifying the level of confidence of the network on each prediction, directly based on the observed distribution of the objects in the input parameter space.
In practice, as already illustrated in Fig.~\ref{missed_wrong_zoom} where misclassified objects stack around the inter-class boundaries, the classification reliability of individual objects is mostly a function of their distance to these boundaries. One strength of the probabilistic output presented in Sect.~\ref{final_ann} is that the probability values provided by the network take advantage of the network ability to combine the boundaries directly in the ten dimensions of the feature space.

We used the probabilistic predictions to measure the degree of confusion of an object between the output classes. This is illustrated by the ternary plots in Fig.~\ref{ternary_plots}\footnote{These plots make use of the Python ternary package \citep{marc_2019_2628066}.}, where the location of the objects corresponds to their predicted probability of belonging to each class. In these plots an object with a high confidence level lies near the peaks. Objects that are in the inner part of the graph are the most confused of the three classes, while objects on the edges illustrate a confusion between only two classes. The sample size obviously plays a role in this representation, but each class clearly shows a level of confusion that is higher than  one specific other class. The graph for all outputs shows that the confusion between CI and Others is the lowest, followed by the confusion between CI and CII YSOs, with the highest confusion level being between the CII and Others classes. Those observations are strongly consistent with our previous analysis based only on the confusion matrix.

The probabilistic predictions can be used to remove objects that are not reliable enough. The misclassified objects show a higher degree of confusion, and therefore a maximum value of membership probability that is lower than that for the  properly classified objects. This characteristic is illustrated by Figs.~\ref{ternary_plots} and \ref{hist_proba}. The latter compares histograms of the highest output probability for properly and incorrectly classified objects. This figure reveals that the great majority of correctly classified YSOs have a membership probability greater than 0.95, whereas most missed or incorrectly classified YSOs have a probability membership below that threshold. In this context applying a threshold on the membership probability will proportionally remove more misclassified objects than properly classified ones, therefore improving the recall and precision of our network. The threshold value is arbitrary, depending on the application. We illustrate this selection effect on the F-C case in Tables~\ref{conf_proba_09}, \ref{conf_proba_095}, and \ref{conf_proba_099}. These tables represent the confusion matrix of the complete Combined dataset after selecting objects with membership probability above 0.9, 0.95, and 0.99, respectively. In the 0.9 case (Table~\ref{conf_proba_09}), 25\% (104) of the CI YSOs were removed, while their recall increased by 4.5\%. In the same way, 8.2\% (218) of the CII YSOs were removed leading to a 1.2\% increase in their recall. Contaminants were less affected;  only 1.8\% of objects were removed, which still increased the recall by 0.6\%. This is an additional demonstration of the CI YSOs being less constrained than the other output classes. In the 0.95 case (Table~\ref{conf_proba_095}), the output classes has lost 31.6\% (131), 12.7 (338), and 2.4\% (579) of objects, respectively. This still improved the recall of the two YSO classes with a 1\% increase for CI and a 0.4\% increase for CII, when compared to the 0.9 case. This result is also the first one to be close to having all quality estimators above 90\% since the CI YSO precision is 89\%, while losing an acceptable fraction of them. The 0.99 case (Table~\ref{conf_proba_099}) is more extreme since almost 50\% (206) of CI YSOs were removed, but the recall of the remaining one reached 97.6\%, which is a 6.3\% improvement over the regular F-C full dataset case. However, the CII YSOs are also strongly affected, with 25.6\% (681) of them removed, and only yielding a 0.4\% improvement in comparison to the 0.95 case.
Another illustration of the fact that this strategy effectively excludes objects that are near the cuts is presented in Fig.~\ref{membership_threshold_comparison} where the objects above or below a $0.9$ membership threshold are plotted separately for a usual set of CMDs. This effect is particularly visible in the ([4.5]-[8],[3.6]-[5.8]) (second frame) and the ([4.5]-[5.8],[3.6]-[4.5]) (forth frame) diagrams. This figure illustrates that a membership probability of less  than 0.9 can be considered   unreliable. 

With the inclusion of this probability in our results, we provide a substantial addition to the original G09 classification, for which it might be more difficult to identify the reliable objects. The membership probability for each object in Orion and NGC 2264 is included in the public catalog presented in Sect.~\ref{sec:publiccat}.

It is important to note that the membership probability output is not a direct physical probability. It is a probability regarding the network knowledge of the problem, which can be biased, incomplete, or both. Therefore, selecting a $0.9$ membership probability does not necessarily correspond to a $90\%$ certainty prediction level. The only usable probability is the one given by the confusion matrix. Consequently, according to Table~\ref{conf_proba_09}, when applying a $0.9$ membership limit, the probability that a predicted class I YSO is correct is estimated to be $87.6\%$; instead, with the same limit class II YSOs are correct in $96.1\%$ of the cases. These two values are not equivalent and   the network output membership probability should not be used as a true estimate of the reliability of an object. It can only be used to compare objects from the same network training, and must be converted into a true quality estimator using the confusion matrix.

\begin{table}[!t]
        \small
        \centering
        \caption{F-C case forwarded on the full dataset with membership probability $> 0.9$. }
        \vspace{-0.1cm}
        \begin{tabularx}{\hsize}{r l |*{3}{m}| r }
        \multicolumn{2}{c}{}& \multicolumn{3}{c}{{Predicted}}&\\
        \cmidrule[\heavyrulewidth](lr){2-6}
        \parbox[l]{0.2cm}{\multirow{6}{*}{\rotatebox[origin=c]{90}{{Actual}}}} & Class & CI YSO & CII YSO & Others & Recall \\
        \cmidrule(lr){2-6}
         &  CI YSO    & 297     & 5       & 8       & 95.8\% \\
         &  CII YSO   & 16      & 2412    & 13      & 98.8\% \\
         &  Others     & 26      & 118     & 23247   & 99.4\% \\
        \cmidrule(lr){2-6}
         &  Precision & 87.6\% & 95.1\% & 99.9\% & 99.3\% \\
        \cmidrule[\heavyrulewidth](lr){2-6}
        \end{tabularx}
   \tablefoot{The selection led to the removal of 104 CI ($-25.1\%$), 218 CII ($-8.2\%$), and 439 Others ($-1.8\%$).}
        \vspace{-0.1cm}
        \label{conf_proba_09} 
\end{table}

\begin{table}[!t]
        \small
        \centering
        \caption{F-C case forwarded on the full dataset with membership probability $> 0.95$. }
        \vspace{-0.1cm}
        \begin{tabularx}{\hsize}{r l |*{3}{m}| r }
        \multicolumn{2}{c}{}& \multicolumn{3}{c}{{Predicted}}&\\
        \cmidrule[\heavyrulewidth](lr){2-6}
        \parbox[l]{0.2cm}{\multirow{6}{*}{\rotatebox[origin=c]{90}{{Actual}}}} & Class & CI YSO & CII YSO & Others & Recall \\
        \cmidrule(lr){2-6}
         &  CI YSO    & 274     & 2       & 7       & 96.8\% \\
         &  CII YSO   & 11      & 2302    & 8       & 99.2\% \\
         &  Others     & 23      & 92      & 23136   & 99.5\% \\
        \cmidrule(lr){2-6}
         &  Precision & 89.0\% & 96.1\% & 99.9\% & 99.4\% \\
        \cmidrule[\heavyrulewidth](lr){2-6}
        \end{tabularx}
        \tablefoot{The selection led to the removal of 131 CI ($-31.6\%$), 338 CII ($-12.7\%$), and 579 Others ($-2.4\%$).}
        \vspace{-0.1cm}
        \label{conf_proba_095} 
\end{table}

\begin{table}[!t]
        \small
        \centering
        \caption{F-C case forwarded on the full dataset with membership probability $> 0.99$.}
        \vspace{-0.1cm}
        \begin{tabularx}{\hsize}{r l |*{3}{m}| r }
        \multicolumn{2}{c}{}& \multicolumn{3}{c}{{Predicted}}&\\
        \cmidrule[\heavyrulewidth](lr){2-6}
        \parbox[l]{0.2cm}{\multirow{6}{*}{\rotatebox[origin=c]{90}{{Actual}}}} & Class & CI YSO & CII YSO & Others & Recall \\
        \cmidrule(lr){2-6}
         &  CI YSO    & 203     & 0       & 5       & 97.6\% \\
         &  CII YSO   & 4       & 1970    & 4       & 99.6\% \\
         &  Others     & 14      & 51      & 22747   & 99.7\% \\
        \cmidrule(lr){2-6}
         &  Precision & 91.9\% & 97.5\% & 99.9\% & 99.7\% \\
        \cmidrule[\heavyrulewidth](lr){2-6}
        \end{tabularx}
        \tablefoot{The selection led to the removal of 206 CI ($-49.8\%$), 681 CII ($-25.6\%$), and 1018 Others ($-4.3\%$).}
        \vspace{-0.1cm}
        \label{conf_proba_099} 
\end{table}

\subsection{Possible method improvements}

Using our approach comes with several caveats, the main one being that we built our labeled dataset from a pre-existing classification that has its own limitations. The membership probability discussed in the previous section provides a first but limited view of the uncertainties inherent to the original classification scheme. One approach to completely releasing our methodology from its dependence on the G09 scheme would consist in building our training set from a more conclusive type of observations, like visible spectroscopy to detect the $H_\alpha$ line that is usually attributed to gas accretion by the protostar \citep{Kun_2009}  or (sub)millimeter interferometry to detect the disks \citep[e.g.,][]{ruiz-rodriguez_2018, alma_disk_yso, tobin_2020}. Alternatively, a large set of photometric bands could be gathered to reconstruct the SED across a wider spectral range, as in \citet{miettinen_protostellar_2018}. Unfortunately, to date, too few objects have been observed that extensively to build a labeled sample large enough to efficiently train most of the ML algorithms.

Another approach would be to use simulations of star-forming regions \citep[e.g.,][]{padoan_2017, vazquez-semadini_2019} and   star-forming cores \citep[e.g.,][]{robitaille_interpreting_2006} to provide a mock census of YSOs and emulate their observational properties. This option would enable us to generate large training catalogs, and would provide additional control on the YSO classes, but at the cost of other kinds of biases coming from the simulation assumptions. An additional difficulty of this approach would be the large variety of contaminant objects, each of which would require a dedicated treatment.

A different strategy could consist in improving the method itself. With feedforward neural networks, as used  in this paper, there may still be improvement possibilities by using deeper networks with, for example, a different activation function, a weight initialization, or a more complex error propagation. By choosing a completely different, unsupervised method, one could work independently of any prior classification. However, there is a risk that the classes identified by the method do not match the classical ones. In particular, the continuous distribution from CI to CII YSOs, and then to main sequence stars is likely to be identified as a single class by such algorithms. A middle-ground could be the semi-supervised learning algorithms such as Deep Belief Networks \citep{Hinton504}. Such algorithms were designed to find a dimensional reduction of the given input feature space that is more suitable to the problem, {basing} its own classes   on the proximity of objects in the feature space. It could then be connected with a regular supervised feedforward neural network layer that would combine the found classes into more usual ones. This approach would reduce the impact of the original classification on the training process, and therefore its impact on the final results.

\section{Conclusion}
We   presented a detailed methodology to use deep neural networks to extract and classify YSO candidates from several star-forming regions using Spitzer infrared data, based on the method described by \citep{gutermuth_spitzer_2009}. { The analysis is based on the ability of ANNs to quantitatively characterize the classification properties and reliability, demonstrating the advantage of our neural network methodology over a CMD-selection scheme like that of G09.} 
We make public the table containing the YSO candidates in Orion and NGC 2264 from the classification of our final and best ANN. The table includes the class membership probability for each object, and is available at CDS.
This study led to the following conclusions.

Deep Neural networks are a suitable solution for performing an efficient YSO classification using the four Spitzer  IRAC bands and the MIPS 24 $\mu$m band. When trained on one cloud only, the prediction performance mostly depends on the size of the sample. Fairly simple networks can be used for this task with just one hidden layer that only consists of 15 to 25 neurons with a classical sigmoid activation function.

The prediction capability of the network on a new region strongly depends on the properties of the region used for training. { The} study revealed the necessity to train the network on a census of star-forming regions to improve the diversity of the training sample. A network trained on a more diverse dataset has been able to maintain a high quality prediction, which is promising for its ability to be applied to new star-forming regions.

The dataset imbalance has a strong effect on the results, not only on the classes of interest, but also for the hidden subclasses considered as contaminants. Carefully rebalancing each subclass in the training dataset, according to its respective feature space coverage complexity and to its proximity with other classes of interest, has shown to be of critical importance. The use of observational proportions to measure the quality of the prediction has been shown to be of major importance to properly assess the quality of the prediction.

This study showed that the network membership probability prediction complements the original G09 classification with an estimate of the prediction reliability. It allows one to select objects based on their proximity to the whole set of classification cuts in a multidimensional space, using a single quantity. { The} identification of objects with a higher degree of confusion highlights parts of the parameter space that might lack constraints and that would benefit from a refinement of the original classification. 

The current study contains various limitations, mainly the lack of additional near star-forming region catalogs that contain the sub-contaminant distinction to construct complete training samples. { Some} subclasses, namely Shocks and PAHs, remain strongly unconstrained due to their scarcity. Identifying additional shocks and resolved PAH emission in Spitzer archive data could significantly improve their classification by our networks, and consequently improve the YSO classification. Attention has also been drawn toward the use of simulations to compile large training datasets that might be used in ensuing studies. 

Finally, our method could be improved by adopting more advanced networks that would probably overcome some difficulties, for example by avoiding local minima more efficiently, and would improve the raw computational performance of the method. Semi-supervised or fully unsupervised methods may also be promising tracks to predict YSO candidates that may surpass the supervised methods in terms of prediction quality. On the other hand, we have highlighted that most of the difficulties come from the training set construction, which is mostly independent of the chosen method. Therefore, future improvements in YSO identification and classification from ML applied to mid-IR surveys will require us to compile larger and more reliable training catalogs, either by taking advantage of current and future surveys from various facilities, like the Massive Young Star-Forming Complex Study in Infrared and X-ray \citep[MYStIX,][]{Feigelson_2013} and the VLA/ALMA Nascent Disk and Multiplicity survey \citep[VANDAM,][]{tobin_2020}, or by synthesizing such catalogs from simulations.

\begin{acknowledgements}
This work was supported by the Programme National "Physique et Chimie du Milieu  Interstellaire" (PCMI) of CNRS/INSU with INC/INP co-funded by CEA and CNES. The authors thank  the French ministry of foreign affairs (French embassy in Budapest) and the Hungarian  national office for research and innovation (NKFIH) for financial support (Balaton program 40470VL/2017-2.2.5-TET-FR-2017-00027). D. Cornu work was supported by the Centre National d'Études Spatiales (CNES) through PhD grant 0102541 and the Région Bourgogne Franche-Comté.
\end{acknowledgements}

%
%

\bibliographystyle{aa}
\bibliography{yso_paper}

\begin{appendix}

\section{Example of tuning of the sample proportions and network parameters}
\label{app:tuning}

\begin{figure}
        \centering
        \includegraphics[width=0.95\hsize]{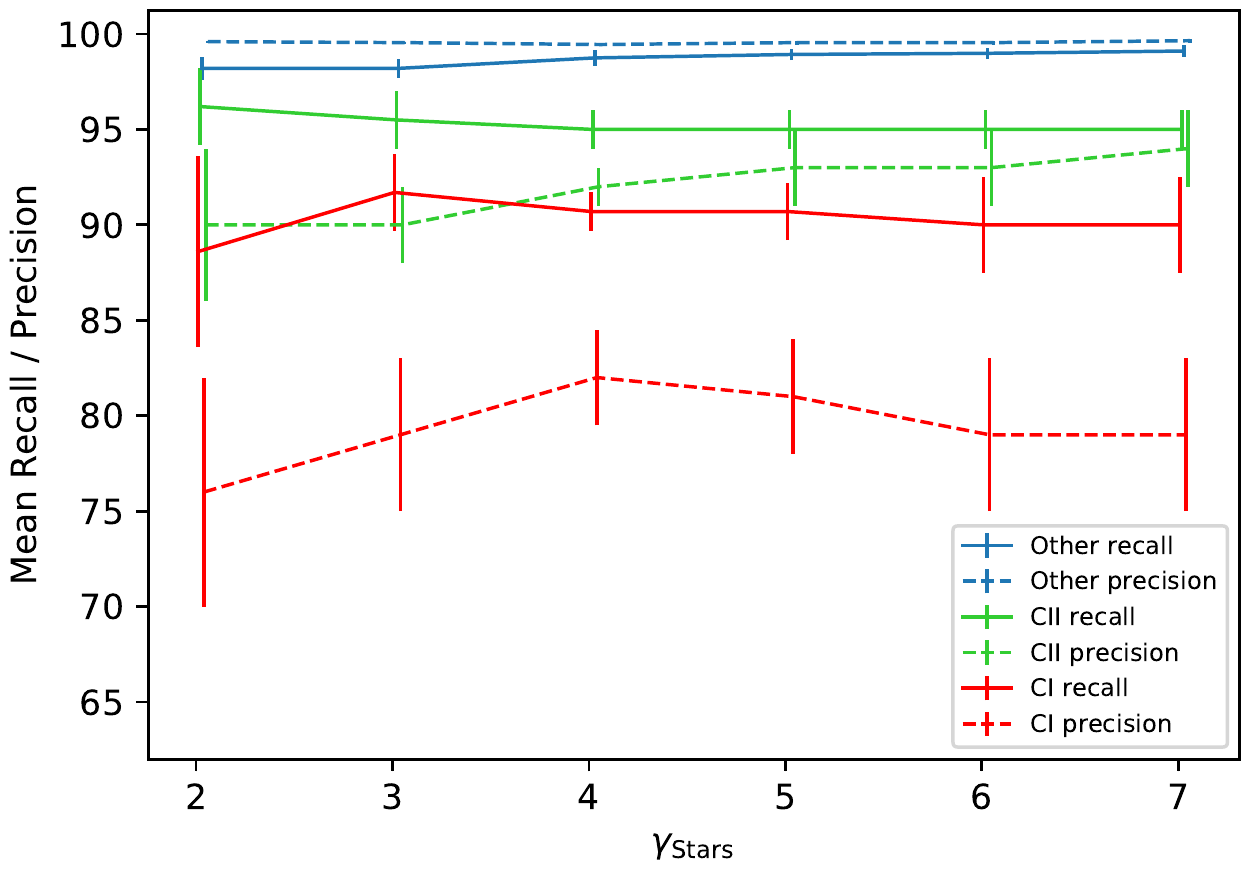}
        \caption{Quality indicators for each network output class as a function of the proportion of stars in the training sample $\gamma_{\rm Stars}$. The lines are the recall (continuous) and the precision (dashed) mean value of convergence over ten trainings, and the error bars represent the typical ranges of convergence value.}
        \label{fig:tuninggamma}
\end{figure}

In Sects.~\ref{train_process} and \ref{network_tuning} we presented the strategy adopted to build the training and test sets, and to set the network parameters. We illustrate this strategy here on the example of the Orion region. We note that although the dataset parameters ($\theta$ and $\gamma_i$) and the network parameters (learning rate $\eta$ and momentum $\alpha$) are conceptually different kinds of parameters, in practice they are mutually dependent. In total, there were at least ten different parameters to optimize simultaneously, and the computation time of an individual training made it very difficult, if not impossible, to search for the optimum automatically. This difficulty to automatize the search for the best parameters is increased by the fact that the importance of an observable depends on our interest. For example, it is more important for us to maximize the recall and precision of CI YSOs than those of the other classes. This is why we adopted a manual iterative procedure to identify satisfying parameters based on the values of precision and recall of the different classes, the priority being given to the CI YSOs, as discussed in Sect.~\ref{conv_crit}. As we show below, the interpretation of the observable variations plays an important role in this procedure.

To illustrate this search we consider here the optimization of the $\gamma_{\rm Stars}$ parameter for the Orion sample, all other parameters being fixed to the values presented in Tables~\ref{sat_factors} and \ref{tab_hyperparam}. Figure~\ref{fig:tuninggamma} shows the variations in recall and precision of the CI, CII, and Others classes when $\gamma_{\rm Stars}$ increases from 2 to 7. We trained the network ten times for each setup in order to get a mean value and a typical range of convergence value for each quality indicator. We note that the fluctuations in the observable values for a given $\gamma_{\rm Stars}$ are mutually dependent. For example, a high recall value for one class is generally obtained at the cost of a smaller recall value in another class for a given training. 

In this figure we observe that for the Others class, which is vastly dominated by Stars, recall and precision values are above 98\% even with the smallest $\gamma_{\rm Stars} = 2$ and slowly increase with $\gamma_{\rm Stars}$. This indicates that the vast majority of genuine Stars are easy to separate, which was expected due to the relatively small proportion of them that are close to a YSO boundary in the feature space. Even so, this small proportion represents a large number of objects {compared to} the number of genuine YSOs. For low values of $\gamma_{\rm Stars}$ the precision of CI and CII YSOs is low and with a high dispersion between repeated training. The recall and precision of CII YSOs mostly follow a linear trend. The recall slowly decreases because more of the network representativity strength is allocated to Stars, and at the same time CII precision increases due to the improvement of the boundary between CII and Stars due to a larger number of star examples. In other words, this {boundary is constrained by} more neurons and data points, but the other boundaries of CII become proportionally less important for the network and are less well constrained.

Examining the variations in the scatter of observable values helps to estimate parameters that provide more reproducible results. The dispersion of CII recall values is smaller for larger $\gamma_{\rm Stars}$, while the dispersion of CII precision values decreases, with a minimum around $\gamma_{\rm Stars} = 4$, and then slowly rises for larger $\gamma_{\rm Stars}$ values. Regarding CI YSOs, there are two regimes: (i) for $\gamma_{\rm Stars}<4$, where the mean values of recall and precision improve overall, while their dispersion decreases when $\gamma_{\rm Stars}$ increases, and (ii) for $\gamma_{\rm Stars} > 4$, where these trends are reversed. This can be explained by the fact that the network first takes advantage of the additional Stars to better constrain the differences between CI and Stars, but at some point CI YSOs become too diluted in the training sample.

For this example we chose to use $\gamma_{\rm Stars} = 4$ since it provides the highest CI precision and the smallest dispersion for both recall and precision for the two YSO classes. In the rest of the paper we adopted a  method similar to that used in  this example; we preferred parameters that maximized the results for CI YSOs and tried to minimize the negative impact on CII YSOs. When possible we also chose parameters that minimized the dispersion of the YSO quality indicators. We note that we  frequently   observed parameters that did not  significantly impact the observables for a large range of values. In this case, we selected parameter values that seemed reasonable. Although these values are not well constrained, it does not impact our results. While the actual parameter values we used in the present study might not  be the optimal ones, we are confident that the best prediction is contained within our margin of error.

\section{Detailed feature space coverage}

\begin{figure*}[!t]
        \centering
        \vspace{0.3cm}
        \begin{subfigure}[t]{0.24\textwidth}
        \caption*{{\hspace{0.3cm}Actual}}
        \includegraphics[width=\textwidth]{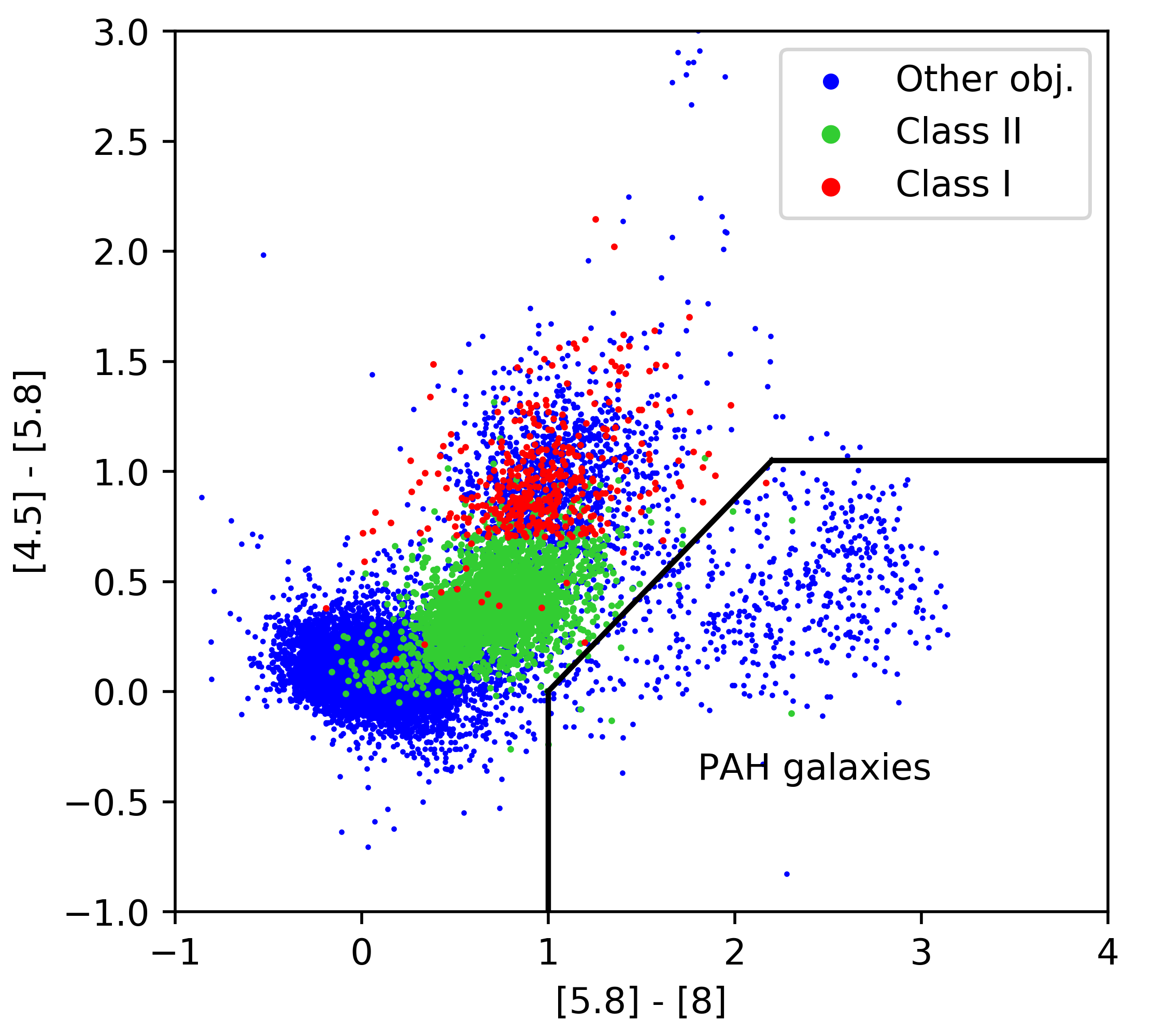}
        \end{subfigure}
        \begin{subfigure}[t]{0.24\textwidth}
        \caption*{{\hspace{0.3cm}Predicted}}
        \includegraphics[width=\textwidth]{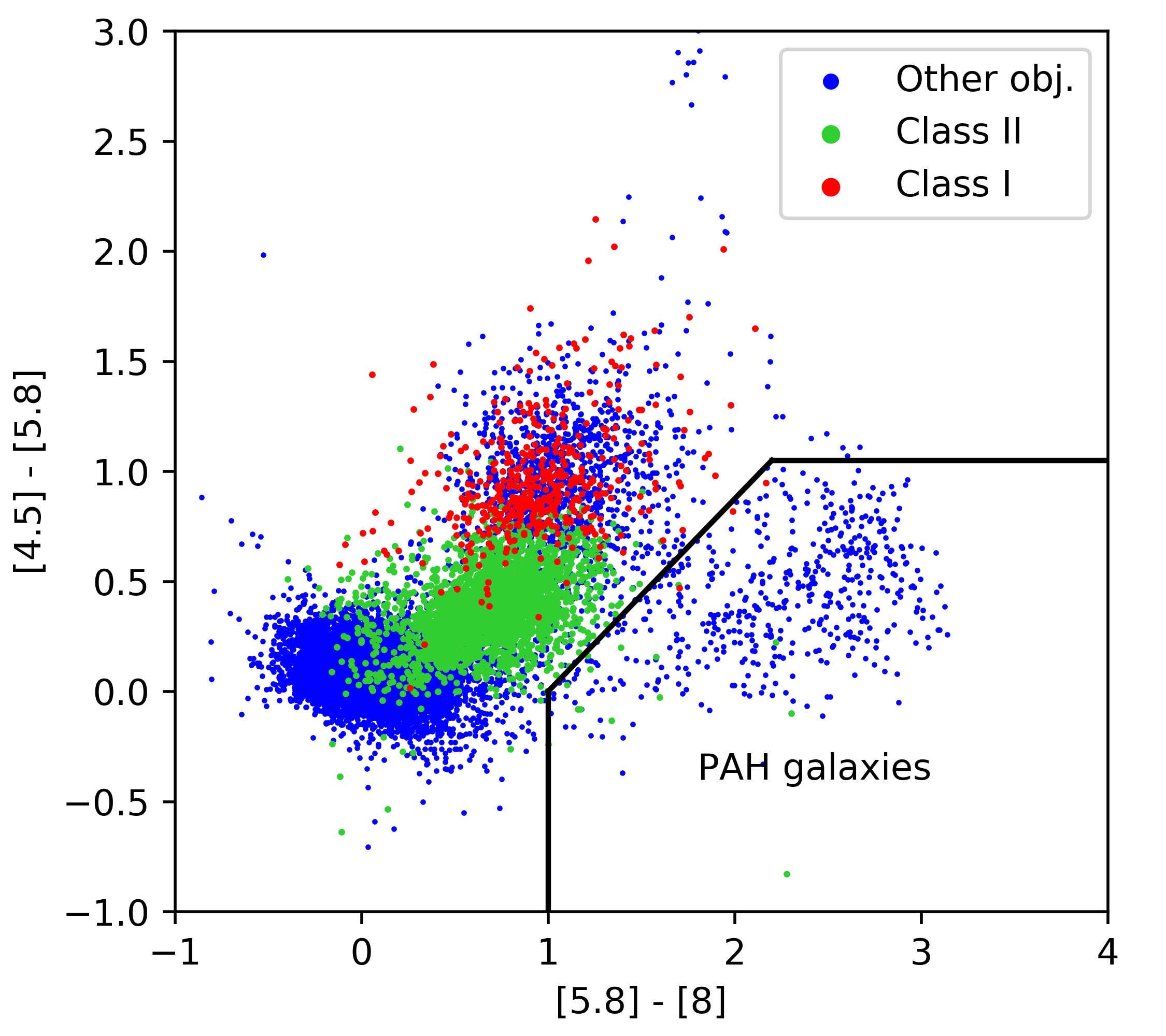}
        \end{subfigure}
        \begin{subfigure}[t]{0.24\textwidth}
        \caption*{{\hspace{0.3cm}Missed}}
        \includegraphics[width=\textwidth]{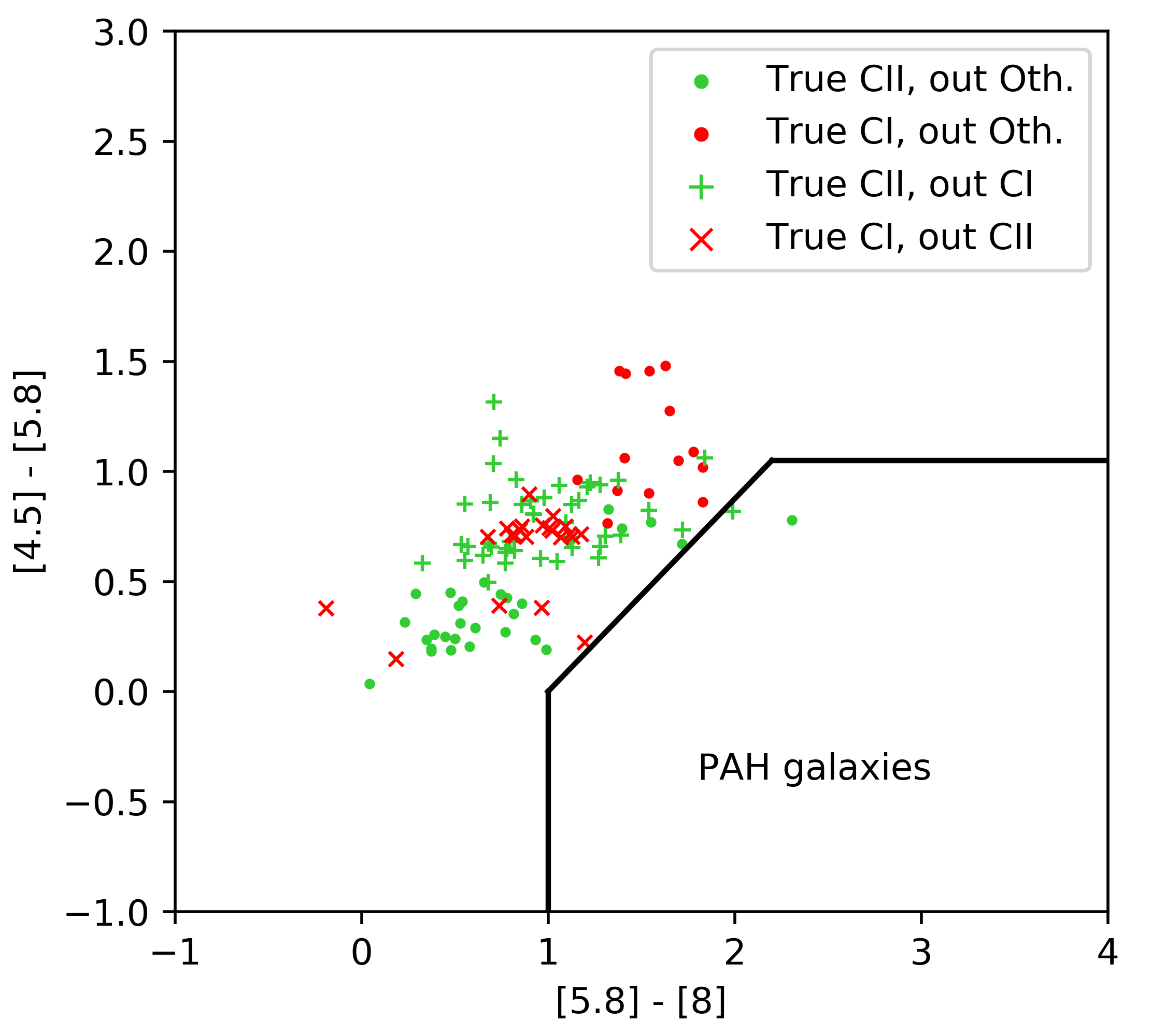}
        \end{subfigure}
        \begin{subfigure}[t]{0.24\textwidth}
        \caption*{{\hspace{0.3cm}Wrong}}
        \includegraphics[width=\textwidth]{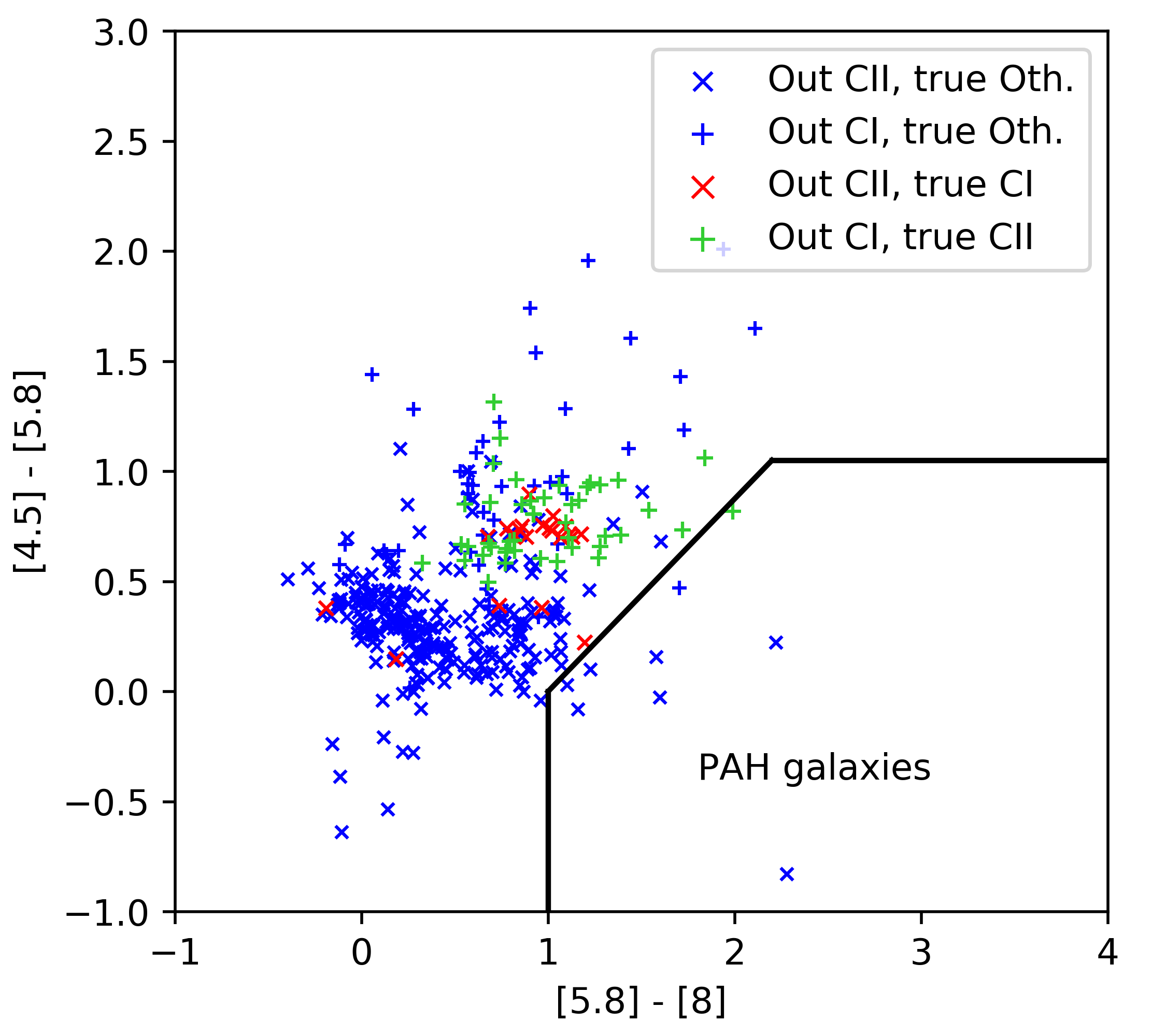}
        \end{subfigure}\\
        \begin{subfigure}[t]{0.24\textwidth}
        \includegraphics[width=\textwidth]{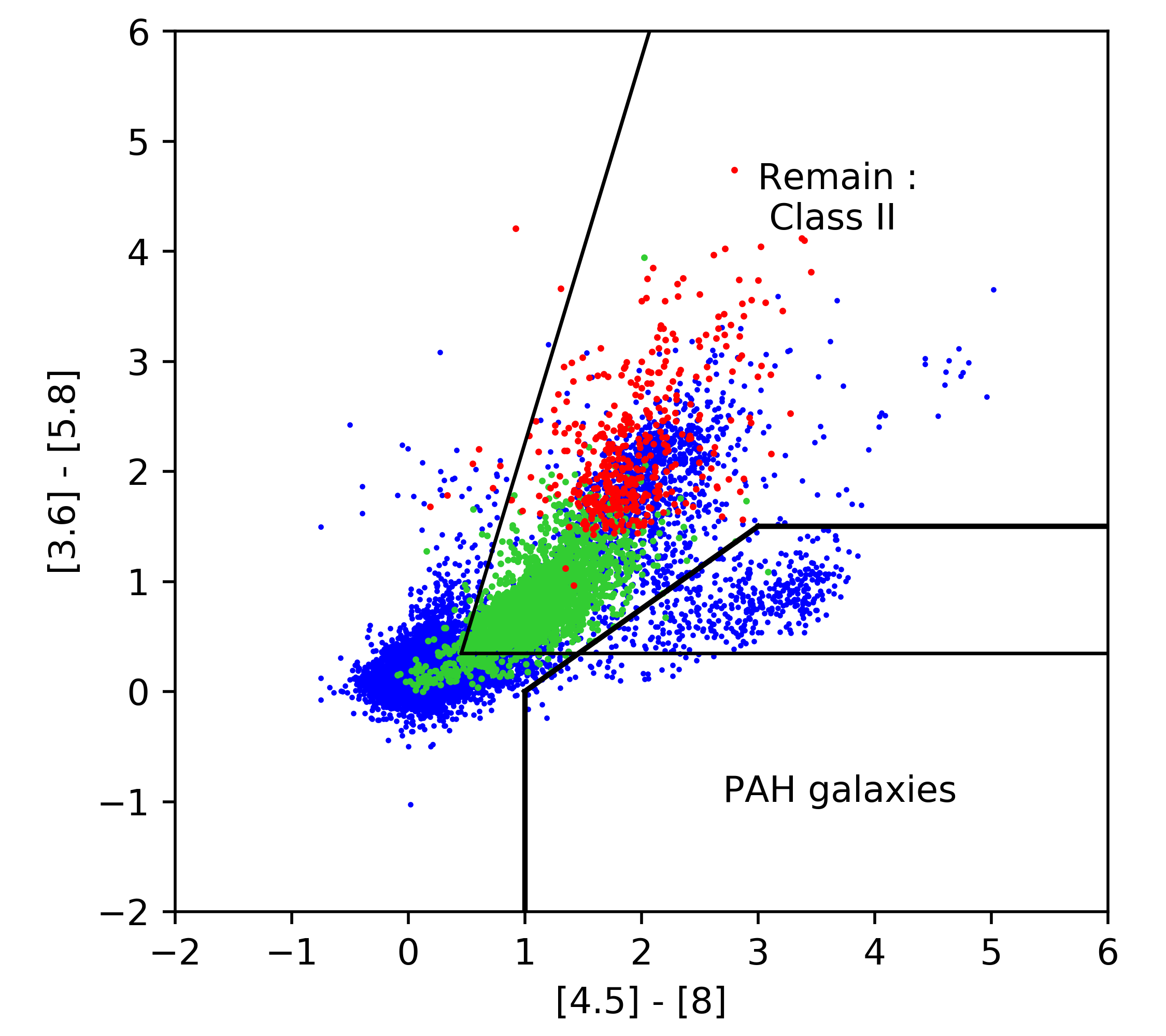}
        \end{subfigure}
        \begin{subfigure}[t]{0.24\textwidth}
        \includegraphics[width=\textwidth]{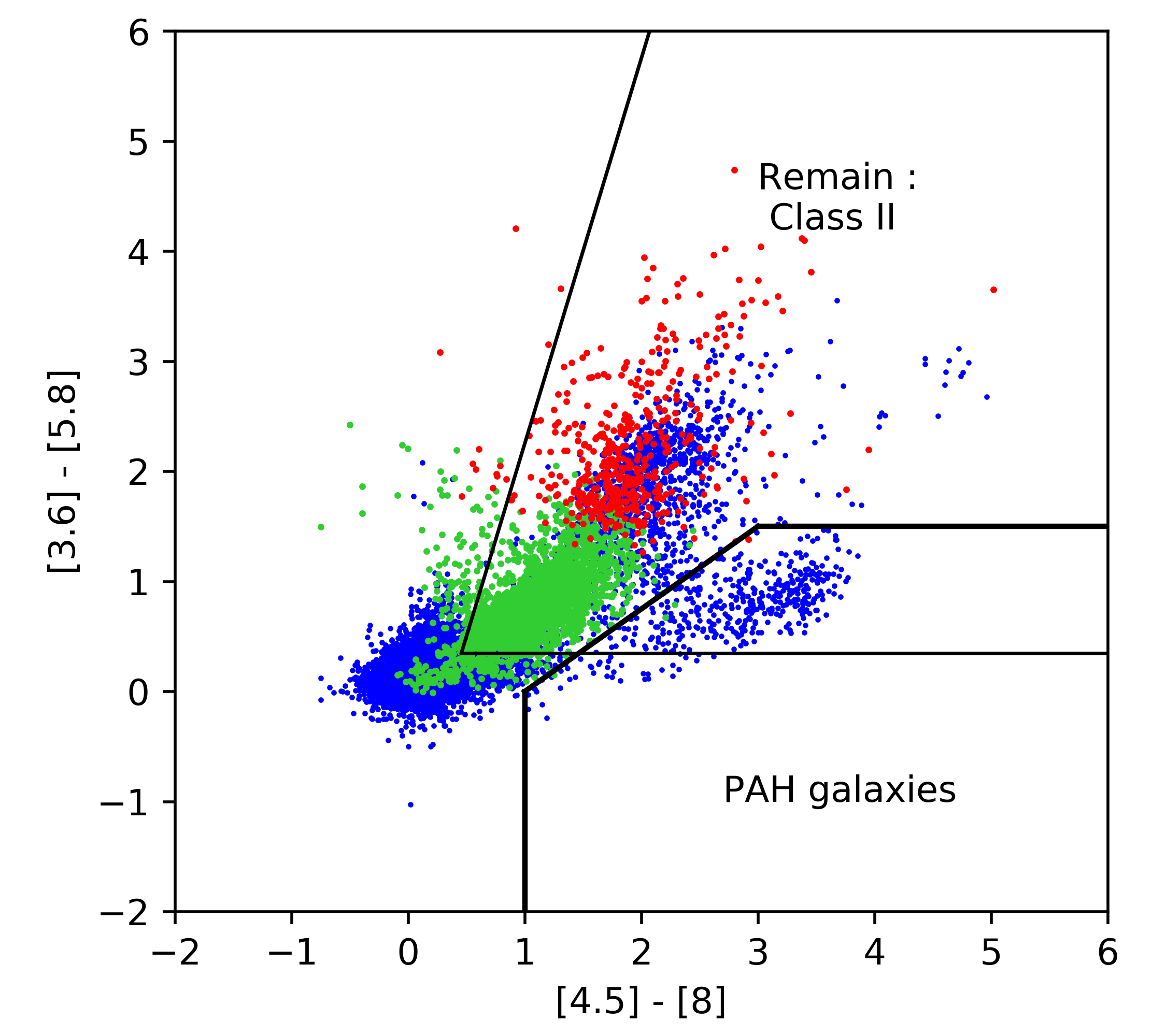}
        \end{subfigure}
        \begin{subfigure}[t]{0.24\textwidth}
        \includegraphics[width=\textwidth]{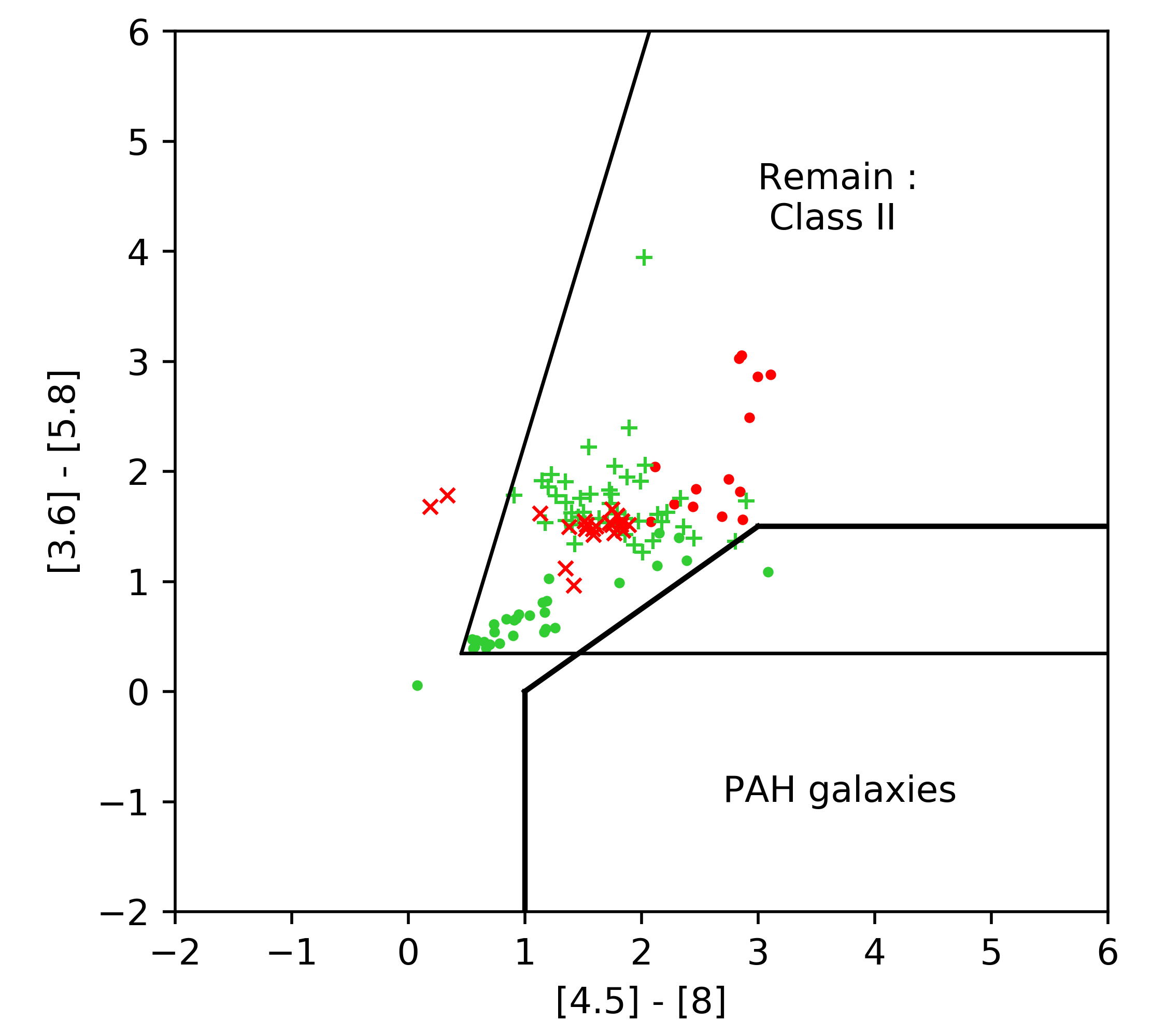}
        \end{subfigure}
        \begin{subfigure}[t]{0.24\textwidth}
        \includegraphics[width=\textwidth]{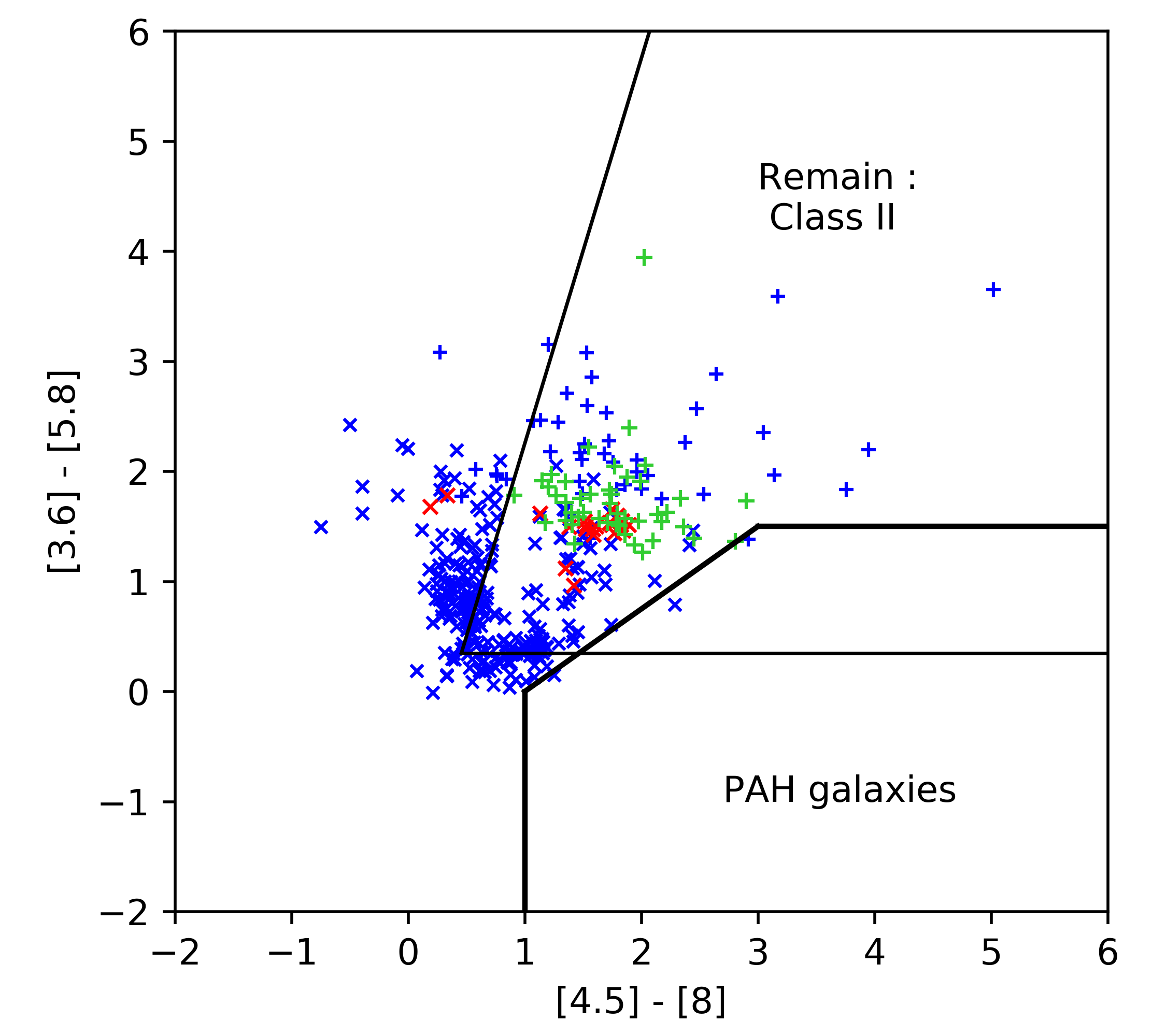}
        \end{subfigure}\\
        \begin{subfigure}[t]{0.24\textwidth}
        \includegraphics[width=\textwidth]{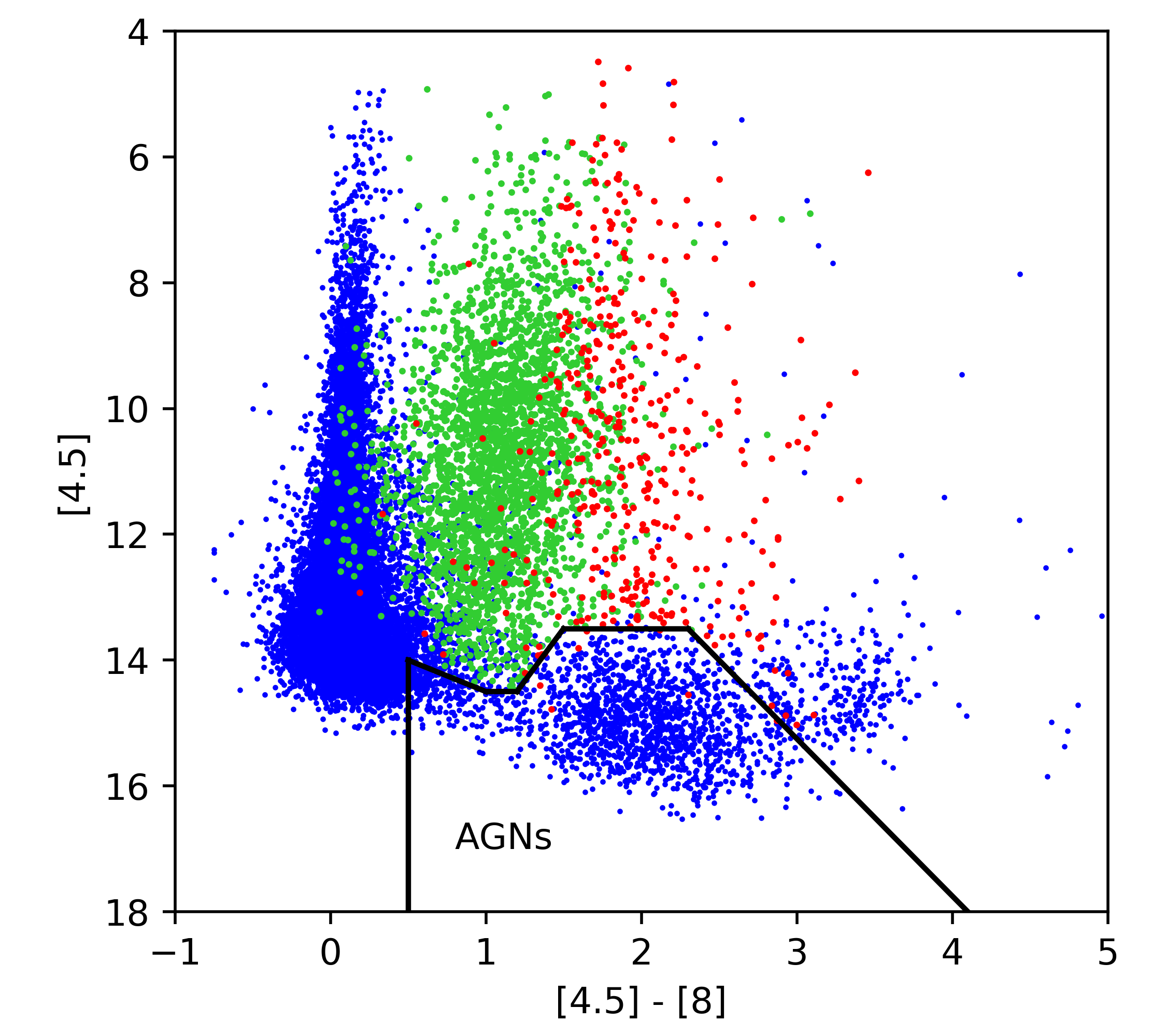}
        \end{subfigure}
        \begin{subfigure}[t]{0.24\textwidth}
        \includegraphics[width=\textwidth]{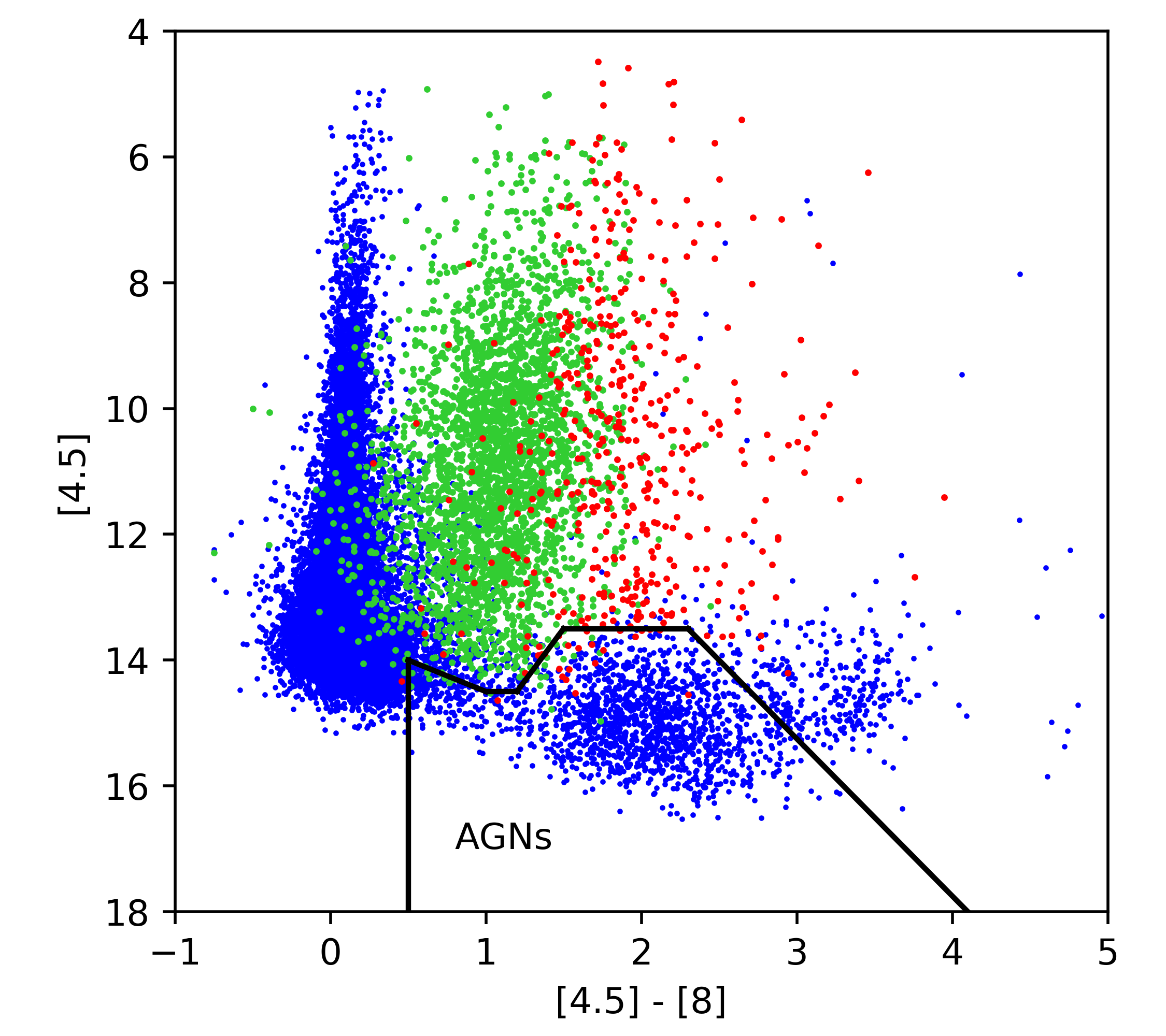}
        \end{subfigure}
        \begin{subfigure}[t]{0.24\textwidth}
        \includegraphics[width=\textwidth]{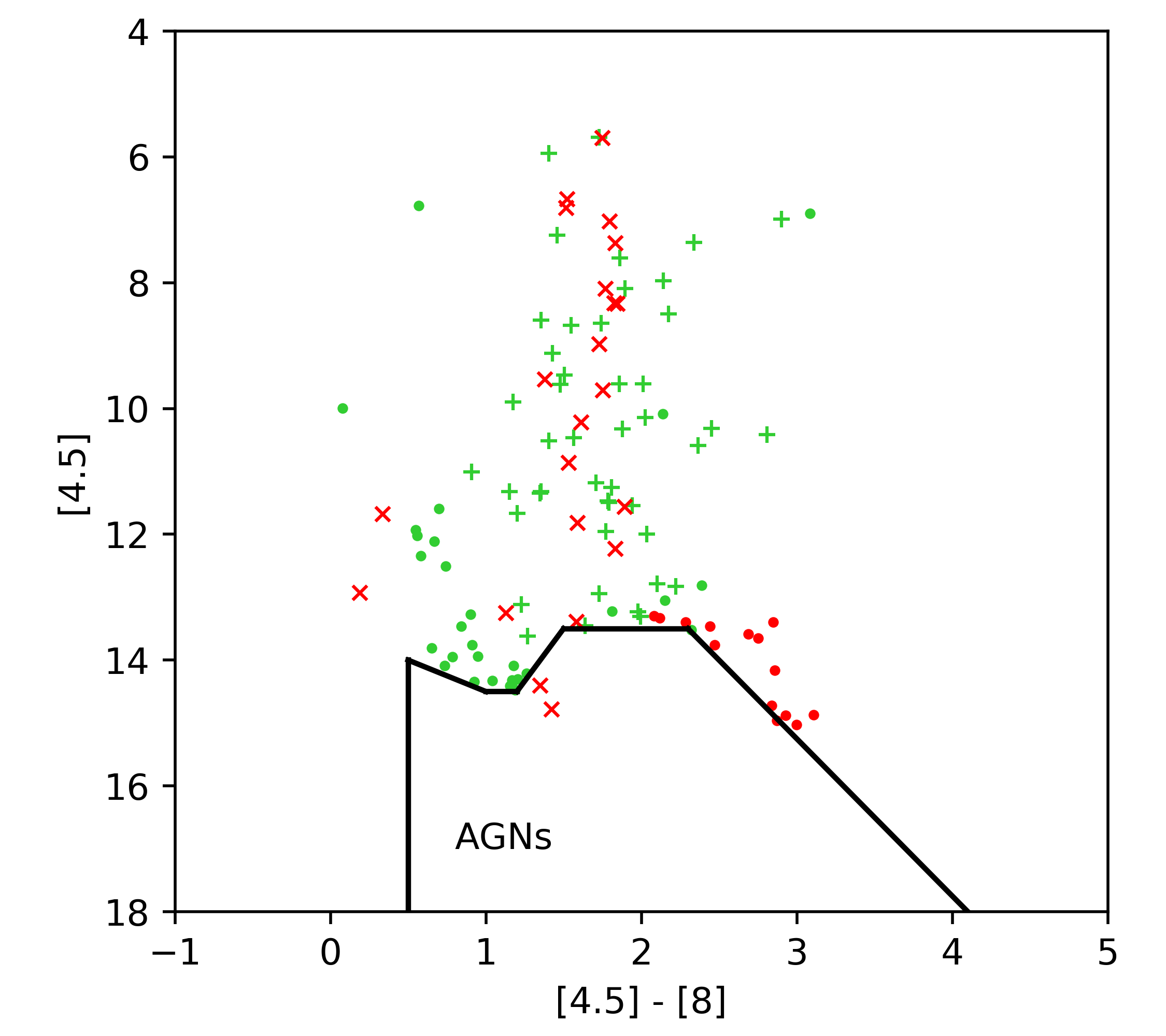}
        \end{subfigure}
        \begin{subfigure}[t]{0.24\textwidth}
        \includegraphics[width=\textwidth]{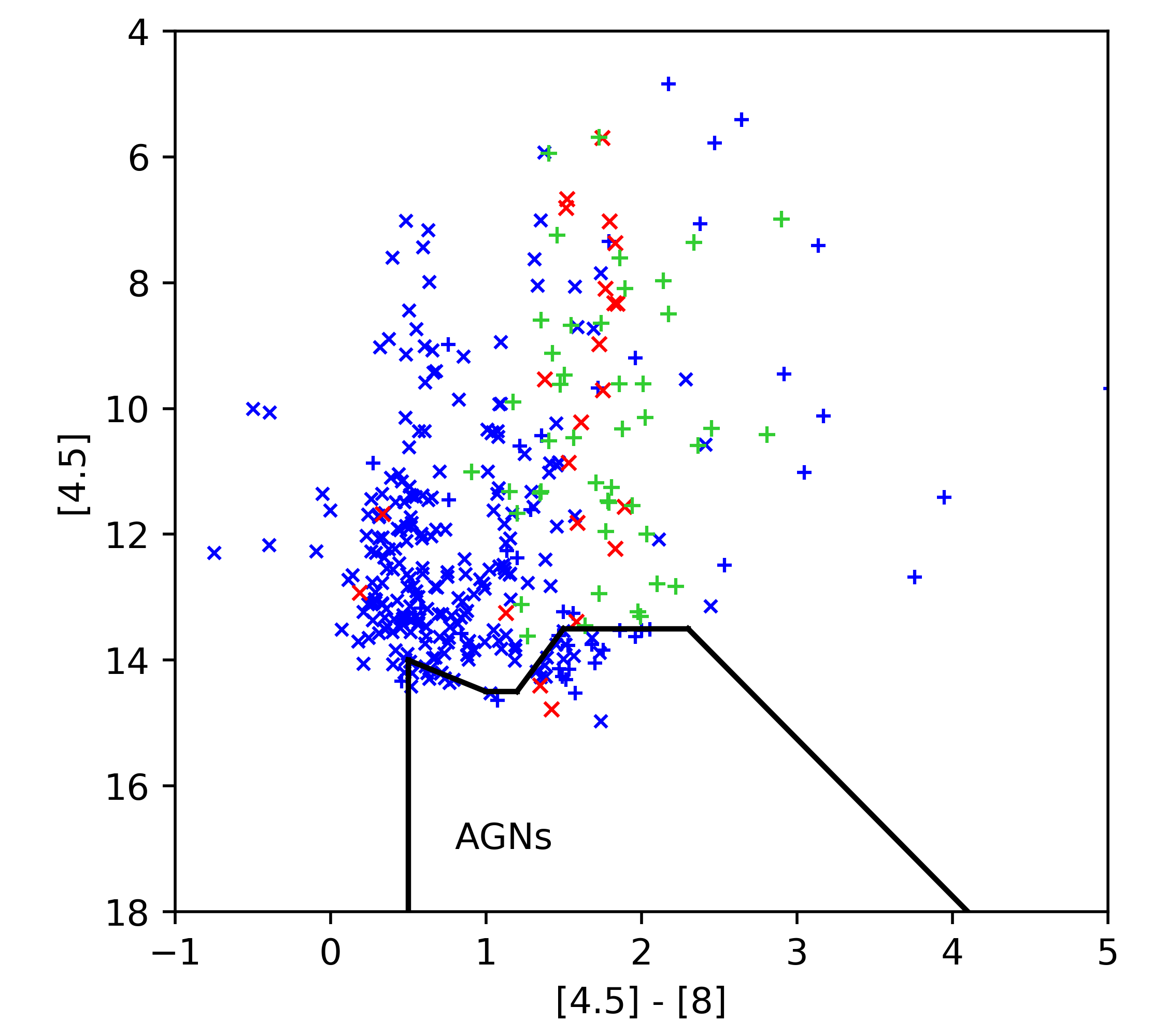}
        \end{subfigure}\\
        \begin{subfigure}[t]{0.24\textwidth}
        \includegraphics[width=\textwidth]{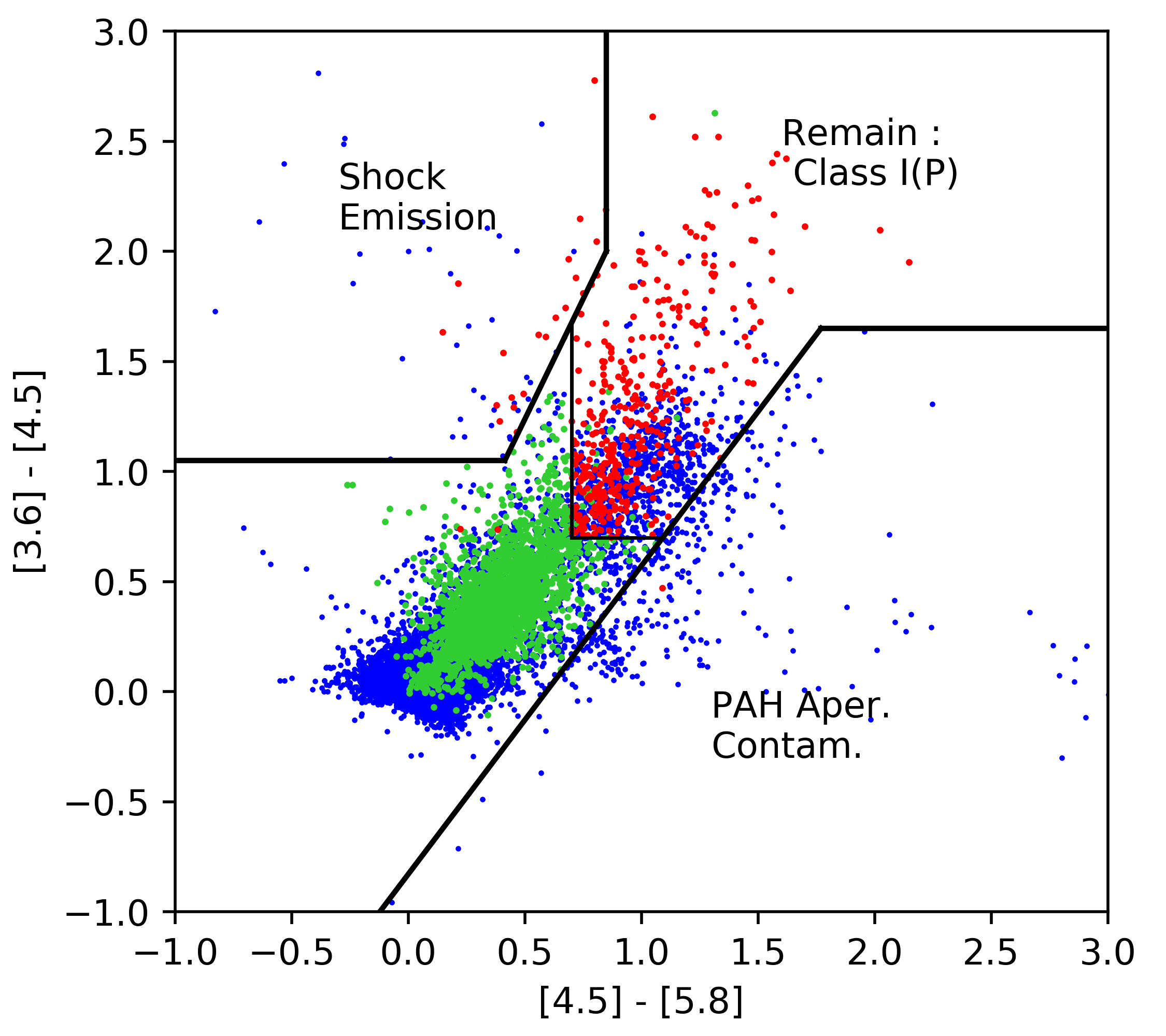}
        \end{subfigure}
        \begin{subfigure}[t]{0.24\textwidth}
        \includegraphics[width=\textwidth]{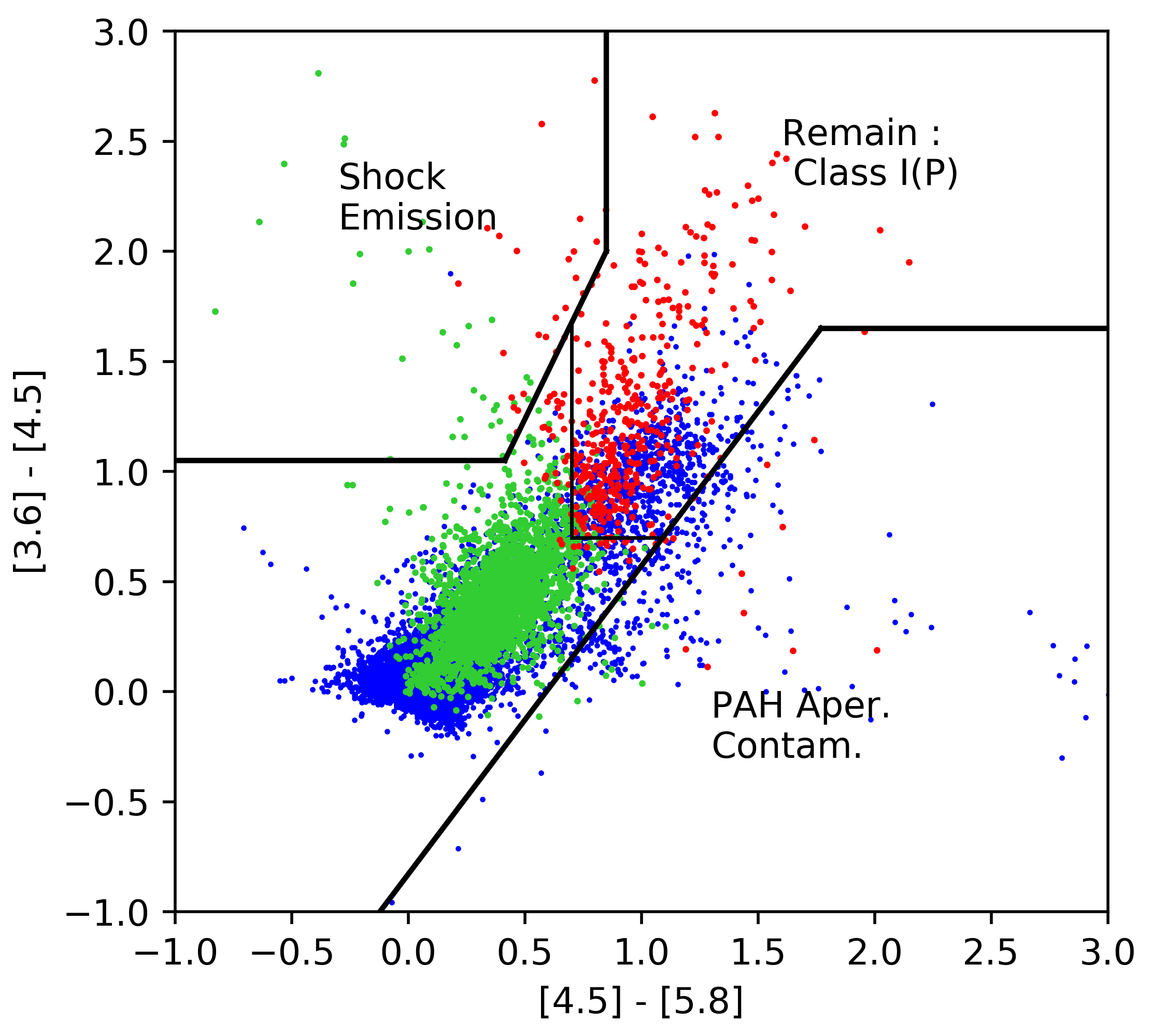}
        \end{subfigure}
        \begin{subfigure}[t]{0.24\textwidth}
        \includegraphics[width=\textwidth]{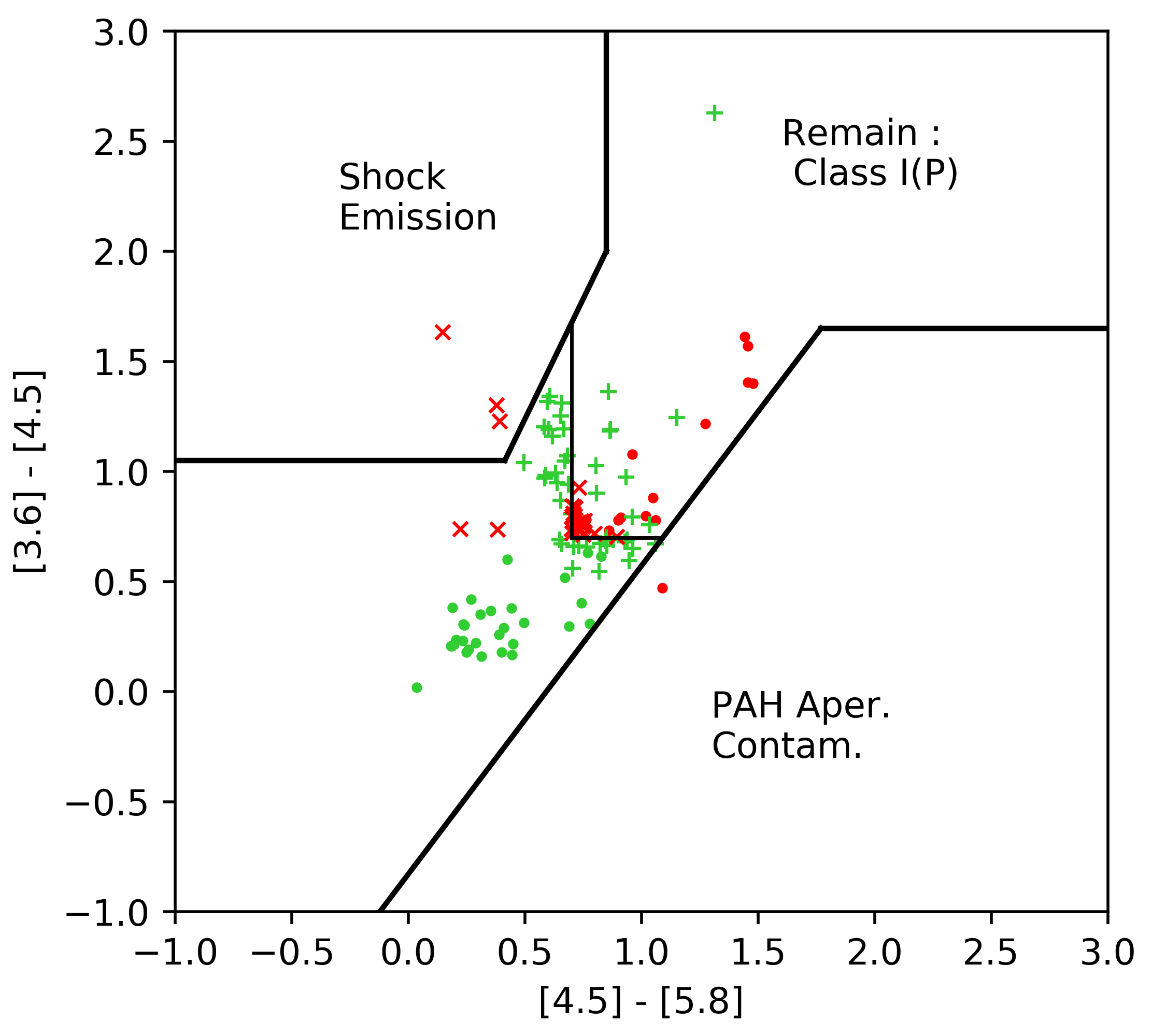}
        \end{subfigure}
        \begin{subfigure}[t]{0.24\textwidth}
        \includegraphics[width=\textwidth]{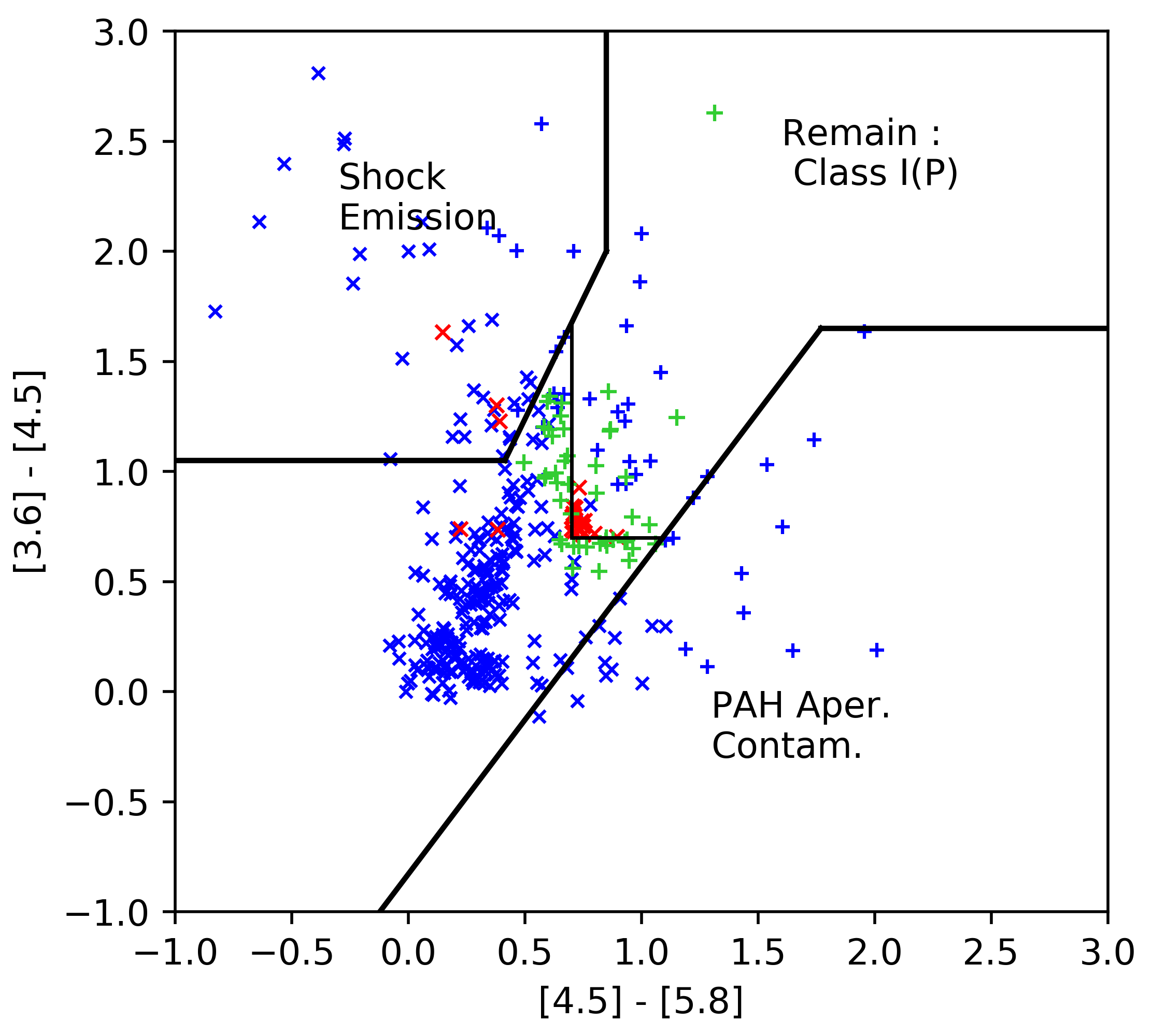}
        \end{subfigure}\\
        \begin{subfigure}[t]{0.24\textwidth}
        \includegraphics[width=\textwidth]{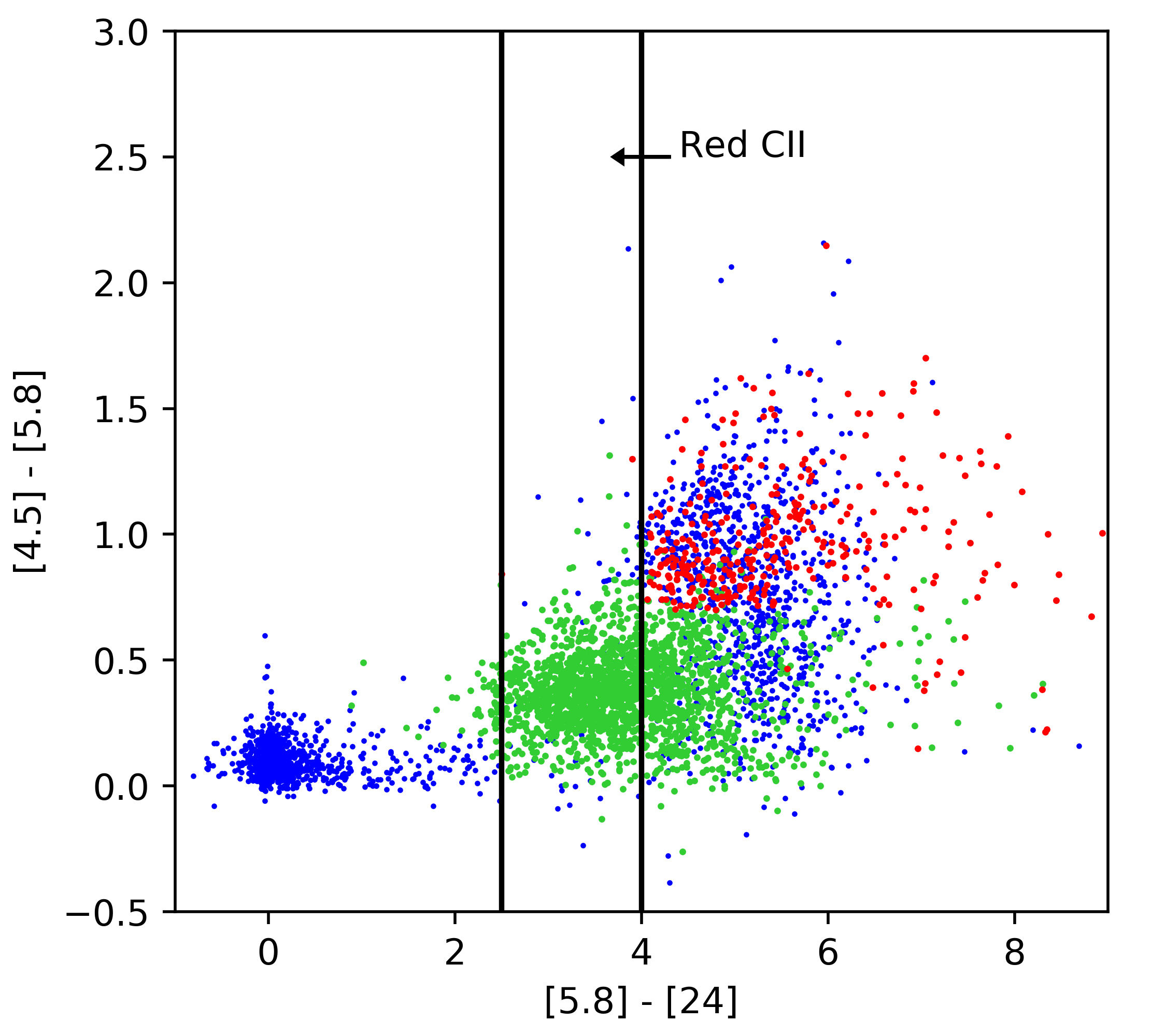}
        \end{subfigure}
        \begin{subfigure}[t]{0.24\textwidth}
        \includegraphics[width=\textwidth]{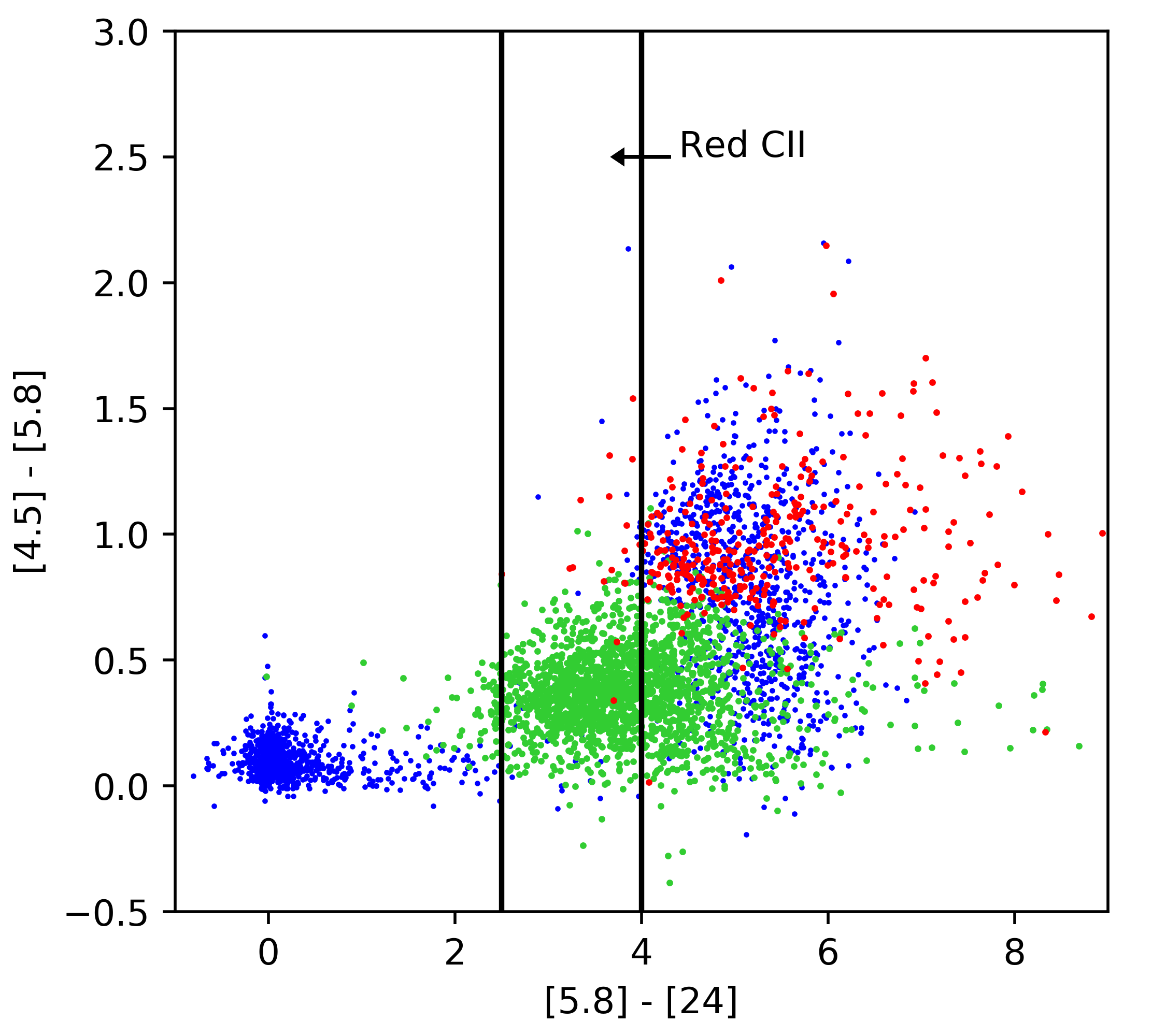}
        \end{subfigure}
        \begin{subfigure}[t]{0.24\textwidth}
        \includegraphics[width=\textwidth]{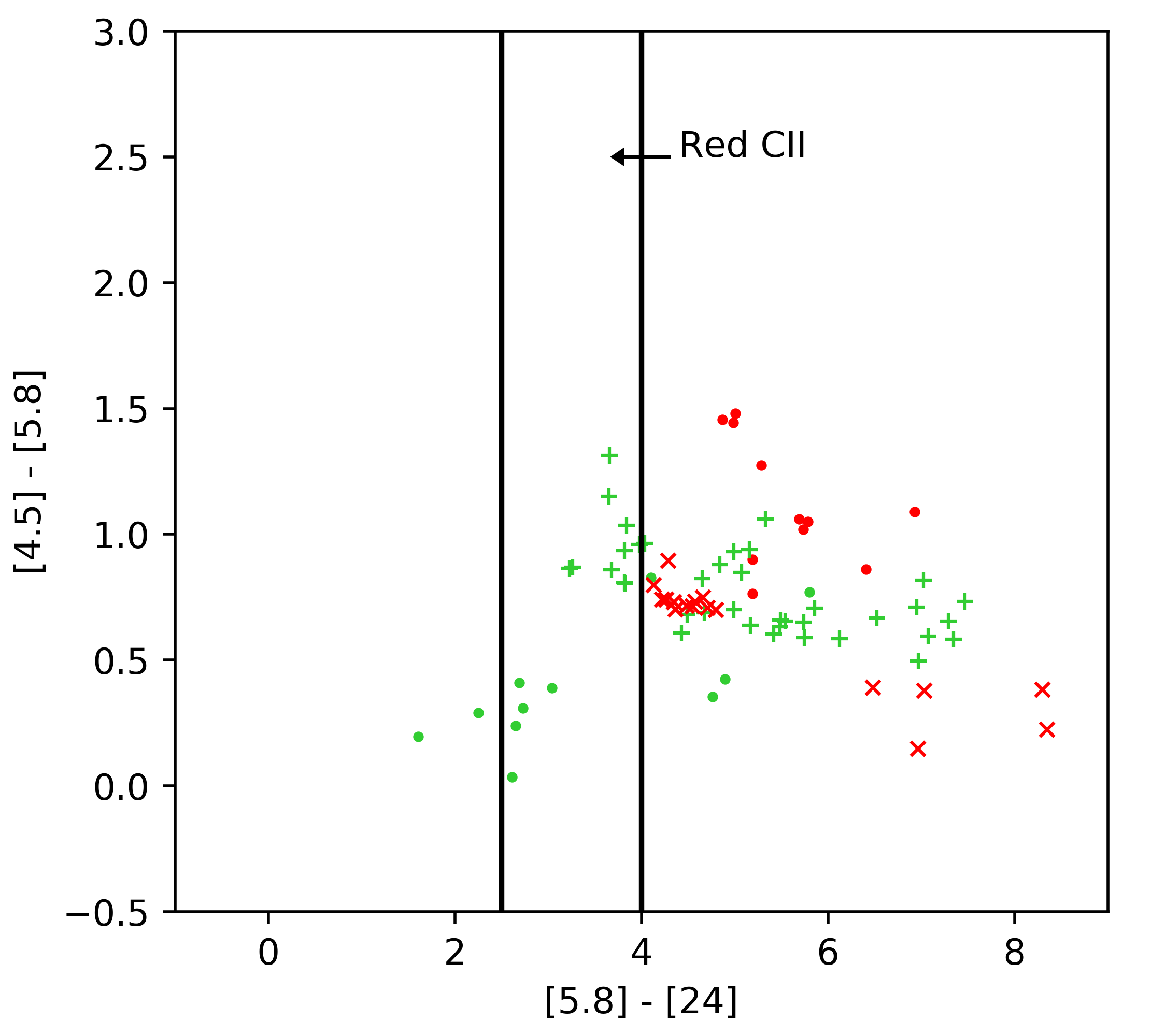}
        \end{subfigure}
        \begin{subfigure}[t]{0.24\textwidth}
        \includegraphics[width=\textwidth]{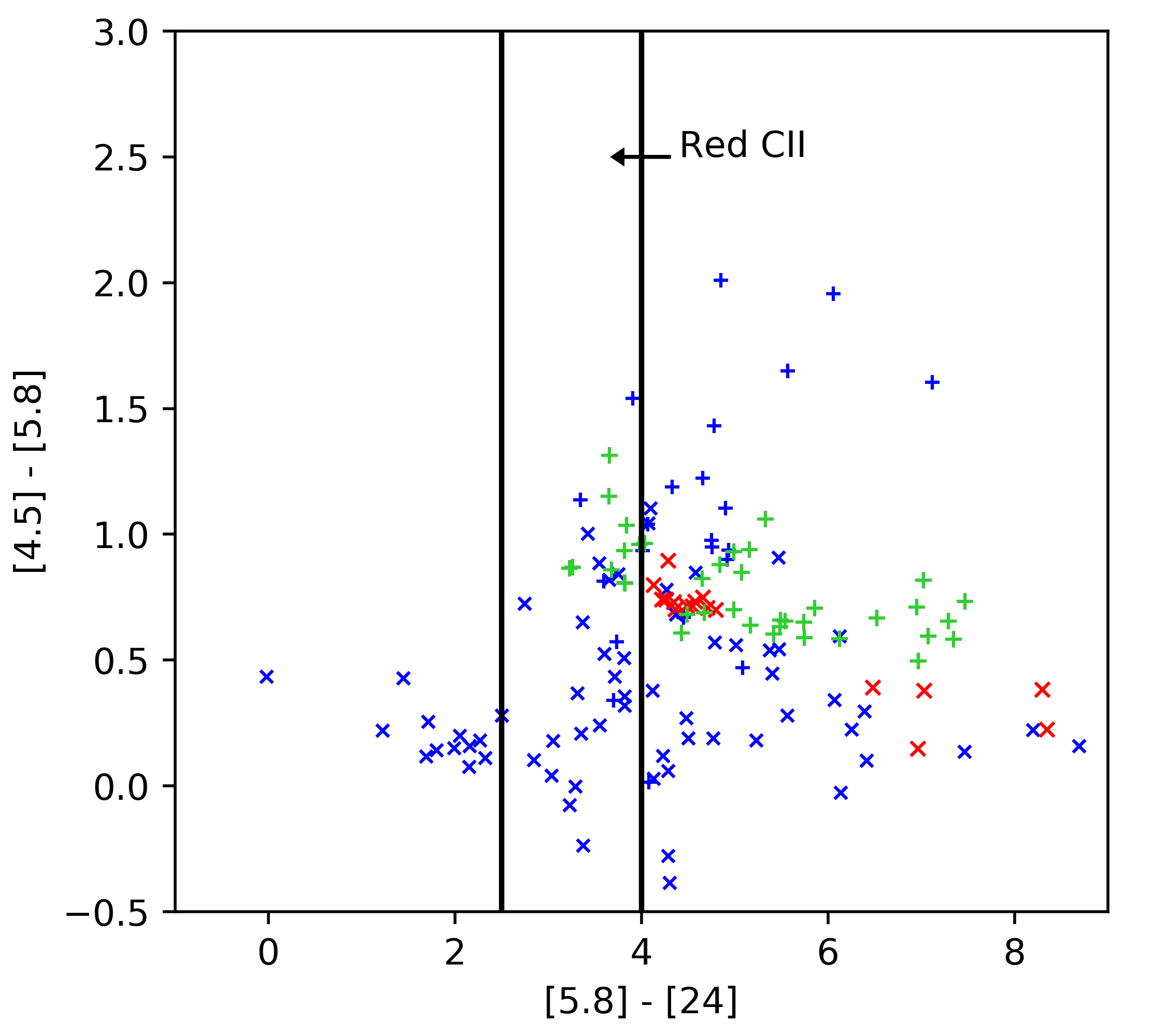}
        \end{subfigure}
        \caption{Input parameter space coverage in the CMDs used for the G09 method in the F-C case on the full dataset regarding different populations. \textit{Actual:} Distribution of genuine classes. CI YSOs are in red, CII YSOs are in green, and Others are in blue. \textit{Predicted:} Prediction given by the network with the same color-coding as for the \textit{actual} frames. { \textit{Missed:} Genuine CI and CII according to the labeled dataset that were misclassified by the network. Green is for genuine CII YSOs, red for genuine CI YSOs. The points and crosses indicate the network output as specified in the legend. \textit{Wrong:} YSO predictions of the network that are known to be incorrect based on the labeled dataset. Green is for genuine CII YSOs, red for genuine CI YSOs, and blue   for genuine contaminants. The two types of crosses indicate the predicted YSO class as specified in the legend.}}
        \label{missed_wrong_space} 
\end{figure*}

Figure~\ref{missed_wrong_space} presents a detailed comparison of the parameter space coverage for several of the classical CMDs used in the G09 classification for different objects in the F-C case (see Table~\ref{results_cases} and Sect.~\ref{1kpc_train}). The first and second columns of the figure represent the distribution of sources for the target class and the predicted class, respectively. The comparison of these two columns illustrates the global quality of our prediction and provides a reference point for the two other columns on the right-hand side, which show the objects that were misclassified by the network. In the {third and fourth columns} the colors encode the genuine class while the symbol shapes encode the prediction class. The third column focuses on genuine YSOs for which the prediction is incorrect. It is then possible to identify feature space regions where increasing the number of objects would significantly improve the recall of the corresponding class. In the fourth column the view is reversed; it focuses on the original class of misclassified objects. It reveals confused boundaries between classes that could be better constrained in order to improve the precision. Naturally, both the CI YSOs that were misclassified as CII and CII YSOs that were misclassified as CI are present in the two representations.

\section{YSO candidates catalog}

Table~\ref{yso_catalog} presents an excerpt of the YSO candidate catalog that is publicly available at the CDS. It is the result of our F-C trained network (see Sect.~\ref{1kpc_train}). The prediction is made only for objects from the Orion and NGC 2264 catalogs using our pre-selection criteria (Sect.~\ref{data_prep}). Our catalog lists the original catalog of each object, all the Spitzer bands and their uncertainties that were used as input features for the network, the target associated with each object using the subclasses, and the prediction of the network using our three classes (CI, CII, Others). Compared to the data published by \citet{megeath_spitzer_2012} for Orion and by \citet{rapson_spitzer_2014} for NGC 2264 applying the G09 method, we provide the membership probability for each object, making it possible to select objects according to the reliability of their classification. The membership probability is given for all the three output classes, enabling subsequent refinement of the classification following the prescriptions from Sect.~\ref{proba_discussion}.\\

\begin{sidewaystable*}
        \tiny
        \centering
        \caption{First 20 and last 20 elements of the catalog of network prediction in the F-C case using the full dataset.}
        \vspace{-0.1cm}
        \begin{tabularx}{0.93\hsize}{*{19}{c}}
        \toprule
        \toprule
        RA & DEC & Catalog & Orig. Class & 3.6   & e3.6  & 4.5   & e4.5  & 5.8   & e5.8  & 8.0   & e8.0  & 24    & e24   & Targ. & Pred. & P(CI) & P(CII) & P(Oth.)\\
        (deg)   & (deg)   &         &             & (mag) & (mag) & (mag) & (mag) & (mag) & (mag) & (mag) & (mag) & (mag) & (mag) &             &             &       &        & \\
        \vspace{-0.3cm}\\
        \toprule
100.792999 & +8.7531472   & 0 & III/F & 10.32 & 0.003 & 10.17 & 0.003  & 10.07 & 0.005  & 10.03 & 0.008 & \dots & \dots & 6 & 2 & 0.0         & 5.7e-5    & 0.9999\\
100.677625 & +8.7556250   & 0 & III/F & 11.62 & 0.003 & 11.60 & 0.004  & 11.55 & 0.016  & 11.56 & 0.035 & \dots & \dots & 6 & 2 & 0.0         & 0.0       & 1.0   \\
100.760958 & +8.7566528   & 0 & III/F & 13.38 & 0.007 & 13.28 & 0.011  & 13.27 & 0.05   & 13.78 & 0.155 & \dots & \dots & 6 & 2 & 0.0         & 0.0       & 1.0   \\
100.757875 & +8.7589389   & 0 & III/F & 12.52 & 0.005 & 12.52 & 0.006  & 12.4  & 0.03   & 12.41 & 0.053 & \dots & \dots & 6 & 2 & 0.0         & 0.0       & 1.0   \\
100.724500 & +8.7606944   & 0 & III/F & 13.71 & 0.009 & 13.66 & 0.013  & 13.6  & 0.069  & 13.67 & 0.148 & \dots & \dots & 6 & 2 & 0.0         & 0.0       & 1.0   \\
100.728917 & +8.7609722   & 0 & III/F & 13.23 & 0.007 & 13.17 & 0.008  & 12.99 & 0.042  & 13.16 & 0.081 & \dots & \dots & 6 & 2 & 0.0         & 0.0       & 1.0   \\
100.744958 & +8.7630750   & 0 & III/F & 11.28 & 0.003 & 11.37 & 0.004  & 11.32 & 0.011  & 11.22 & 0.027 & \dots & \dots & 6 & 2 & 0.0         & 0.0       & 1.0   \\
100.667167 & +8.7653722   & 0 & III/F & 13.5  & 0.015 & 13.36 & 0.029  & 13.39 & 0.075  & 13.43 & 0.11  & \dots & \dots & 6 & 2 & 2.0e-6      & 1.4e-4    & 0.9998\\
100.670250 & +8.7691222   & 0 & III/F & 8.64  & 0.002 &  8.54 & 0.002  & 8.4   & 0.002  & 8.36  & 0.002 & 8.29  & 0.043 & 6 & 2 & 0.0         & 5.9e-4    & 0.9994\\
100.792083 & +8.7692694   & 0 & III/F & 13.39 & 0.007 & 13.40 & 0.010  & 13.41 & 0.052  & 13.01 & 0.09  & \dots & \dots & 6 & 2 & 0.0         & 4.6e-5    & 0.9999\\
100.769292 & +8.7704556   & 0 & III/F & 12.93 & 0.006 & 12.87 & 0.008  & 12.89 & 0.038  & 12.7  & 0.058 & \dots & \dots & 6 & 2 & 0.0         & 3.0e-6    & 0.9999\\
100.757708 & +8.7710500   & 0 & III/F & 10.58 & 0.002 & 10.65 & 0.003  & 10.59 & 0.007  & 10.56 & 0.012 & \dots & \dots & 6 & 2 & 0.0         & 0.0       & 1.0   \\
100.811250 & +8.7714556   & 0 & III/F & 7.75  & 0.002 &  7.83 & 0.002  & 7.63  & 0.002  & 7.63  & 0.002 & \dots & \dots & 6 & 2 & 0.0         & 8.2e-3    & 0.9917\\
100.768208 & +8.7728194   & 0 & III/F & 13.83 & 0.009 & 13.80 & 0.014  & 13.99 & 0.088  & 13.56 & 0.117 & \dots & \dots & 6 & 2 & 0.0         & 0.0       & 1.0   \\
100.773667 & +8.7744222   & 0 & III/F & 11.68 & 0.004 & 11.87 & 0.004  & 11.61 & 0.012  & 11.61 & 0.032 & \dots & \dots & 6 & 2 & 0.0         & 0.0       & 1.0   \\
100.672208 & +8.7765889   & 0 & III/F & 13.37 & 0.007 & 13.31 & 0.010  & 13.32 & 0.055  & 13.21 & 0.089 & \dots & \dots & 6 & 2 & 0.0         & 1.0e-6    & 0.9999\\
100.768375 & +8.7775694   & 0 & III/F & 12.49 & 0.005 & 12.52 & 0.006  & 12.44 & 0.026  & 12.41 & 0.053 & \dots & \dots & 6 & 2 & 0.0         & 0.0       & 1.0   \\
100.697292 & +8.7783972   & 0 & III/F & 10.78 & 0.003 & 10.79 & 0.003  & 10.74 & 0.007  & 10.57 & 0.014 & \dots & \dots & 6 & 2 & 0.0         & 3.0e-6    & 0.9999\\
100.684208 & +8.7784639   & 0 & III/F & 12.87 & 0.005 & 12.85 & 0.007  & 12.81 & 0.033  & 12.74 & 0.065 & \dots & \dots & 6 & 2 & 0.0         & 0.0       & 1.0   \\
100.792542 & +8.7796389   & 0 & AGN   & 16.18 & 0.047 & 5.12  & 0.035  & 14.37 & 0.141  & 13.00 & 0.079 & \dots & \dots & 3 & 2 & 4.9e-5      & 2.0e-6    & 0.9999\\
\dots & \dots & \dots & \dots & \dots & \dots & \dots & \dots & \dots & \dots & \dots & \dots & \dots & \dots & \dots & \dots & \dots & \dots & \dots   \\
86.8015397 & -0.7217830   & 1 & Other & 13.79 & 0.011 & 13.73 & 0.017  & 13.66 & 0.106  & 13.54 & 0.153 & \dots & \dots & 6 & 2 & 0.0         & 1.0e-6    & 0.9999\\
86.7227924 & -0.7204420   & 1 & Other & 11.88 & 0.005 & 11.85 & 0.006  & 11.87 & 0.026  & 11.73 & 0.036 & \dots & \dots & 6 & 2 & 0.0         & 0.0       & 1.0   \\
86.7296191 & -0.7189594   & 1 & Other & 14.54 & 0.019 & 14.14 & 0.026  & 13.84 & 0.147  & 10.98 & 0.022 & 8.44  & 0.110 & 2 & 2 & 0.0         & 0.0       & 1.0   \\
86.6185832 & -0.7163786   & 1 & Other &  9.60 & 0.002 &  9.60 & 0.003  &  9.57 & 0.006  &  9.53 & 0.007 & \dots & \dots & 6 & 2 & 0.0         & 2.1e-5    & 0.9999\\
86.8822281 & -0.7111607   & 1 & Other & 13.16 & 0.009 & 13.13 & 0.012  & 13.02 & 0.056  & 13.00 & 0.110 & \dots & \dots & 6 & 2 & 0.0         & 2.0e-6    & 0.9999\\
86.8187251 & -0.7086041   & 1 & Other & 13.40 & 0.010 & 13.32 & 0.011  & 13.24 & 0.072  & 13.34 & 0.123 & \dots & \dots & 6 & 2 & 0.0         & 0.0       & 1.0   \\
86.8938200 & -0.7075397   & 1 & Other & 11.36 & 0.004 & 11.38 & 0.005  & 11.27 & 0.017  & 11.29 & 0.027 & \dots & \dots & 6 & 2 & 0.0         & 0.0       & 1.0   \\
86.7451751 & -0.7074037   & 1 & Other & 11.61 & 0.004 & 11.54 & 0.004  & 11.51 & 0.024  & 11.53 & 0.024 & \dots & \dots & 6 & 2 & 0.0         & 0.0       & 1.0   \\
86.6627309 & -0.7060398   & 1 & Other & 12.99 & 0.008 & 12.93 & 0.008  & 12.86 & 0.056  & 12.95 & 0.064 & \dots & \dots & 6 & 2 & 0.0         & 0.0       & 1.0   \\
86.6652294 & -0.7036116   & 1 & Other & 11.52 & 0.004 & 11.47 & 0.004  & 10.59 & 0.013  & 11.45 & 0.022 & \dots & \dots & 5 & 2 & 0.0         & 0.0       & 1.0   \\
86.6478710 & -0.7028939   & 1 & Other & 11.40 & 0.004 & 11.44 & 0.005  & 11.35 & 0.018  & 11.28 & 0.027 & \dots & \dots & 6 & 2 & 0.0         & 0.0       & 1.0   \\
86.6734974 & -0.7025317   & 1 & Other & 12.68 & 0.006 & 12.64 & 0.009  & 12.60 & 0.045  & 12.48 & 0.064 & \dots & \dots & 6 & 2 & 0.0         & 1.0e-6    & 0.9999\\
86.6593266 & -0.6985667   & 1 & Other & 13.54 & 0.010 & 13.50 & 0.015  & 13.55 & 0.095  & 13.36 & 0.147 & \dots & \dots & 6 & 2 & 0.0         & 0.0       & 1.0   \\
86.8586910 & -0.6948155   & 1 & Other & 12.77 & 0.007 & 12.75 & 0.010  & 12.63 & 0.043  & 12.72 & 0.077 & \dots & \dots & 6 & 2 & 0.0         & 0.0       & 1.0   \\
86.6522543 & -0.6913875   & 1 & Other &  8.95 & 0.007 &  8.95 & 0.002  &  8.87 & 0.004  &  8.83 & 0.004 & 8.94  & 0.168 & 6 & 2 & 1.0e-6      & 3.1e-3    & 0.9969\\
86.7718531 & -0.6903623   & 1 & Other & 13.83 & 0.014 & 13.78 & 0.020  & 13.63 & 0.104  & 13.65 & 0.184 & \dots & \dots & 6 & 2 & 0.0         & 1.0e-6    & 0.9999\\
86.8164438 & -0.6901289   & 1 & Other & 13.39 & 0.010 & 13.29 & 0.011  & 13.12 & 0.066  & 13.08 & 0.084 & \dots & \dots & 6 & 2 & 1.0e-6      & 2.5e-5    & 0.9999\\
86.7218342 & -0.6863691   & 1 & Other & 14.15 & 0.015 & 13.90 & 0.022  & 13.08 & 0.076  & 10.32 & 0.014 & 8.32  & 0.095 & 2 & 2 & 0.0         & 0.0       & 1.0   \\
86.6604287 & -0.6855791   & 1 & Other & 10.51 & 0.003 & 10.46 & 0.002  & 10.40 & 0.009  & 10.42 & 0.010 & \dots & \dots & 6 & 2 & 0.0         & 0.0       & 1.0   \\
86.8976492 & -0.6839529   & 1 & Other &  8.97 & 0.006 &  8.97 & 0.002  &  8.85 & 0.004  &  8.81 & 0.004 & 8.82  & 0.152 & 6 & 2 & 1.0e-6      & 3.0e-3    & 0.9970\\
        \bottomrule
        \end{tabularx}
        \tablefoot{The full catalog is publicly available at the CDS. The columns are: 
        (1-2) the source coordinates (J2000); 
        (3) the original catalog (0: \citet{megeath_spitzer_2012}, 1: \citet{rapson_spitzer_2014});
        (4) the original classification;
        (5-14) IRAC and MIPS magnitudes and corresponding uncertainties;
        (15) the target classification obtained with our simplified G09 scheme (0: CI YSOs, 1: CII YSOs, 2: Galaxies, 3: AGNs, 4: Shocks, 5: PAHs, 6: Stars);
        (16) the classification predicted by the ANN in the F-C case (0: CI YSOs, 1: CII YSOs, 2: contaminants);
        (17-19) the corresponding membership probabilities.
        }
        \vspace{-0.3cm}
        \label{yso_catalog}
\end{sidewaystable*}

\end{appendix}

\end{document}